\documentclass[twocolumn,showpacs,aps,prd,nobibnotes,nofootinbib,notitlepage,floatfix]{revtex4-1}

\usepackage{natbib}
\usepackage{graphicx,color}
\usepackage{amsmath,amssymb}
\usepackage{verbatim}

\usepackage{float}
\usepackage{wasysym}

\usepackage{hyperref,url}
\usepackage{multimedia}
\usepackage{enumerate}

\newcommand{\bra}[1]{\left( #1 \right)}

\newcommand{\apjl}{Astrophys.\ J.\ Lett.\ }
\newcommand{\mnras}{Mon.\ Not.\ R.\ Astron.\ Soc.\ }

\newcommand{\beq}{\begin{equation}}
\newcommand{\eeq}{\end{equation}}
\newcommand{\beqn}{\begin{eqnarray}}
\newcommand{\eeqn}{\end{eqnarray}}

\newcommand{\llabel}[1]{\label{#1}}              
\newcommand{\labeq}[2]{ \begin{equation} \llabel{#1}
{#2}
\end{equation}}

\begin{document}

\title{Accretion disks around binary black holes of unequal mass: \\ GRMHD simulations of postdecoupling and merger}

\author{Roman Gold${}^{1,2}$, Vasileios Paschalidis${}^{1,3}$, Milton Ruiz${}^1$, Stuart L. Shapiro${}^{1,4}$}
\affiliation{
${}^1$Department of Physics, University of Illinois at Urbana-Champaign, Urbana, IL~61801\\
${}^2$Department of Physics \& Joint Space-Science Institute, University of Maryland, College Park, MD~20742 \\
${}^3$Department of Physics, Princeton University, Princeton, NJ~08544 \\
${}^4$Department of Astronomy \& NCSA, University of Illinois at Urbana-Champaign, Urbana, IL~61801
\vspace{-10pt}
}
\author{Zachariah B. Etienne${}^{5}$ and Harald P. Pfeiffer${}^{6,7}$}
\affiliation{
${}^5$Department of Mathematics, West Virginia University, Morgantown, WV 26506 \\
${}^6$Canadian Institute for Theoretical Astrophysics, University of Toronto, Toronto, ON M5S 3H8, Canada\\
${}^7$Canadian Institute for Advanced Research, Toronto, ON M5G 1Z8, Canada
}

\bibliographystyle{apsrev4-1}

\begin{abstract}
We report results from simulations in general relativity of magnetized
disks accreting onto merging black hole binaries, starting from
relaxed disk initial data. The simulations feature an effective, rapid
radiative cooling scheme as a limiting case of future treatments with
radiative transfer. Here we evolve the systems after binary-disk
decoupling through inspiral and merger, and analyze the dependence on
the binary mass ratio with $q\equiv m_{\rm bh}/M_{\rm BH}=1,1/2,$ and
$1/4$. We find that the luminosity associated with local cooling is
larger than the luminosity associated with matter kinetic outflows,
while the electromagnetic (Poynting) luminosity associated with bulk
transport of magnetic field energy is the smallest. The cooling
luminosity around merger is only marginally smaller than that of a
single, non-spinning black hole. Incipient jets are launched
independently of the mass ratio, while the same initial disk accreting
on a single non-spinning black hole does not lead to a jet, as
expected. For all mass ratios we see a transient behavior in the
collimated, magnetized outflows lasting $2-5 \left( M/10^8M_\odot \right)
\rm days$ after merger: the outflows become increasingly magnetically
dominated and accelerated to higher velocities, boosting the Poynting
luminosity. These sudden changes can alter the electromagnetic
emission across the jet and potentially help distinguish mergers of
black holes in AGNs from single accreting black holes based on jet
morphology alone.
\end{abstract}

\maketitle

\section{Introduction and Motivation}

Mergers of supermassive black holes (SMBHs) and near-Eddington
accretion of gas \cite{Soltan1982} are both central ingredients in
theoretical models of the assembly of the SMBH population (see
e.g.~\cite{Tanaka:2008bv}). These models show a steadily growing
consistency with data from quasar surveys
\cite{Volonteri:2002vz,Tanaka:2008bv}, indicating that SMBH mergers
are not just a likely outcome of galaxy mergers, but {\it necessary}
to explain the BH mass distribution in the Universe.  When these
mergers occur following the collision of galaxies, they are expected
to be immersed in a magnetized plasma and surrounded by stars
\cite{Begelman:1980vb,Mayer:2013jja}. Due to both their gravitational
and electromagnetic (EM) radiation in their late evolutionary stages,
such systems are unique probes of spacetime and relativistic plasmas,
and are therefore interesting systems in General Relativity (GR),
relativistic astrophysics and multi-messenger astronomy.  Here we give
only a brief introduction to this topic and refer to the more detailed
discussion and additional references in \cite{Gold:2013zma}.

The first gravitational wave (GW) detectors \cite{LIGO_web}, sensitive
enough to observe GWs directly, will come online soon, but they
will not be sensitive to SMBH binaries. 
However, pulsar timing arrays (PTAs) \cite{Hobbs:2009yy} could detect
GWs both from individual SMBH binaries \cite{Sesana:2008xk} and the
stochastic background from unresolved SMBH binaries
\cite{Sesana:2012ak,Sesana:2008mz} within this decade.  Thanks to
encouraging improvements in data analysis
\cite{Taylor:2014iua,Wang:2014ava} and new discoveries in pulsar
timing, PTAs may soon detect (sub-pc) SMBH binaries in the
universe. Identifying EM signals from binary SMBHs will improve our
understanding of the cosmic evolution of SMBHs, especially if a
simultaneous GW signal from the same source is
detected~\cite{Sesana:2011zv,Tanaka:2011af,Tanaka:2013oju}.  Data from
EM surveys like the Sloan Digital Sky Survey (SDSS)
\cite{Volonteri:2009nh,Ju:2013hna} will probe larger redshifts with
time.
In addition, current and future EM detectors such as PanStarrs
\cite{Kaiser:2002zz}, the LSST \cite{Abell:2009aa} and WFIRST
\cite{Green:2012mj} will provide us with unprecedented data of
transient phenomena.

In general, but especially until GWs from SMBH binaries are detected,
it is crucial to have a thorough theoretical understanding of the full
nonlinear dynamics and radiation properties of these systems to make
the most out of EM observations. A key theoretical task is to
predict observational features that distinguish accretion flows onto
single versus binary SMBHs. From the point of view of transient signals,
the most interesting regime is when the black hole binary
merges. Modeling these systems through the late inspiral and merger
phases requires a fully relativistic calculation, i.e.~taking into
account the dynamical black hole spacetime, as well as treating
magnetized plasmas and radiative transfer relativistically.

Theoretical modeling of circumbinary disks is still in its infancy and
has shown that the evolution of a circumbinary disk is roughly
composed of three phases: i) the early inspiral {\it predecoupling}
phase, during which the disk viscous timescale ($t_{\rm vis}$) is
shorter than the gravitational wave timescale ($t_{\rm GW}$), and the
disk relaxes to a quasi-equilibrium state
as the BHBH slowly inspirals; ii) the {\it postdecoupling} phase,
during which $t_{\rm vis}> t_{\rm GW}$ and the binary decouples from
the disk before the disk can relax; iii) the post-merger or {\it
  re-brightening/afterglow} phase during which the disk begins to
refill the hollow left behind by the BHBH and accretion ramps up onto
the remnant BH. Simple, analytic considerations reveal that for
geometrically thick disks, {\it binary-disk decoupling} occurs during
the late stages of the binary inspiral. Analytic and semi-analytic
models focus on the geometrically-thin, optically thick disk case (see
e.g.~\cite{Haiman:2009te,Tanaka:2009iy,Shapiro:2009uy,Liu:2010mh,Kocsis:2012ui,Shapiro:2013qsa}
and references therein).
These 1D (semi)analytic studies make simplifying assumptions such as
the adoption of an azimuthally averaged formula for the binary tidal
torques, which may overestimate the tidal-torque barrier
\cite{Barausse:2014oca}, and also misses non-axisymmetric features such
as the formation of accretion streams.
Additional features, such as the presence of an inner cavity (hollow)
of lowered density near the binary, were revealed in hydrodynamic
studies in Newtonian gravity in 3D
\cite{Artymowicz:1994bw,Cuadra:2008xn,Rodig:2011jz,Roedig:2012nc} and
2D \cite{MacFadyen:2006jx,D'Orazio:2012nz,Farris:2014qma}. This
picture has been refined by the first (Newtonian) ideal
magnetohydrodynamic (MHD) treatments in 3D \cite{Shi:2011us} and in
Post-Newtonian gravity \cite{Noble:2012xz,Zilhao:2014ida}. Infalling
clouds onto and the subsequent disk formation around BHBHs has
recently been studied via Newtonian smoothed particle hydrodynamic
simulations in \cite{Dunhill:2014oka}. The dynamics of EM fields in
force-free electrodynamics in GR, but without modeling the disk
itself, has been studied in
\cite{Mosta:2009rr,Neilsen:2010ax,Palenzuela:2010xn,Palenzuela:2010nf,Alic:2012df}. GR
evolutions of geometrically thick disks have been achieved in
\cite{Bode:2009mt,Bogdanovic:2010he,Bode:2011tq,Farris:2011vx} and
only quite recently magnetohydrodynamic simulations of these systems
have been performed \cite{Farris:2012ux,Gold:2013zma} (see also
\cite{Giacomazzo:2012iv} for treatments of BHBHs inspiraling in
magnetized gaseous clouds). The overall conclusions are that both the
clearing of an inner cavity and the binary disk decoupling is at best
only partial: Some gas remains near the BHs all the way through
inspiral and merger. Note that these features in 3D MHD turbulent
disks are drastically different from findings in hydrodynamical,
geometrically thin disks \cite{MacFadyen:2006jx}, but see
\cite{Zilhao:2014ida}.
The transient
induced on the disk following the merger has been modeled
approximately by boosting either the BH or the disk, with the BH mass
suddenly reduced to mimic the energy loss through GWs, see
e.g.~\cite{Corrales:2009nv,Rossi:2009nk,Anderson:2009fa,
Megevand:2009yx,Zanotti:2010xs,Ponce:2011kv}. The
work presented here constitutes a substantial advancement over the
above treatments, because it (i) features a magnetized disk, which is
self-consistently evolved through pre- and postdecoupling, and finally
through merger and (ii) takes into account the dynamical spacetime in
full GR, with no artificial boundary conditions imposed to mimic the
role of a BH horizon.

In this paper we focus on the \textit{postdecoupling} phase including
the BHBH merger, as an extension of our results in the predecoupling
regime \cite{Gold:2013zma}. We consider the BHBH binary mass ratios
$q=m_{\rm bh}/M_{\rm BH}=$ 1:1, 1:2 and 1:4. We use relaxed matter and
magnetic field initial data, starting from the predecoupling epoch
obtained in \cite{Gold:2013zma}.  We consider geometrically thick
disks resembling slim disks \cite{Abramowicz1988,lrr-2013-1}. In our
models no physical scale is set by microphysics, resulting in the
scale freedom of our results both with binary ADM mass and the disk
density. However, we have in mind disks that accrete near the
Eddington limit and are dominated by radiation pressure. A key
purpose of this paper is to develop and test computational machinery
that will be required for a GRMHD treatment of the BHBH-disk problem
with full radiative transport.

The paper is structured as follows: In Sec. \ref{sec:methods} we
summarize the adopted techniques before reporting our results in
Sec. \ref{sec:results}. Section \ref{sec:q-trend} focuses on the
dependence on mass ratio, while Sec. \ref{sec:decoupling-merger} focuses
on universal features independent of the binary mass ratio.
Finally we discuss astrophysical implications of our results in
Sec. \ref{sec:implications} and conclude in Sec.
\ref{sec:conclusions}. Geometrized units, where $G=1=c$, are adopted 
throughout unless stated otherwise.

\section{Methods}
\label{sec:methods}

Our BHBH-disk models adopt the following set of assumptions and
simplifications: a) The non-spinning black hole binaries are initially
in quasi-circular orbits, b) the disk self-gravity is neglected
because we assume it is small compared to the gravity of the BHBH
binary, c) ideal MHD describes well the plasma in the disk, d) the
same effective emissivity employed in \cite{Gold:2013zma} [Eq.~(A2)]
(see also \cite{Paschalidis:2011ez}), with the same cooling time
scale, is adopted to model rapid cooling as a limiting case of
realistic cooling. See \cite{Gold:2013zma} for a more detailed
discussion and motivation for these simplifications.

\subsection{Initial data}
\label{sec:initialdata}

\subsubsection{Metric initial data}
\label{sec:tests1}
For the initial black hole binary spacetime geometry we adopt
conformal-thin-sandwich (CTS) solutions which correspond to
quasi-equilibrium black hole binaries in quasicircular orbits
~\cite{Pfeiffer:2002iy,Cook:2004kt,Caudill:2006hw,BSBook}. These
solutions possess a helical Killing vector. The CTS initial data have
been generated using the spectral techniques described in
\cite{Pfeiffer:2002wt} as implemented in the Spectral Einstein Code
(SpEC) \cite{SpECwebsite,Szilagyi:2014fna} (see also
\cite{Mroue:2013xna}). We list the initial data parameters describing
our spacetimes in Tab.~\ref{tab:cts_evolve}.

\begin{center}
 \begin{table}[th]
  \caption{CTS initial data parameters for the BHBH vacuum spacetime.
    Columns show mass ratio ($q$), ADM mass ($M_{\rm ADM}$), ADM
    angular momentum ($J_{\rm ADM}$), BH irreducible masses
    ($M^i_{irr}, \ i=1,2$), and apparent horizon radii ($r^i_{hor}$)
    for the two black holes. Diagnostics generating these quantities,
    but computed from independent vacuum test simulations, agree with
    these values to within one part in $10^4$.
\label{tab:cts_evolve}}
  \begin{tabular}{ccccccc} \hline\hline

 $q$   & $M_{\rm ADM}$ & $J_{\rm ADM}$ &  $M^1_{\rm irr}$ &  $M^2_{\rm irr}$ & $r^1_{\rm hor}$ & $r^2_{\rm hor}$ \\ \hline 
$1:1$  & 0.98989 & 0.96865 & 0.50000 & 0.50000 & 0.42958 & 0.42958 \\ 
$1:2$  & 0.99097 & 0.85933 & 0.66667 & 0.33333 & 0.60192 & 0.27312 \\ 
$1:4$  & 0.99342 & 0.61603 & 0.80000 & 0.20000 & 0.75140 & 0.15832 \\ 
\hline\hline 
  \end{tabular}
 \end{table}
\end{center}

We stress that the spectrally accurate, CTS initial data for the
spacetime metric have been mapped directly onto our computational
grids without requiring the lower-order interpolation from the
spherical auxiliary grids used in \cite{Gold:2013zma}.

\subsubsection{Matter and B-field initial data}
\label{sec:tests2}
For the magnetized fluid we use as initial data the relaxed end state
obtained in \cite{Gold:2013zma}, which started from equilibrium disk
models around single BHs as in
\cite{Chakrabarti:1985,DeVilliers:2003gr,Farris:2011vx,Farris:2012ux}
with an adiabatic index $\Gamma=4/3$, appropriate for thermal radiation
pressure-dominated disks. These disks correspond to
accretion flows driven by MHD turbulence, which is self-consistently
triggered by the magnetorotational instability \cite{Balbus:1991ay}.
These solutions are interpolated onto grids appropriate for a
spacetime evolution, that have the same spatial extent but contain
additional levels of refinement.


\subsection{Evolution equations and methods}

We use the GRMHD AMR code developed by the Illinois Numerical
Relativity Group 
\cite{Duez:2005sf,Etienne:2010ui,Etienne:2011re}, which adopts the
Cactus/Carpet infrastructure
\cite{Goodale2002a,Carpet-cactusweb,Carpet-carpetweb}, and includes an
effective radiative cooling scheme. This code has been extensively
tested and used in the past to study numerous systems involving
compact objects and/or magnetic fields (see
e.g.~\cite{Paschalidis2011a,Paschalidis:2011ez,Etienne:2011ea,Etienne:2012te,Paschalidis:2012ff,Etienne:2013qia,Paschalidis:2013jsa}),
including black hole binaries in gaseous media
\cite{Farris:2009mt,Farris:2011vx,Farris:2012ux} (see
\cite{Gold:2013zma} for additional references and details).

For the metric evolution we solve the BSSN equations
\cite{Shibata:1995,Baumgarte:1998te} coupled to the moving-puncture
gauge conditions, see Eqs.~(9)-(16) in \cite{Etienne:2007jg}. For the
1:4 case we use the spatially varying damping coefficient $\eta$
appearing in the shift condition, as was done in the case of the
\verb!LEAN!-code NRAR runs \cite{Hinder:2013oqa}. See
\cite{Lousto:2010ut,Muller:2010zze} for a motivation of similar
strategies.

We adopt a number of diagnostics to analyze accretion disks onto
binary black holes. For brevity here we describe only those
diagnostics that characterize the outgoing flow of energy which
include: a) the Poynting luminosity $L_{\rm (EM)}=\oint_S
T_{0,}{}^r_{\rm (EM)}dS$, where $T_{\mu,}{}^\nu_{\rm (EM)}$ is the EM
stress-energy tensor.  $L_{\rm (EM)}$ measures the outgoing flux of
large scale EM energy.  b) The cooling luminosity $L_{\rm
  cool}=\Lambda u_0\alpha\sqrt{\gamma}d^3x$, where $\Lambda$ is the
cooling emissivity, $u^\mu$ is the fluid four-velocity, and $\gamma$
is the determinant of the 3-metric. $L_{\rm cool}$ measures the total
thermal emission. c) The kinetic luminosity $L_{\rm kin}=\oint_S
T_{0,}{}^r_{\rm (fluid)}dS$ computed for unbound ($E=-u_0-1>0$)
material, where $T_{\mu,}{}^\nu_{\rm (fluid)}$ is the perfect fluid
stress-energy tensor. $L_{\rm kin}$ measures the outgoing flux of
kinetic energy carried by unbound matter. We also compute the gas
Lorentz factors (measured by a normal observer) of the plasma
$W=\alpha u^0$ in the funnel region before and after the outflow
settles following merger. Here, $\alpha$ is the lapse function. For
definitions of all other diagnostics we adopt in this work see
\cite{Farris:2012ux,Gold:2013zma}.

\begin{center}
 \begin{table*}[th]
  \caption{List of grid parameters for all models. Equatorial
    symmetry is imposed in all cases. The computational mesh consists
    of three sets of nested AMR grids, one centered on each BH and one
    in between (with $7$ levels of refinement for all cases), with the outer
    boundary at $240$M in all cases. From left to right the columns
    indicate the case name, mass ratio $q$, 
    the coarsest grid spacing $\Delta x_{\rm max}$,
    number of AMR levels around the primary (BH) and the secondary
    (bh), and the half length of each AMR box centered on each BH. The
    grid spacing of all other levels is $\Delta x_{\rm
      max}/2^{n-1},\ n = 1, 2,\ldots$, where $n$ is the
    level number such that $n = 1$ corresponds to the coarsest
    level. A dash ``--''~indicates ``not applicable''.
    \label{tab:models}}
  \begin{tabular}{cccccl} \hline\hline
Case name   & $q$ & $\Delta x_{\rm max}$ &\ levels(BH)\ &\ levels(bh)\ & Grid hierarchy \\ \hline
1:1 & 1:1 & $6.0M$              & 9 & 9  & $240M/2^{n-1},\ n =2,\ldots 5$, $240M/2^{n},\ n =6,\ldots,9$ \\
1:2 & 1:2 & $6.0M$              & 9 & 10 & $240M/2^{n-1},\ n =2,\ldots 5$, $240M/2^{n},\ n =6,\ldots,10$ \\
1:4 & 1:4 & $6.\overline{486}M$ & 9 & 11 & $240M/2^{n-1},\ n =2,\ldots 5$, $240M/2^{n},\ n =6,\ldots, 11$\\ 
0   & 0   & $6.0M$              & 6 & -- & $240M/2^{n-1},\ n =2,\ldots 5$, $240M/2^{n},\ n =6$ \\ \hline\hline
  \end{tabular}
 \end{table*}
\end{center}
The grids are similar to those in \cite{Gold:2013zma}, with additional
finer AMR levels centered on each BH and increased resolution in
between the BHs. The higher resolution is needed for a reliable metric
evolution through inspiral and merger. The regridding procedure makes
use of the Cactus/Carpet interpolation routines and is identical to
the procedure used in \cite{Farris:2012ux,Gold:2013zma}.

In Tab.~\ref{tab:models} we list the distinguishing characteristics
and grid-hierarchy of the different cases we consider in this
work. The labels are chosen to designate the mass ratio, e.g. the
label 1:1 means mass ratio $q=1$. We also evolve the same initial disk
model with a single, non-spinning BH (case 0) to normalize some of our
results and for comparisons to the binary cases.

We stress that the study of these systems over the duration of all epochs
(from predecoupling to re-brightening) requires some of the longest
GRMHD evolutions in full GR performed to date: The inner disk
structure relaxes during the predecoupling epoch approximately on a
viscous time scale at the inner disk edge, given by
\begin{align}
  \label{eq:tvisc}
  \! \! \frac{t_{\rm vis}}{M}\! =\! \frac{2R_{\rm in}^2}{3\nu M}\! \sim 6500 \left(\frac{R_{\rm in}}{18M}\right)^{3/2} \!\left(\frac{\alpha_{\rm ss}}{0.13}\right)^{-1}\! \left(\frac{H/R}{0.3}\right)^{-2}\!\!\!.
\end{align}
Here $R$ is the disk (cylindrical) radius, $H$ is the disk scale
height, and $\nu$ is the effective kinematic viscosity driven by MHD
turbulence, which can be expressed as $\nu(R) \equiv (2/3)\alpha_{\rm ss}
(P/\rho_0) \Omega_K^{-1}\approx (2/3)\alpha_{\rm ss} (R/M)^{1/2}(H/R)^2 M$. We
have assumed hydrostatic equilibrium in the vertical direction to
derive an approximate relationship between $P/\rho_0$ and $H/R$ (see
\cite{shapiro_book_83}). The effective viscosity in our disks can be
fit (approximately) to an ``$\alpha$-disk'' law for purposes of
analytic estimates, and we use typical $\alpha_{\rm ss}$ values found in our
evolutions. In \cite{Gold:2013zma} we empirically found that the
relaxation at the inner edge of the same, geometrically thick disks
takes $\sim 5000M$, which is consistent with the order-of-magnitude
estimate of Eq.~\eqref{eq:tvisc} \footnote{Due to the steep, inverse
  dependence of the viscous time scale on scale height, this time
  scale becomes prohibitively long for thinner disks}.
The subsequent inspiral occurs on a GW timescale \cite{Peters:1964}
\begin{align}
\label{eq:tgw}
  \frac{t_{GW}}{M} \sim 3000 \left({\frac{a}{10M}}\right)^4 \tilde{\eta}^{-1},
\end{align}
where $a$ is the initial binary separation, and $\tilde{\eta}\equiv
4\eta \equiv 4q/(1+q)^2$ is the symmetric mass ratio. The normalization
of $a$ is close to our initial binary coordinate separation and is
chosen to be close to the decoupling radius as shown in
\cite{Gold:2013zma}. Note that an equal-mass binary ($q=1$) has
$\tilde{\eta} = 1$, while $q=1/4$ yields $\tilde{\eta} \sim 0.64$.

The inspiral epoch is the shortest epoch, but requires the highest
resolution in order to track the inspiral reliably. The duration until
the remnant disk viscously refills the inner cavity and accretes onto
the merger remnant is largely determined by the radial matter
distribution at decoupling and is expected to occur on a viscous time
[Eq.~\eqref{eq:tvisc}].

The disparity between the duration of the predecoupling (inspiral and
merger) epoch and the dynamical (light-crossing) time scale across
the horizon, where the latter determines the smallest time step in our
explicit time integrations, makes these evolutions expensive and very
time-consuming.

In summary, the minimum total simulation time for the computationally
least expensive case (1:1) is $> 15,000M$. We have simulated the
predecoupling (see \cite{Gold:2013zma}), inspiral and merger epoch for
a total of $\sim 13,000M$. The postdecoupling phase of our most
expensive case (1:4) took approximately $2$ months (of wallclock time)
to finish at a cost of $\sim 200,000$ CPU hours.






\section{Results}
\label{sec:results}


\subsection{Trend with mass ratio}
\label{sec:q-trend}

In this section we discuss the dependence of our multiple diagnostics
on the binary mass ratio $q$. Results independent of $q$ are presented
in Secs.~\ref{sec:decoupling-merger} and \ref{sec:implications}.  For
additional definitions of diagnostic quantities see
\cite{Gold:2013zma}.

\begin{center}
 \begin{table}
  \caption{Columns show case name, the total accretion rate $\dot{M}$
    at merger $t_m$ normalized to the mean accretion rate for a single
    BH with the same cooling prescription $\langle\dot{M}_{q=0}
    \rangle$,
    $\epsilon_{\rm EM}\equiv L_{\rm EM}/\dot{M}_{q=0}$, $\epsilon_{\rm
      cool}\equiv L_{\rm cool}/\dot{M}_{q=0}$, $\epsilon_{\rm
      kin}\equiv L_{\rm kin}/\dot{M}_{q=0}$, and the 99th percentile
    of the gas Lorentz factors $W=\alpha u^0$ in the funnel region
    after the outflow settles following merger.  $L_{\rm EM}$ and
    $L_{\rm kin}$ are computed through surface integrals over a
    spherical surface of coordinate radius $90M$. See also the
    description at the end of Sec.~\ref{sec:methods}.
    Values (except for $W$) are reported at
    merger. Based on the resolution study presented in
    \cite{Gold:2013zma} we estimate the error of the quantities listed
    in the table to be $\sim 50\%$.
      \label{tab:results}}
  \begin{tabular}{cccccc} \hline\hline
  Case & $\dot{M} / \langle\dot{M}_{q=0}\rangle$\footnote{$\langle\dot{M}_{q=0}\rangle=3.05\left(\frac{\rho_{0}}{10^{-11}\rm gr/cm^3}\right)\left(\frac{M}{10^8M_\odot}\right)^2\rm M_\odot\ yr^{-1}$} & $\epsilon_{\rm EM}$
  &$\epsilon_{\rm cool}$ & $\epsilon_{\rm kin}$ & $W$ \\\hline
 1:1   & $0.025$ & $0.00081$& $0.059$ & $0.010$ & $2.4$ \\
 1:2   & $0.027$ & $0.00063$& $0.046$ & $0.020$ & $2.0$ \\
 1:4   & $0.064$ & $0.00278$& $0.042$ & $0.029$ & $1.6$ \\
 0     & $1.0$   & $0.002$  & $0.115$ & $0.04$  & $1.0$\footnote{Near the funnel walls we find a wind-like outflow with substantially smaller velocities $W\sim 1.1$ than the funnel regions in the binary cases.} \\\hline\hline
  \end{tabular}
 \end{table}
\end{center}



\subsubsection{Evolution of the density}
\label{sec:sigma}

\begin{figure*}[t]
  \centering
  \includegraphics[width=0.32\textwidth]{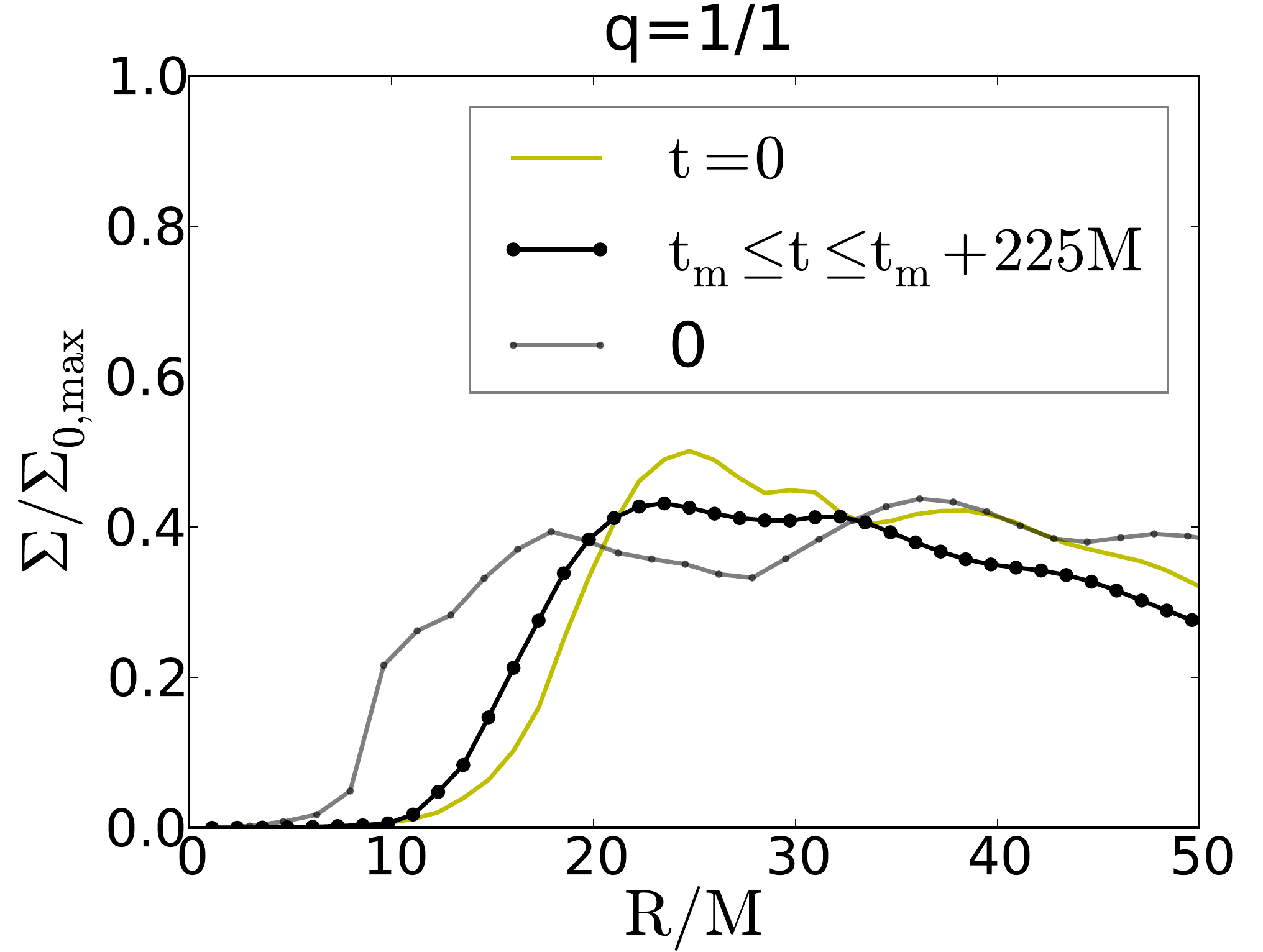}
  \includegraphics[width=0.32\textwidth]{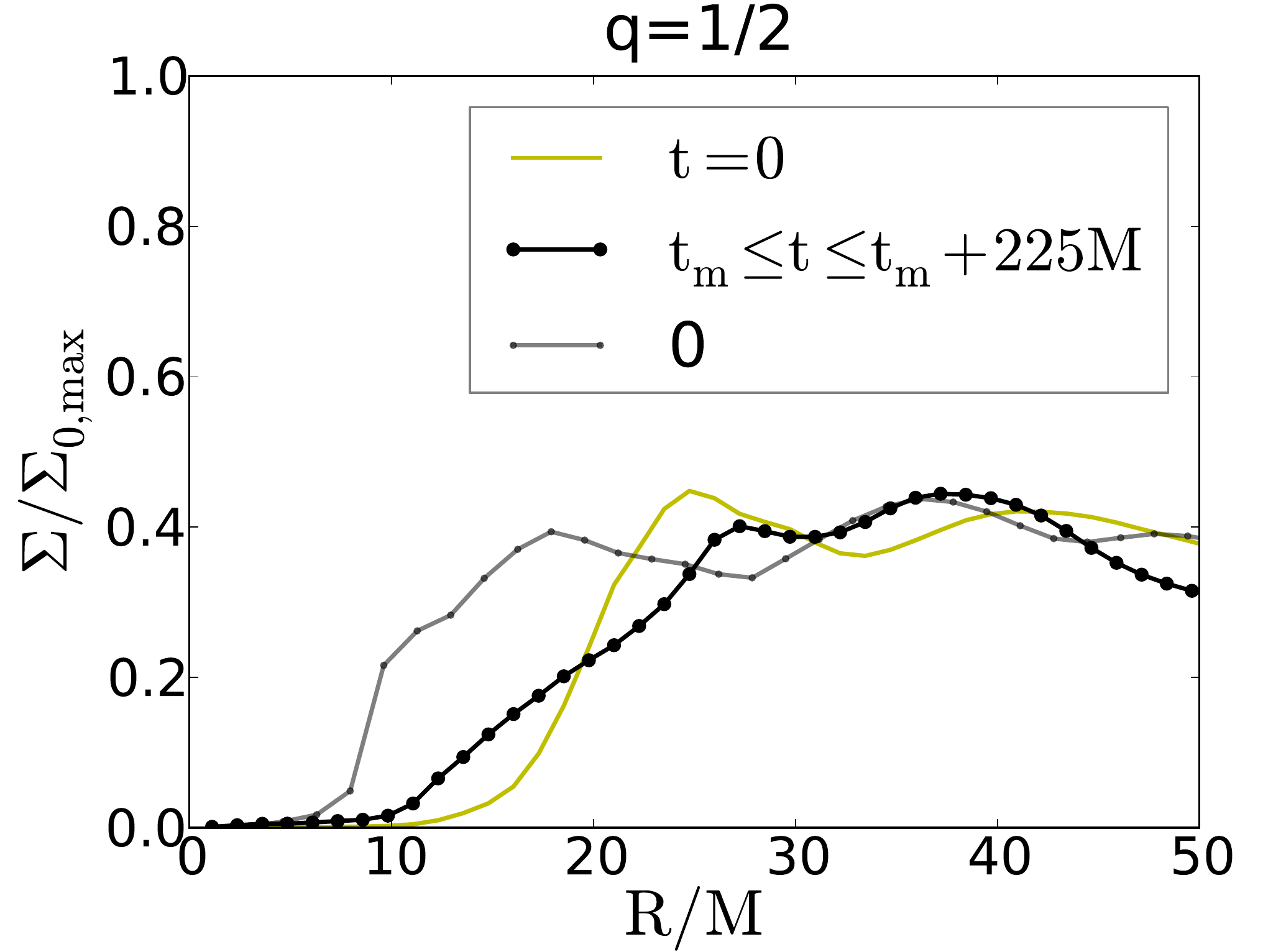}
  \includegraphics[width=0.32\textwidth]{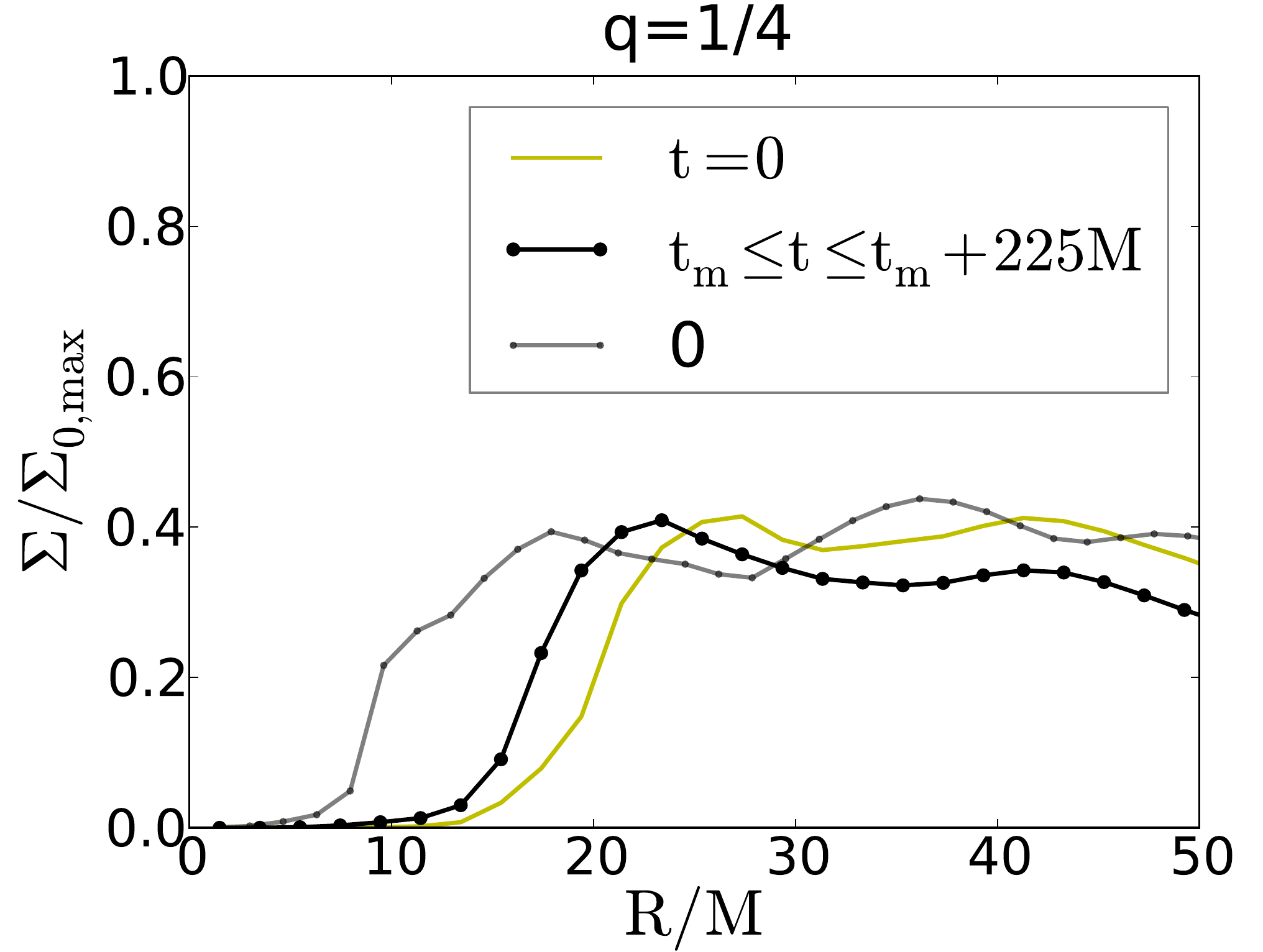}
            \caption{Surface density $\Sigma$ at decoupling (yellow
              solid lines) and immediately after merger, time-averaged
              over $225M$ (corresponding to the initial binary orbital
              period) (black solid line with circles). $\Sigma(r)$ is
              normalized to $\Sigma_{0,max}$ -- the maximum value of
              the surface density in the initial hydrostatic
              equilibrium solution used in \cite{Gold:2013zma}.
              In each panel the profile for the reference stationary single BH
              case is shown (gray solid line with dots). 
            \label{fig:sigma}} 
\end{figure*}

\begin{figure*}[t]
  \centering
  \includegraphics[width=0.99\textwidth]{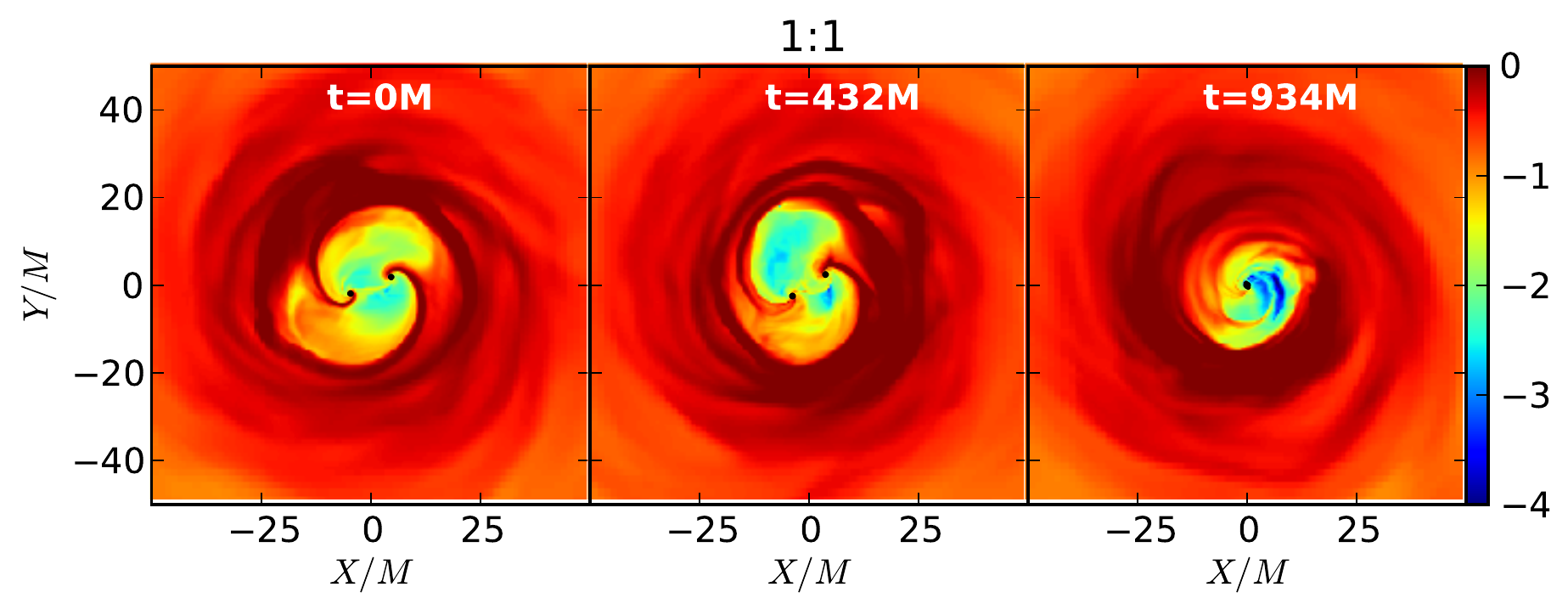}
  \includegraphics[width=0.99\textwidth]{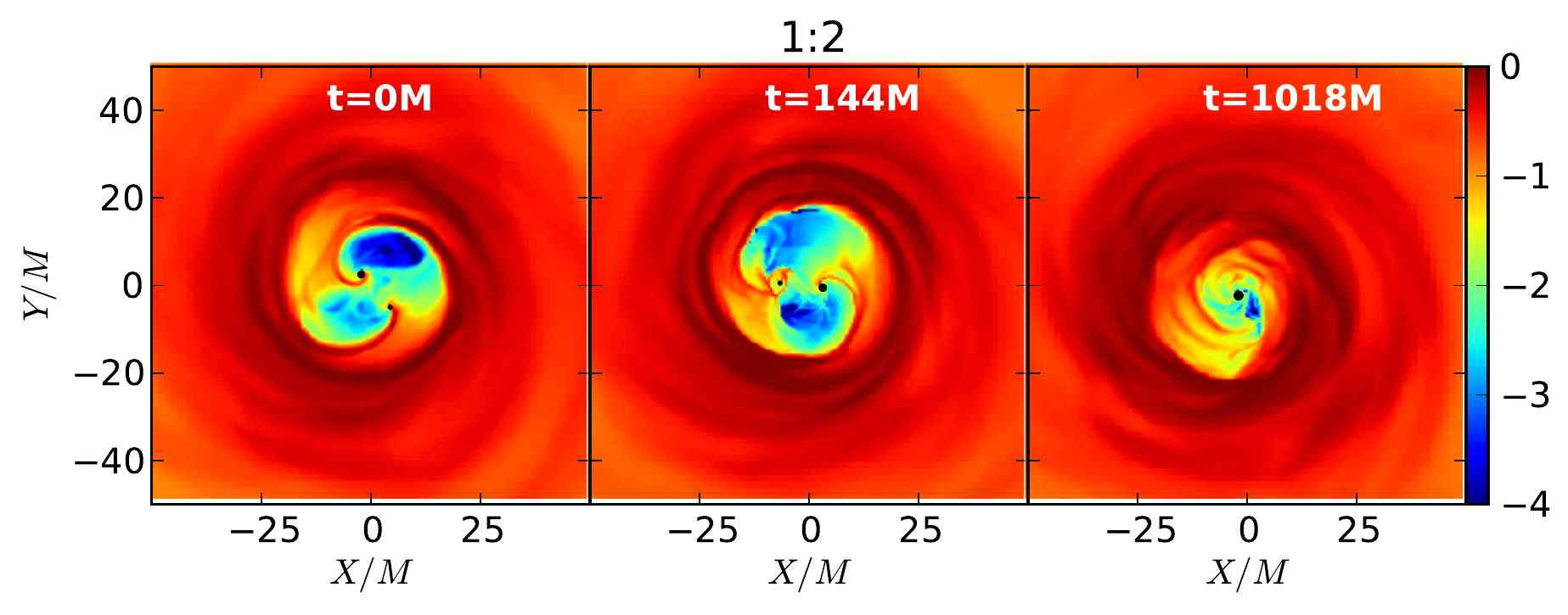}
  \includegraphics[width=0.99\textwidth]{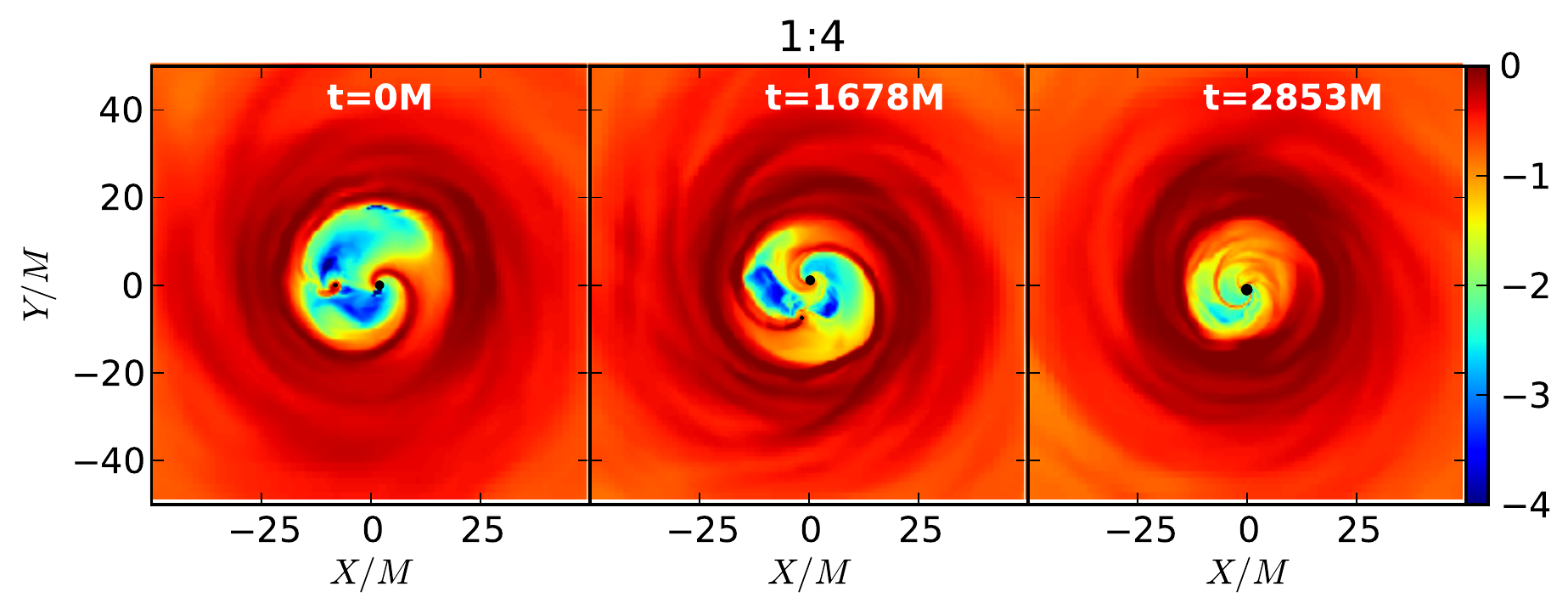}
            \caption{Contours of rest-mass density $\rho_0$ normalized to the
              initial maximum $\rho_{\rm max}$ (log scale) in the
              equatorial plane corresponding to the time at decoupling
              (left), an intermediate time prior to merger (middle),
              and the moment following merger (right). Upper panels:
              Mass ratio 1:1; $\rho_{\rm max} \simeq 2.1 \times 10^{-11}
              \bra{\frac{L_{\rm b}}{L_{\rm Edd}}}\bra{\frac{M}{10^8
                  M_\odot}}^{-1} \bra{\frac{\epsilon}{0.13}}^{-1} \rm
              g\ cm^{-3\ }$.  Middle panels: Mass ratio 1:2; $\rho_{\rm max}
              \simeq 4.2 \times 10^{-11} \bra{\frac{L_{\rm b}}{L_{\rm
                    Edd}}}\bra{\frac{M}{10^8 M_\odot}}^{-1}
              \bra{\frac{\epsilon}{0.08}}^{-1}\rm g\ cm^{-3}$.  Lower
              panels: Mass ratio 1:4; $\rho_{\rm max} \simeq 3.75 \times 10^{-11}
              \bra{\frac{L_{\rm b}}{L_{\rm Edd}}}\bra{\frac{M}{10^8
                  M_\odot}}^{-1} \rm g\ cm^{-3}$. $L_{\rm b} = L_{\rm
                EM}+L_{\rm cool}$ is the bolometric radiative
              luminosity at decoupling. \label{fig:snapshots}}
\end{figure*}

During the predecoupling phase, which furnishes the initial data for
our postdecoupling evolutions, our disk models contain some matter in
the inner cavity mainly in the form of dense accretion streams. The
surface density $\Sigma(r)$ depends on the mass ratio
\cite{Gold:2013zma}. The distribution of material at decoupling
determines the subsequent evolution through inspiral, merger and
re-brightening. The evolution of $\Sigma(r)$ is shown in
Fig.~\ref{fig:sigma}, and the rest-mass density $\rho_0$ on the
equatorial plane is plotted in Fig.~\ref{fig:snapshots}. We observe
accretion streams of dense gas attached to the BHs, as reported in
\cite{Gold:2013zma}, during the inspiral and through merger. During
inspiral there are always two diametrically opposite accretion streams
as in the earlier predecoupling phase \cite{Gold:2013zma}. We
therefore call this an $m=2$ mode by analogy to the terminology used
for spiral density waves in other accretion disk studies,
e.g.~\cite{MacFadyen:2006jx}. As already observed in
\cite{Gold:2013zma} there is an asymmetry between the two accretion
streams (one stream is larger than the other) as $q$ departs from
unity.

Spiral arms are observed throughout merger. However, the pre-merger
$m=2$ mode ceases to dominate (see Fig. \ref{fig:snapshots}) for
any case. Instead higher modes are excited which mostly decay over a
few $100M$ after merger.  Following merger matter begins drifting
inward for all binary cases. A ``lump'' feature reported in earlier
work \cite{Shi:2011us,Noble:2012xz,Gold:2013zma} is also seen for the
1:1 case (see the densest regions in the upper panel in
Fig. \ref{fig:snapshots}) but is hardly noticeable in the other cases.

Several studies
\cite{Sesana:2011zv,Farris:2013uqa,Yan:2014jva,Farris:2014iga,Farris:2014qma}
investigated ``mini disks'' around each BH. While there is no
universal definition of a mini disk, we consider a {\it persistent}
mini disk to be a coherent density structure within the Hill sphere of
each BH with an accretion time scale longer than a binary orbital
period. For the systems under consideration here, we do not find {\it
  persistent} mini disks. Instead, occasionally matter piles up around
the individual BHs before it is accreted. A necessary condition for
{\it persistent} ``mini disks'' to form is that the Hill sphere (or
Roche lobe) $r_{\rm Hill}$ be significantly larger than the innermost
stable circular orbit $r_{\rm ISCO}$ around the individual BH $r_{\rm
  Hill}>r_{\rm ISCO}$. A simple Newtonian estimate for the secondary
BH yields $r_{\rm Hill} \equiv (a/2) (q/3)^{1/3} \sim 3.5M \bra{a/10M}
q^{1/3}$ and $r_{\rm ISCO}\sim 3.0M q \sim r_{\rm Hill}$. These simple
estimates demonstrate why we only see {\it transient}
mini-disks. Given the mass ratio $q$ we expect ``mini disks'' around
the (non-spinning) secondary BH to form at binary separations $a_{\rm
  mini-disk} \gg 8.7M q^{2/3}$. The initial binary separations in our
evolutions are $d\sim 10M$ and thus too close for persistent
mini-disks to form. Note that for the geometrically thick disks we
consider here, the decoupling separation ($a_d$) is given by
\cite{Gold:2013zma} \labeq{}{ a_d\simeq 12M\bigg(\frac{\alpha_{\rm
      ss}}{0.13}\bigg)^{-2/5}\bigg(\frac{H/R}{0.3}\bigg)^{-4/5}\tilde\eta^{2/5}.
}
Thus, initial separations larger than the decoupling radius may be
necessary for persistent mini disks to form in geometrically thick
accretion flows.

\begin{figure}[ht]
  \centering
  \includegraphics[width=0.49\textwidth]{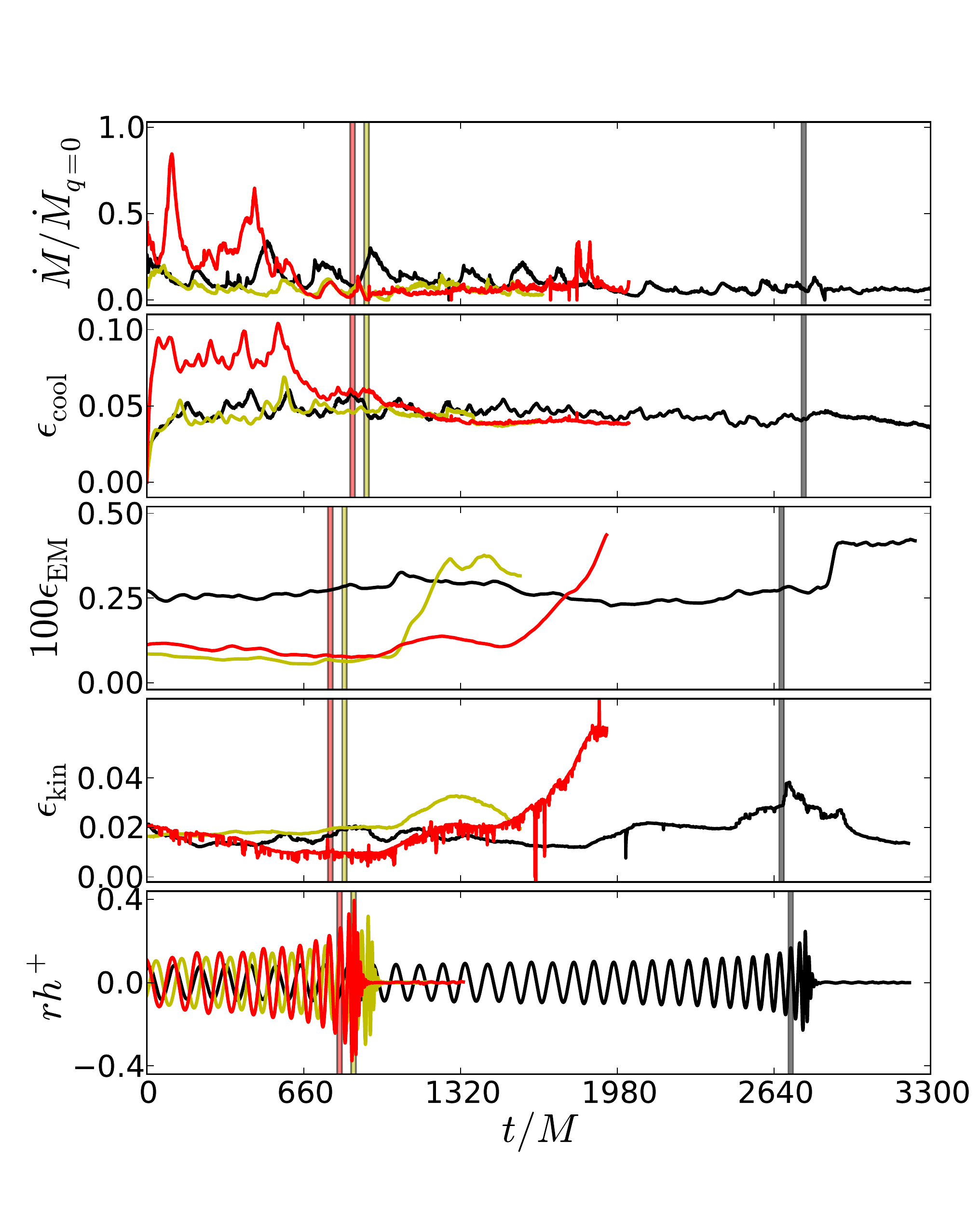}
            \caption{Total accretion rate, luminosities normalized to
              the single BH accretion rate, and GW strain ($+$
              polarization) as functions of time. For $\epsilon_{\rm
                EM}$, $\epsilon_{\rm kin}$ and $h^+$, $t$ is the {\it
                retarded} time. Red solid line: 1:1; yellow solid
              line: 1:2; black solid line: 1:4. The vertical lines
              indicate the corresponding merger times.
            \label{fig:multimessenger}}
\end{figure}

\begin{figure*}[t]
  \centering
  \includegraphics[width=0.31\textwidth]{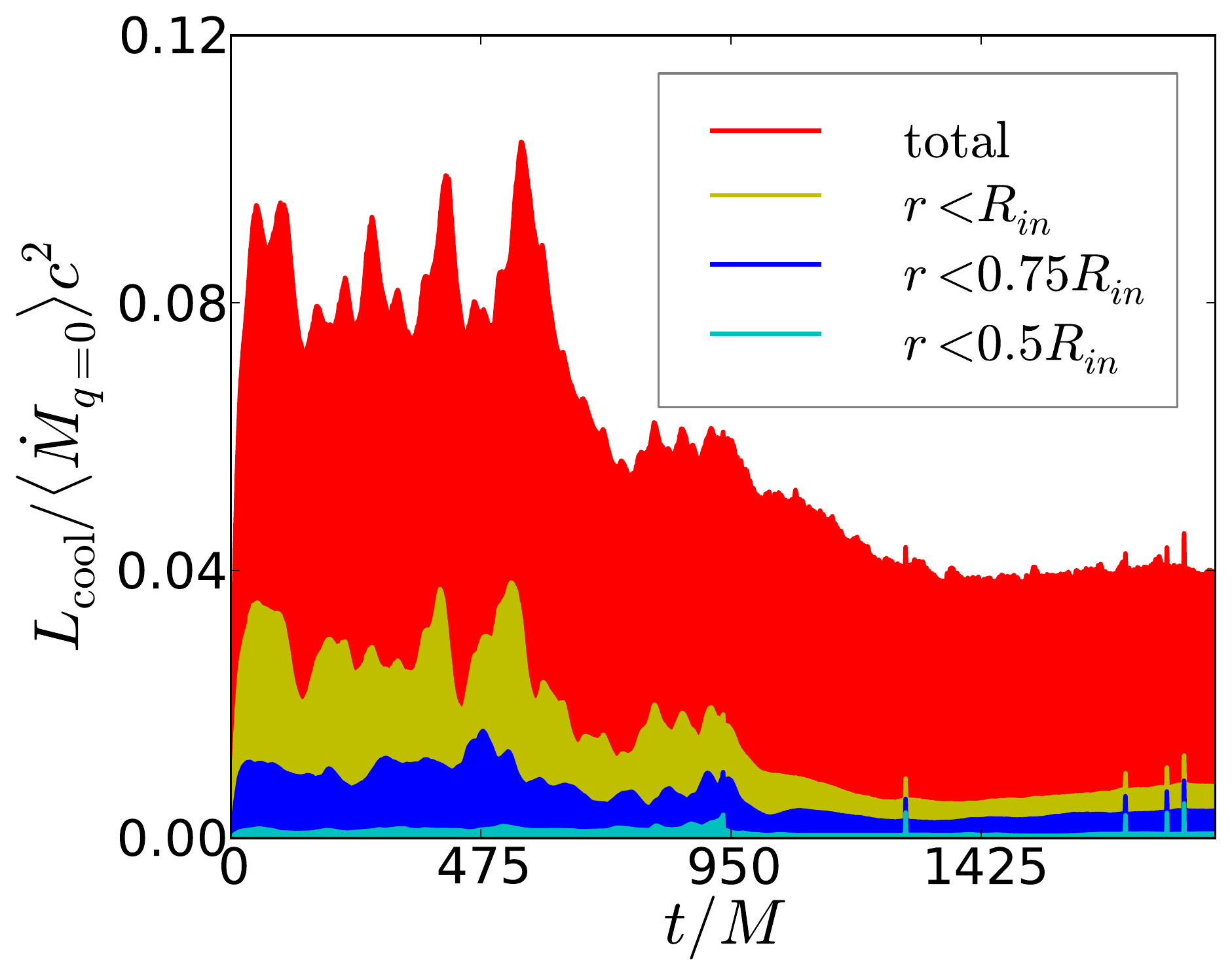}
  \includegraphics[width=0.32\textwidth]{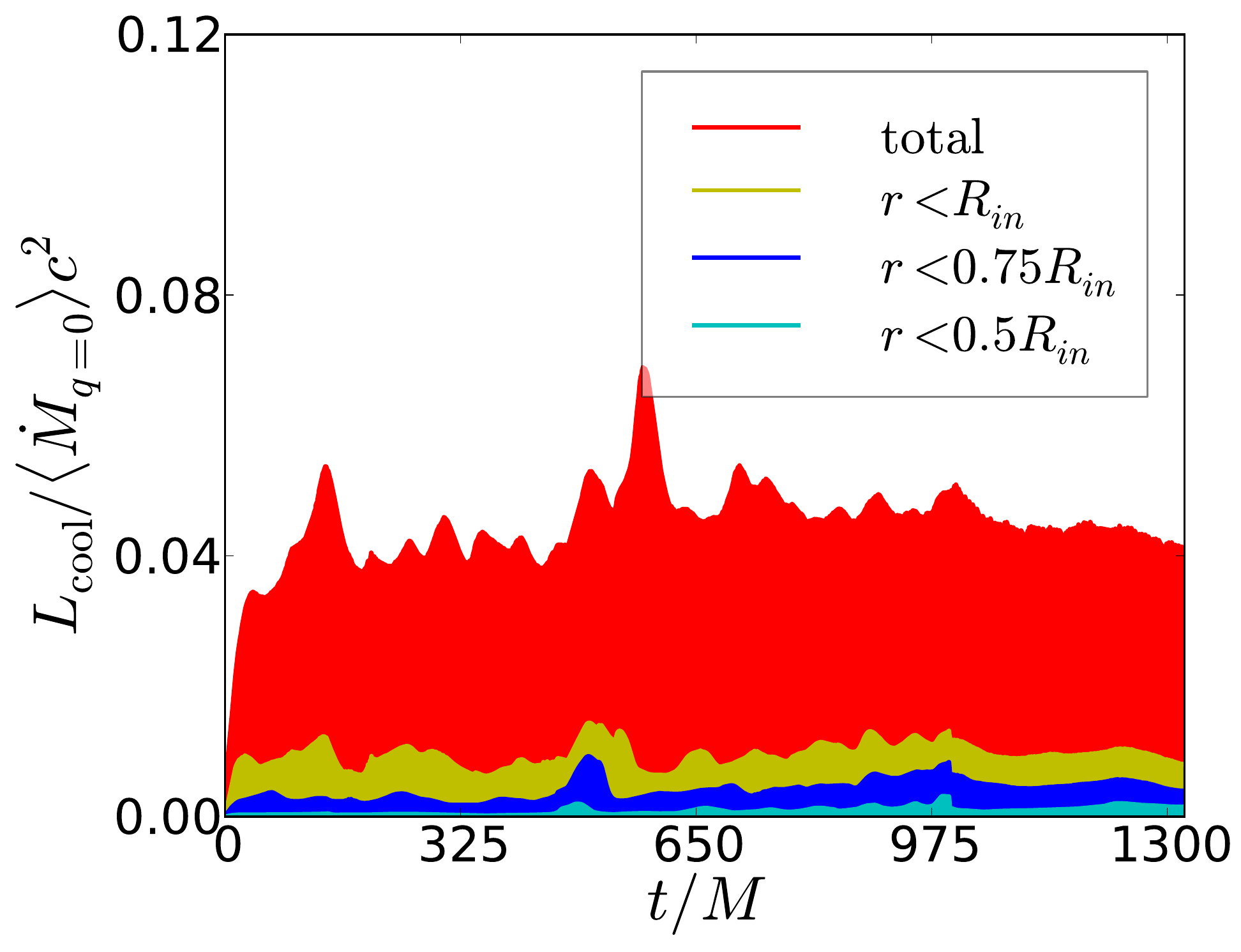}
  \includegraphics[width=0.32\textwidth]{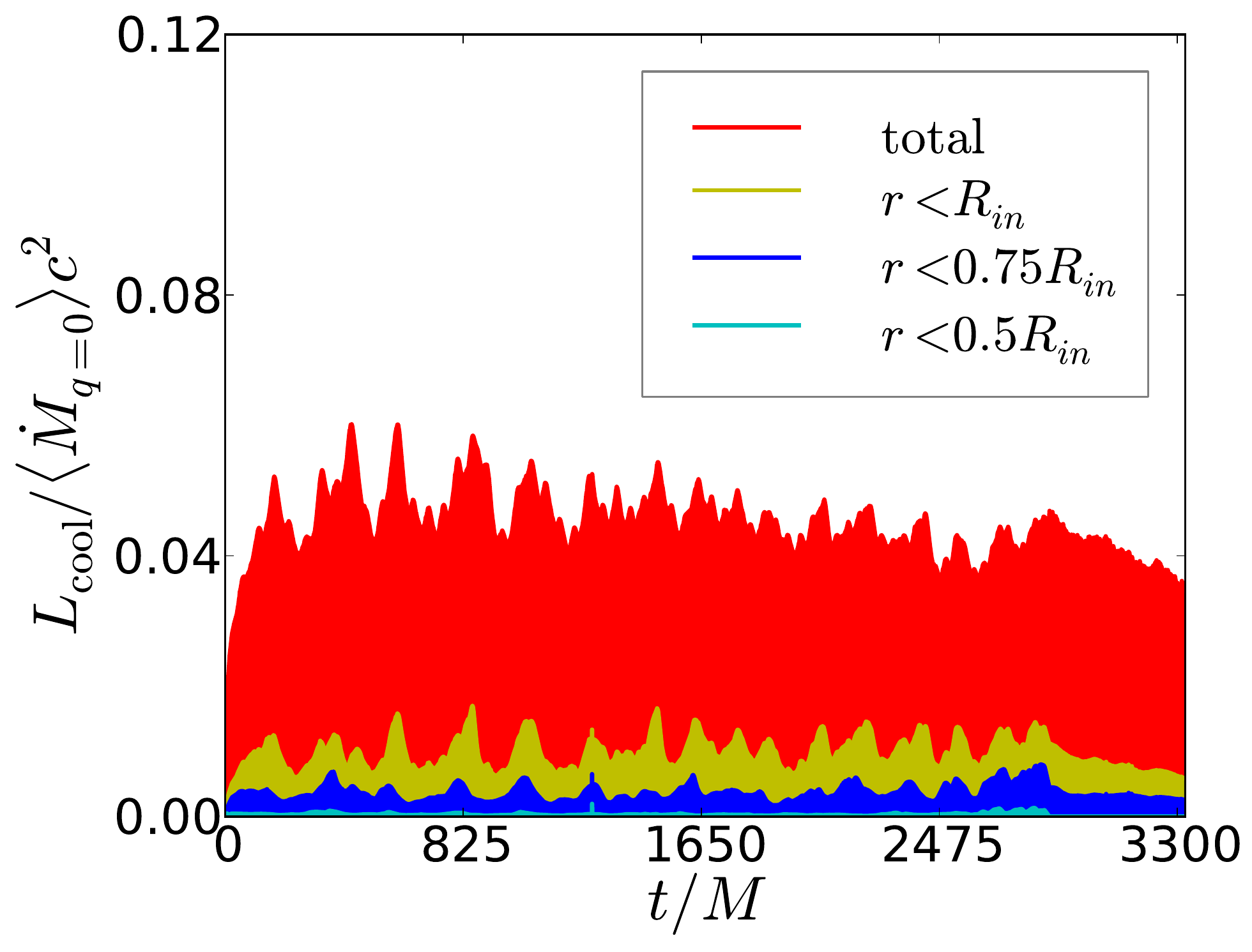}
            \caption{Contributions to the cooling luminosity $L_{\rm
                cool}$ from various spherical regions. Left panel:
              1:1. Middle panel: 1:2. Right panel: 1:4. 
            \label{fig:cavity}}
\end{figure*}

\subsubsection{Accretion rates}
\label{sec:mdot}
We show accretion rates as a function of time, together with our
luminosity estimates and the gravitational wave signal in
Fig.~\ref{fig:multimessenger}. In the 1:1 case the total accretion
rate drops as the inspiral proceeds, as expected. At merger the
accretion rate is more than an order of magnitude below the value at
decoupling. 
The 1:1 and 1:2 cases accrete at a similar rate near merger, see
Tab.~\ref{tab:results}, which is not true over the entire evolution,
see Fig.~\ref{fig:multimessenger}. Despite the stronger tidal torques
in the 1:1 case, it can be seen from the snapshots in
Fig.~\ref{fig:snapshots} that the density in the accretion streams,
which give the dominant contribution to $\dot{M}$, reach higher values
in the equal mass case. In \cite{Gold:2013zma} we found that the 1:1
case accreted at a slightly larger rate than the 1:2 case during the
predecoupling evolutions (see Tab. III in \cite{Gold:2013zma}), but
the difference in $\dot{M}$ between the two cases decays close to
merger.

Within the first $1000M \sim 6 \, \bra{M/10^8M_\odot}{\rm days}$ after
merger, spikes in the accretion rate from matter residing near the
remnant appear in the 1:1 case. This is accompanied by a gradual rise
over a longer time scale which will eventually lead to re-brightening
for all mass ratio cases. In the 1:4 case the accretion rate at merger
is $\dot{M}/\langle \dot{M}_{q=0} \rangle \sim 0.064$. This value is
the largest among the cases we study (see Tab. \ref{tab:results}), but
still substantially smaller than $1$.

\subsubsection{Luminosities}
We report the contribution to $L_{\rm cool}$ from within spheres of
different radii and the total contribution from the disk in
Fig.~\ref{fig:cavity}.  While in the 1:1 case $L_{\rm cool}$ drops
after about $500M \sim 3 \, \bra{M/10^8M_\odot}{\rm days}$ prior to
merger, there is no such feature for the 1:2 case. As the tidal
torques in the 1:1 case are stronger than the other mass ratios, this
behavior is likely due to the decline in tidal heating in the 1:1 case
as the inspiral proceeds. In the 1:2 case the inner cavity
contribution to $L_{\rm cool}$ (see Fig.~\ref{fig:cavity}, middle
panel) shows instead, a gradual rise, but without any prominent
feature during merger. Also in the 1:4 case we observe no prominent
feature in $L_{\rm cool}$ during merger in contrast to the thin disk
case \cite{Shapiro:2013qsa}.

In contrast to $L_{\rm cool}$, the binary-disk decoupling and the
merger {\it are} reflected in $L_{\rm EM}$ (see also the merger
aftermath feature in $L_{\rm EM}$ in \cite{Farris:2012ux}). From the
beginning of the inspiral $L_{\rm EM}$ slowly drops before rising
after merger.  In Fig.~\ref{fig:multimessenger} we also plot the GW
signal. One can compare the GW signal to different luminosity ``light
curves'' and the accretion rates. For 1:1 we find a delay of $\sim
800M \sim 4.6 \bra{M/10^8M_\odot} \rm days$ between the peak in GWs and the rise in $L_{\rm EM}$. For 1:2
this delay is significantly shorter $\sim 300M \sim 1.7 \bra{M/10^8M_\odot}\rm days$ and even shorter for
1:4, $\sim 200M \sim 1.2 \bra{M/10^8M_\odot}\rm days$. The shortening of this delay may be explained by the
fact that due to the decreasing tidal-torque barrier as $q$ decreases,
there is more material near the BHs in the 1:2 and 1:4 cases, which is
immediately available to be launched through the funnel. Despite the
increase after merger, $L_{\rm EM}$ always remains lower than $L_{\rm
  cool}$.  We further report a ``kinetic'' luminosity $L_{\rm kin}$
associated with matter outflows, which includes only unbound material
($E=-u_0-1>0$), identical to $L_{\rm gas}$ used in
\cite{Gold:2013zma}. We find in general $L_{\rm cool}>L_{\rm
  kin}>L_{\rm EM}$. We give values at merger in Tab.
\ref{tab:results}. 
We normalize luminosities by the accretion rate of the single BH case,
because it is not clear what a fair comparison to the time-dependent,
instantaneous binary accretion rate would be. Note that actual
efficiencies, i.e. luminosities normalized to the binary accretion
rate would be much higher.
The ratios $L_{\rm EM}/L_{\rm
  kin}$ range from $0.03$ to $0.09$ and are similar to the values
found in Tab. 2 of \cite{McKinney:2004ka}, e.g. $0.034$ for the
non-spinning case, where the same ratio is designated by
$\dot{E}^{(\rm EM)}/\dot{E}^{(\rm MA)}$. Even in the single BH case,
differences with \cite{McKinney:2004ka} are expected due to different
disk models, our $u_0$-based outflow diagnostic, their absence of
radiative cooling, and possibly different locations where the ratio
$\dot{E}^{(\rm EM)}/\dot{E}^{(\rm MA)}$ is evaluated (on the horizon in
\cite{McKinney:2004ka} vs far away from the black hole in our case).


\subsubsection{Outflows and jets}
\begin{figure*}[t]
  \begin{center}
      \includegraphics[width=0.39\textwidth]{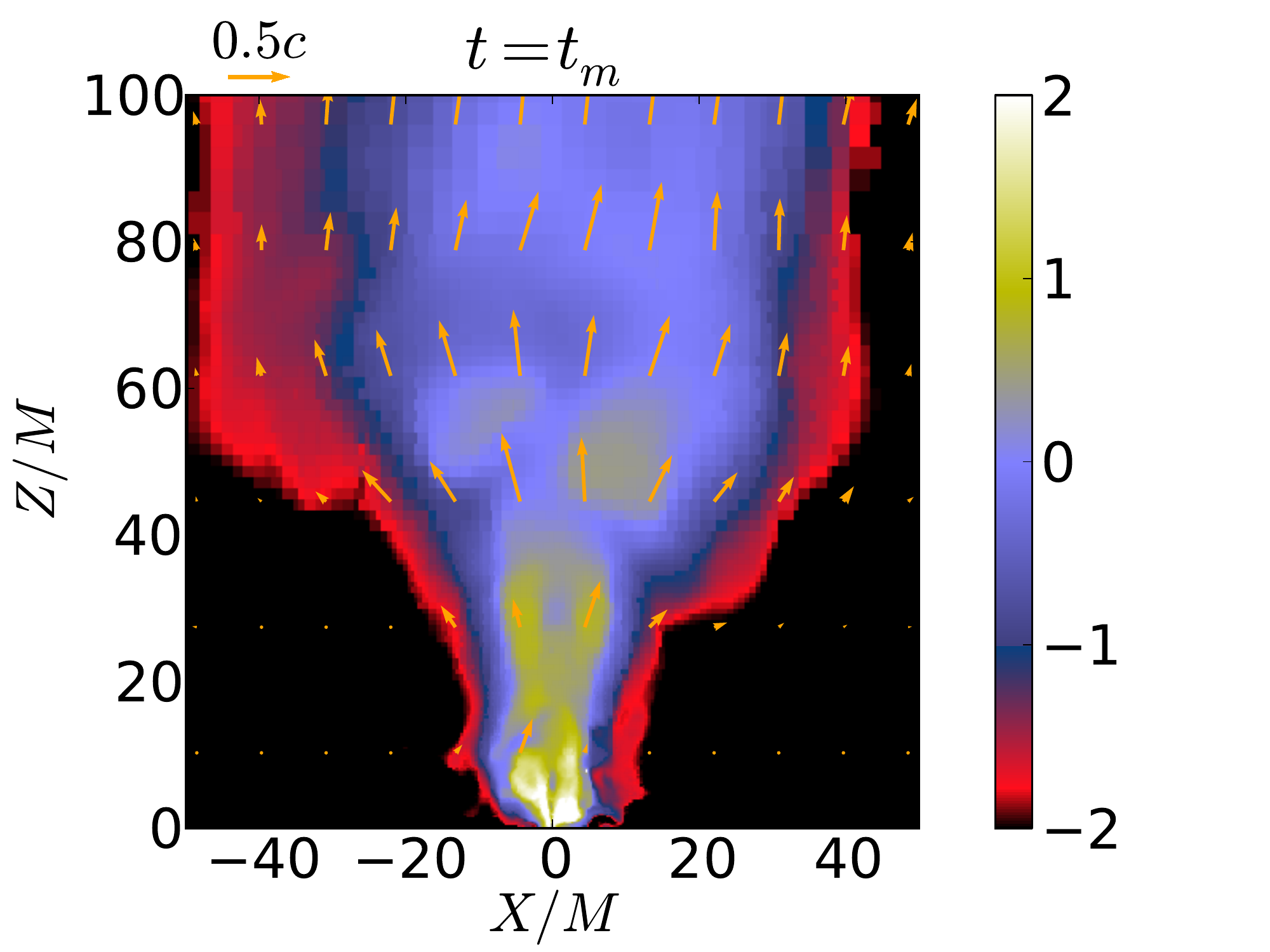} 
      \hspace{-1.8cm}
      \includegraphics[width=0.39\textwidth]{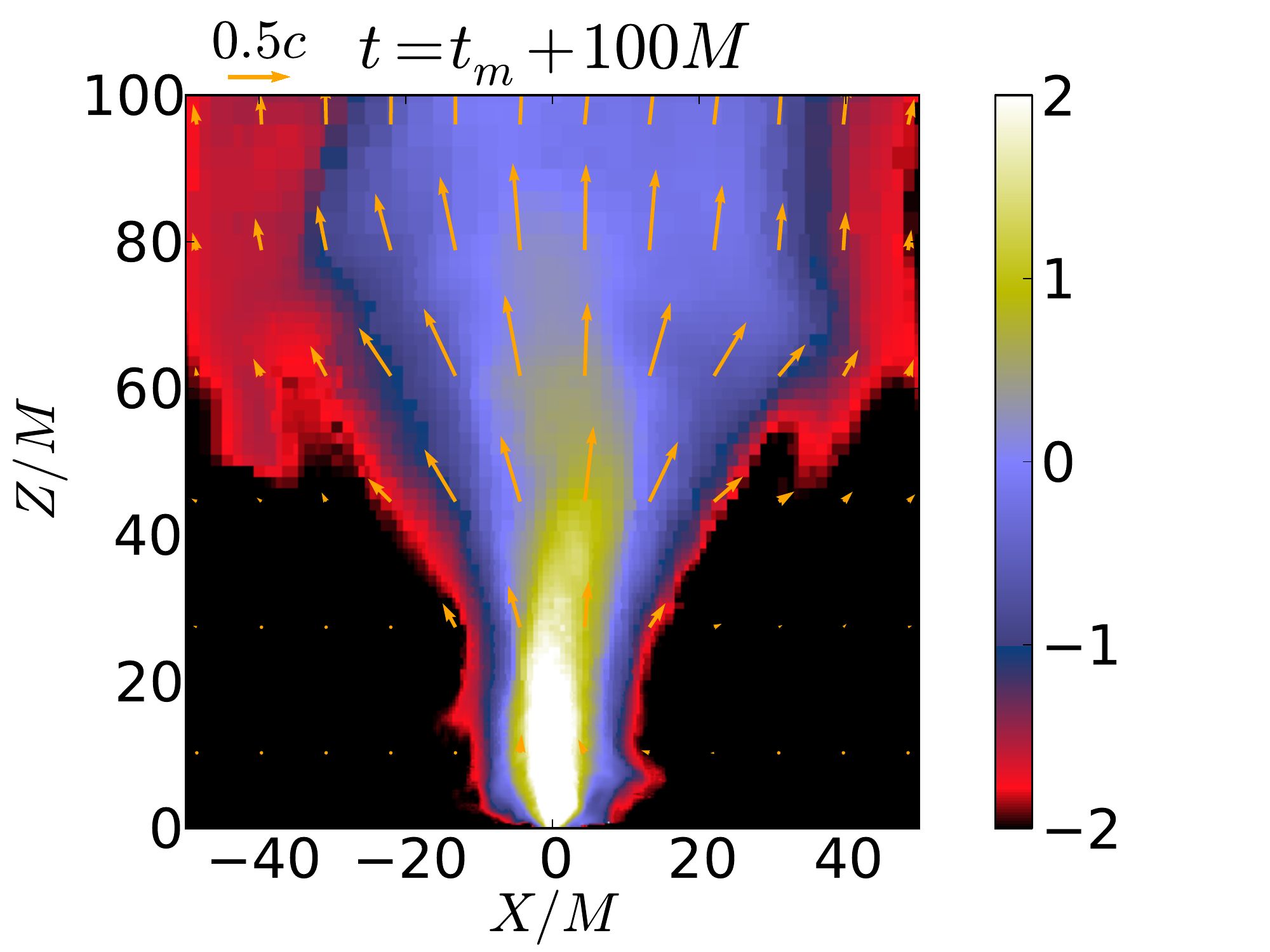} 
      \hspace{-1.8cm}
      \includegraphics[width=0.39\textwidth]{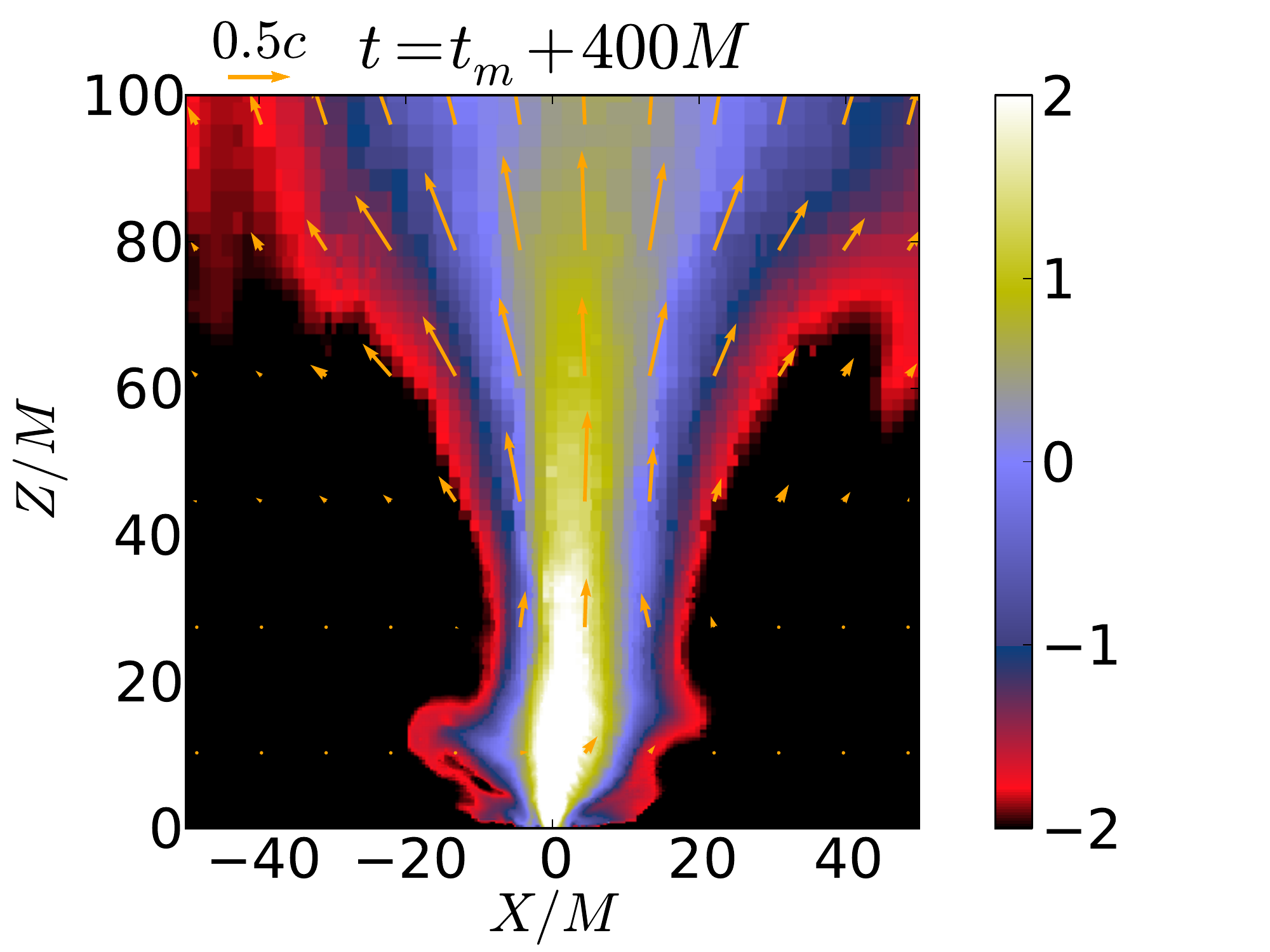} \\ 
      \includegraphics[width=0.39\textwidth]{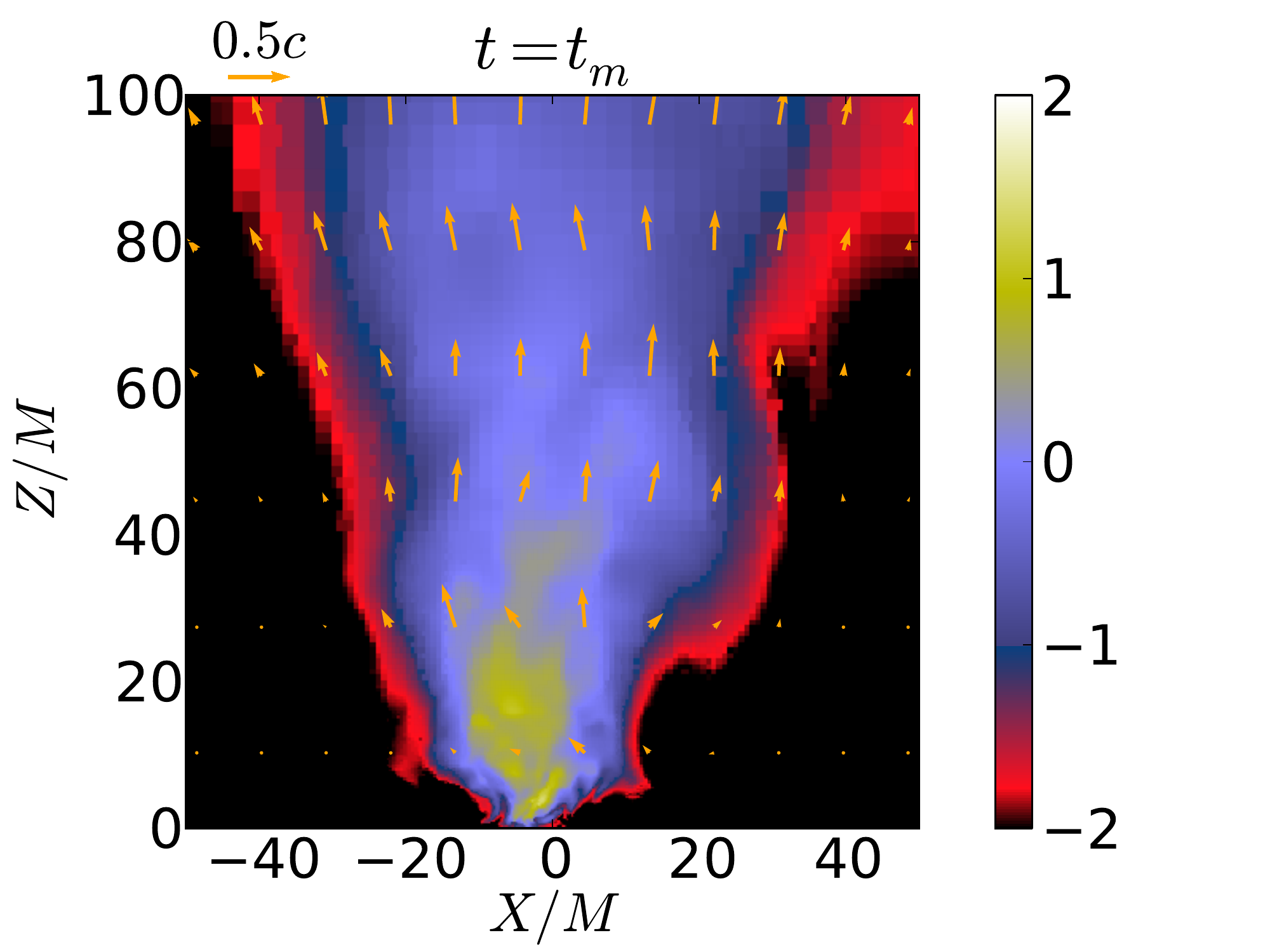} 
      \hspace{-1.8cm}
      \includegraphics[width=0.39\textwidth]{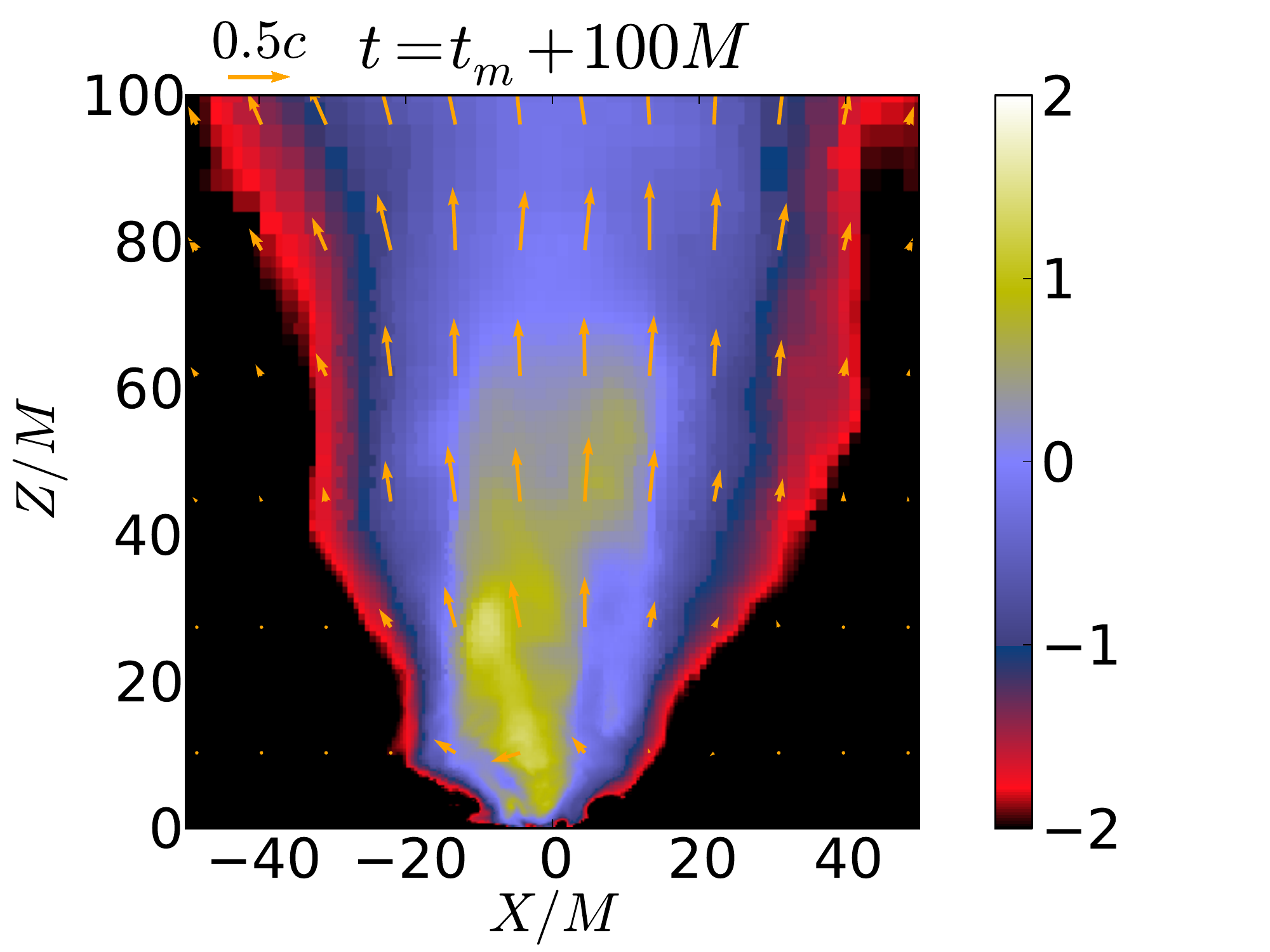} 
      \hspace{-1.8cm}
      \includegraphics[width=0.39\textwidth]{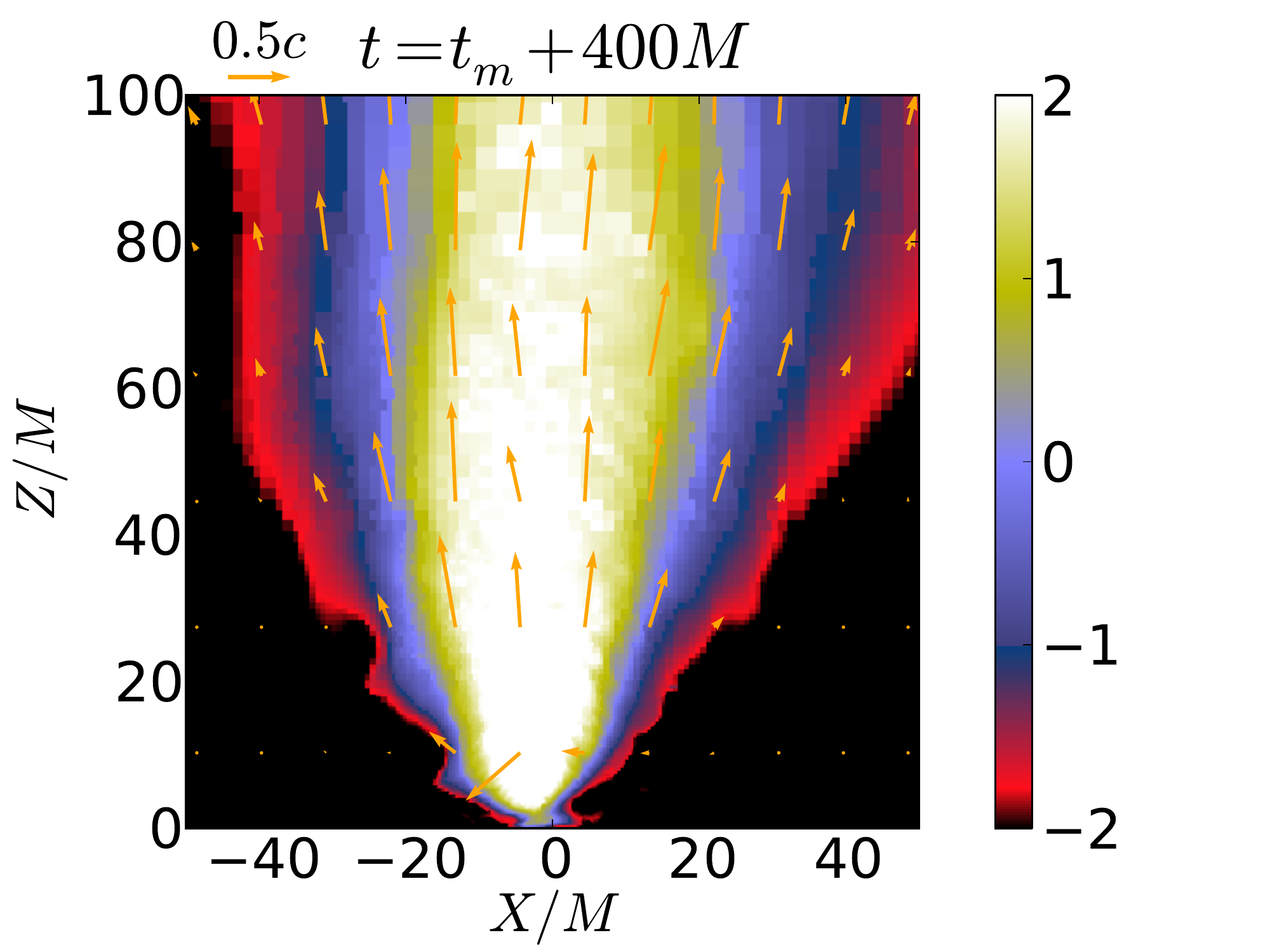} \\ 
      \includegraphics[width=0.39\textwidth]{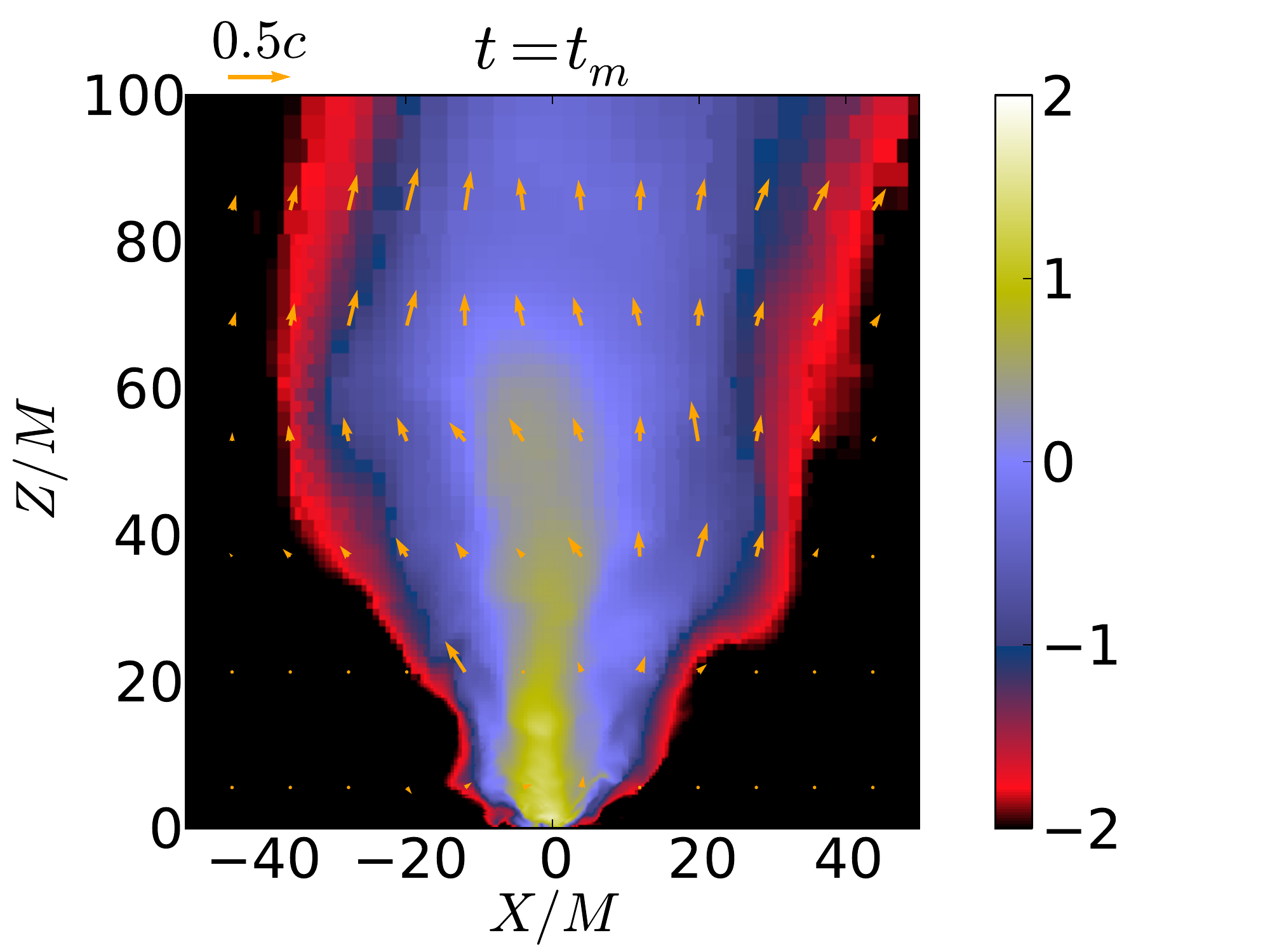} 
      \hspace{-1.8cm}
      \includegraphics[width=0.39\textwidth]{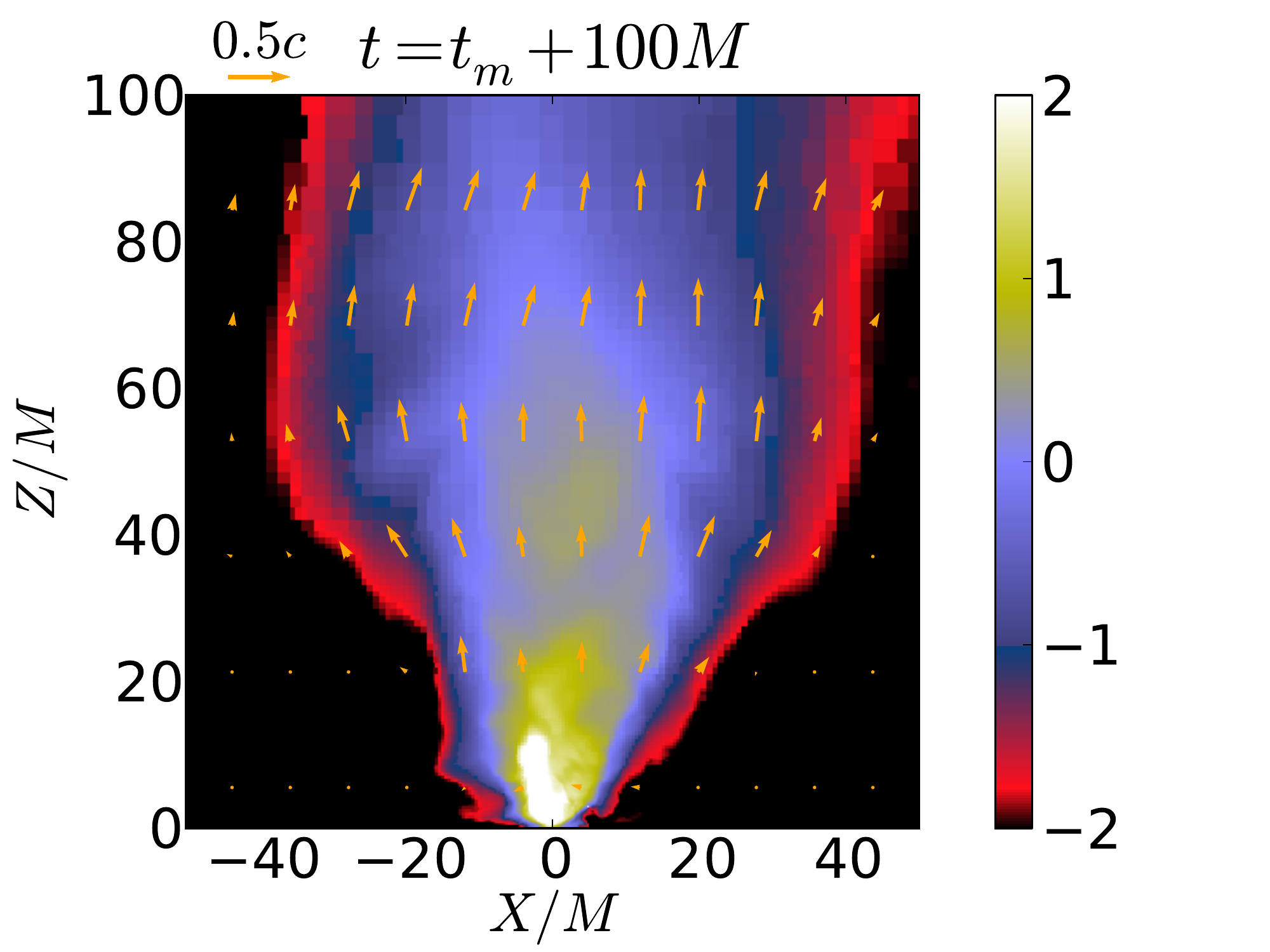} 
      \hspace{-1.8cm}
      \includegraphics[width=0.39\textwidth]{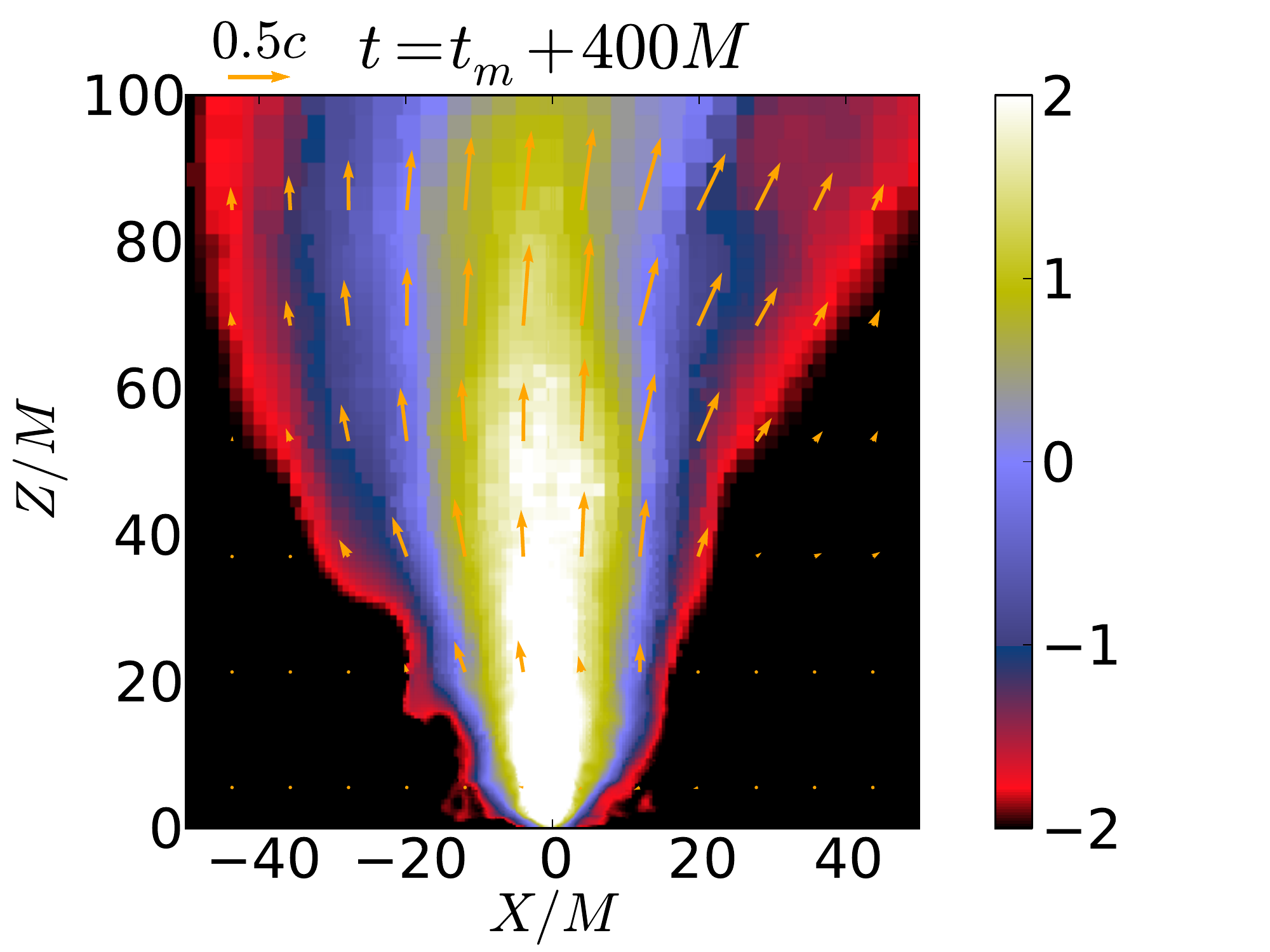} 
      \caption{Contours of $b^2/2\rho_0$, (log color scale) in a
        meridional slice, at merger (left panels), $100M$ after merger
        (middle panels), $400M$ after merger (right panels). Upper
        panels: 1:1. Middle panels: 1:2. Lower panels:
        1:4. \label{fig:outflow}}
  \end{center}
\end{figure*}

\begin{figure*}[h]
  \begin{center}
      \includegraphics[trim =0cm 0cm 1.1cm 0cm,clip=True,width=0.45\textwidth]{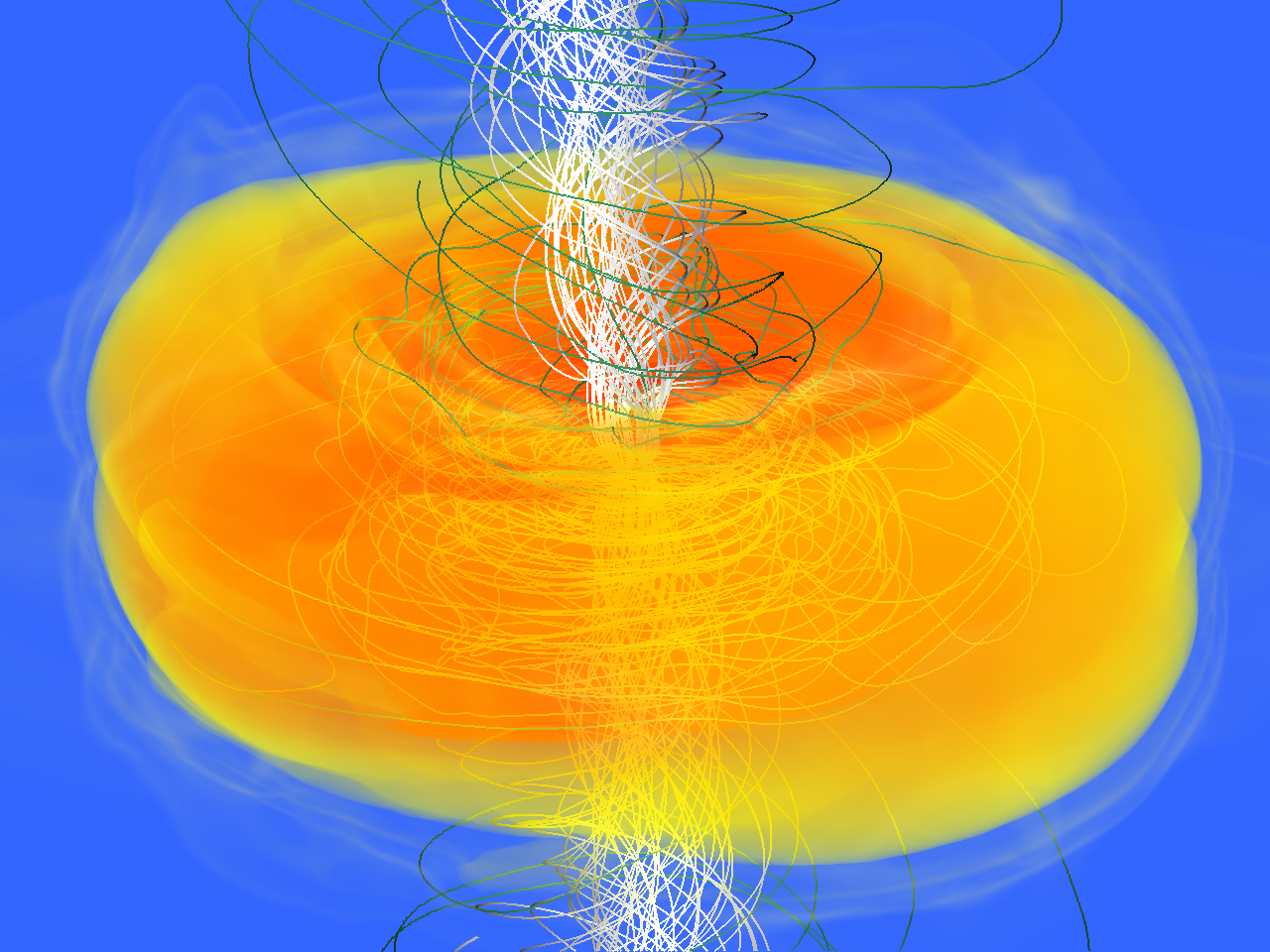}
      \includegraphics[trim =0cm 0cm 1.1cm 0cm,clip=True,width=0.522\textwidth]{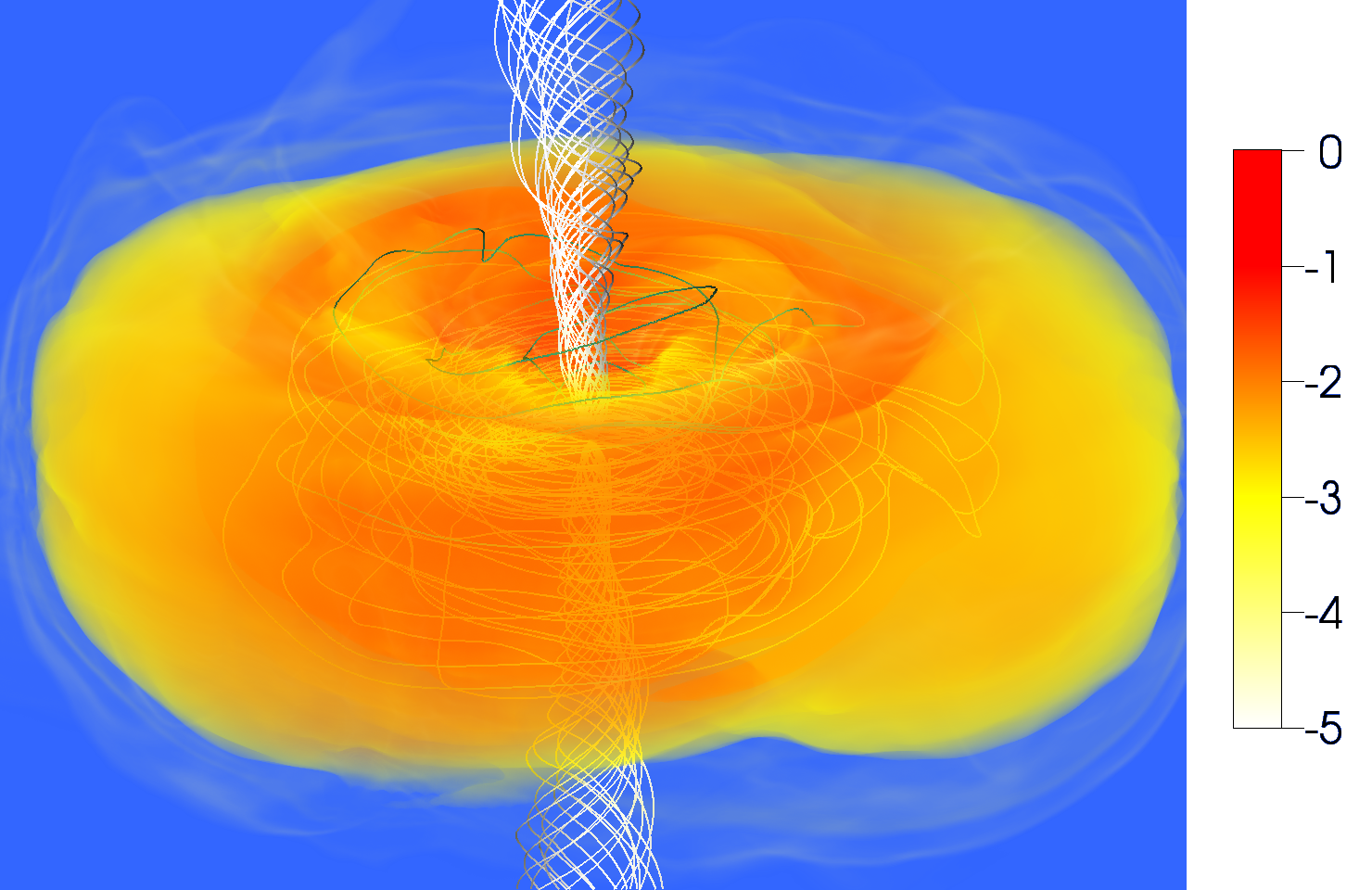}\\
      \includegraphics[trim =0cm 0cm 1.1cm 0cm,clip=True,width=0.45\textwidth]{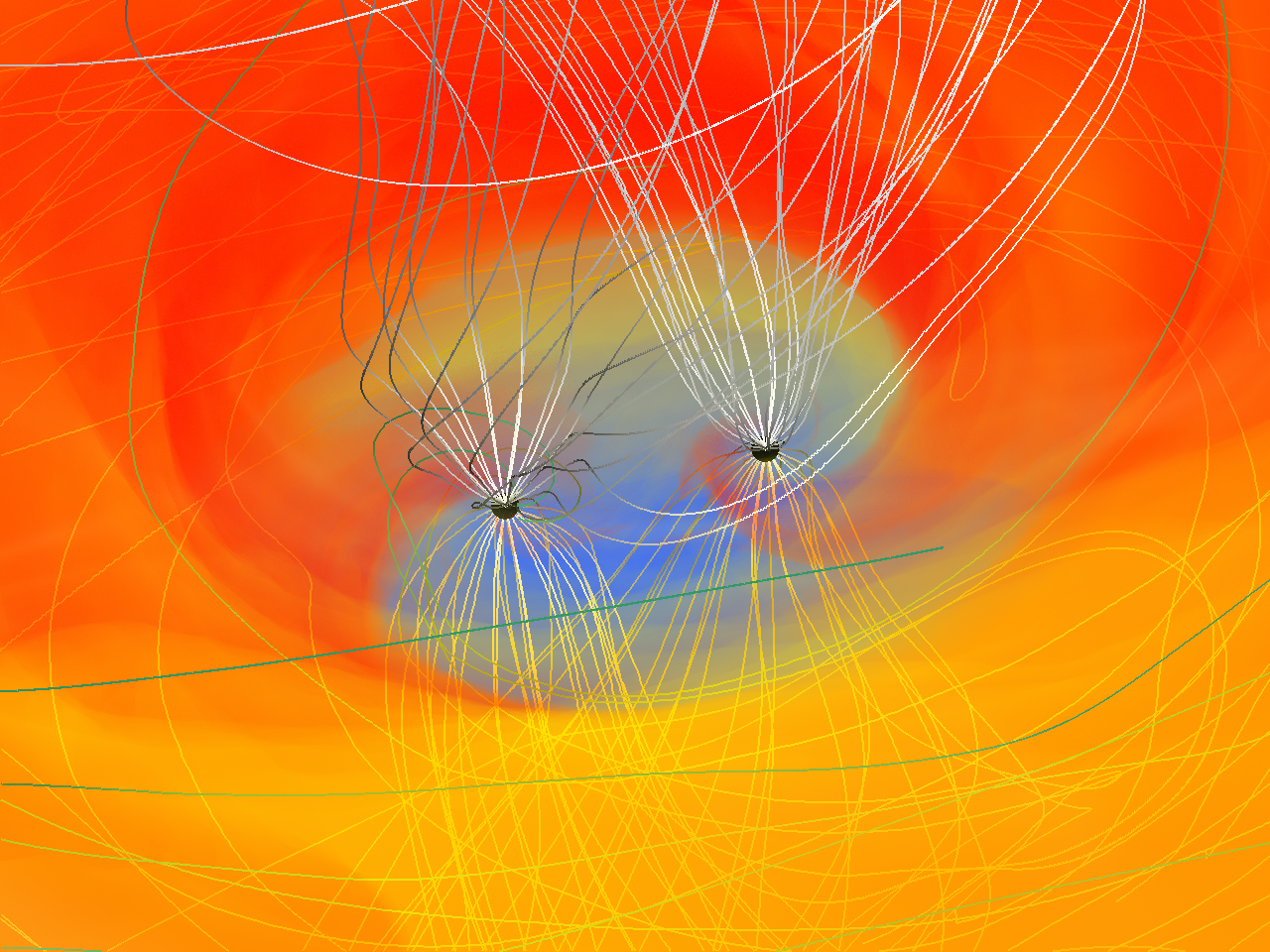}
      \includegraphics[trim =0cm 0cm 1.1cm 0cm,clip=True,width=0.522\textwidth]{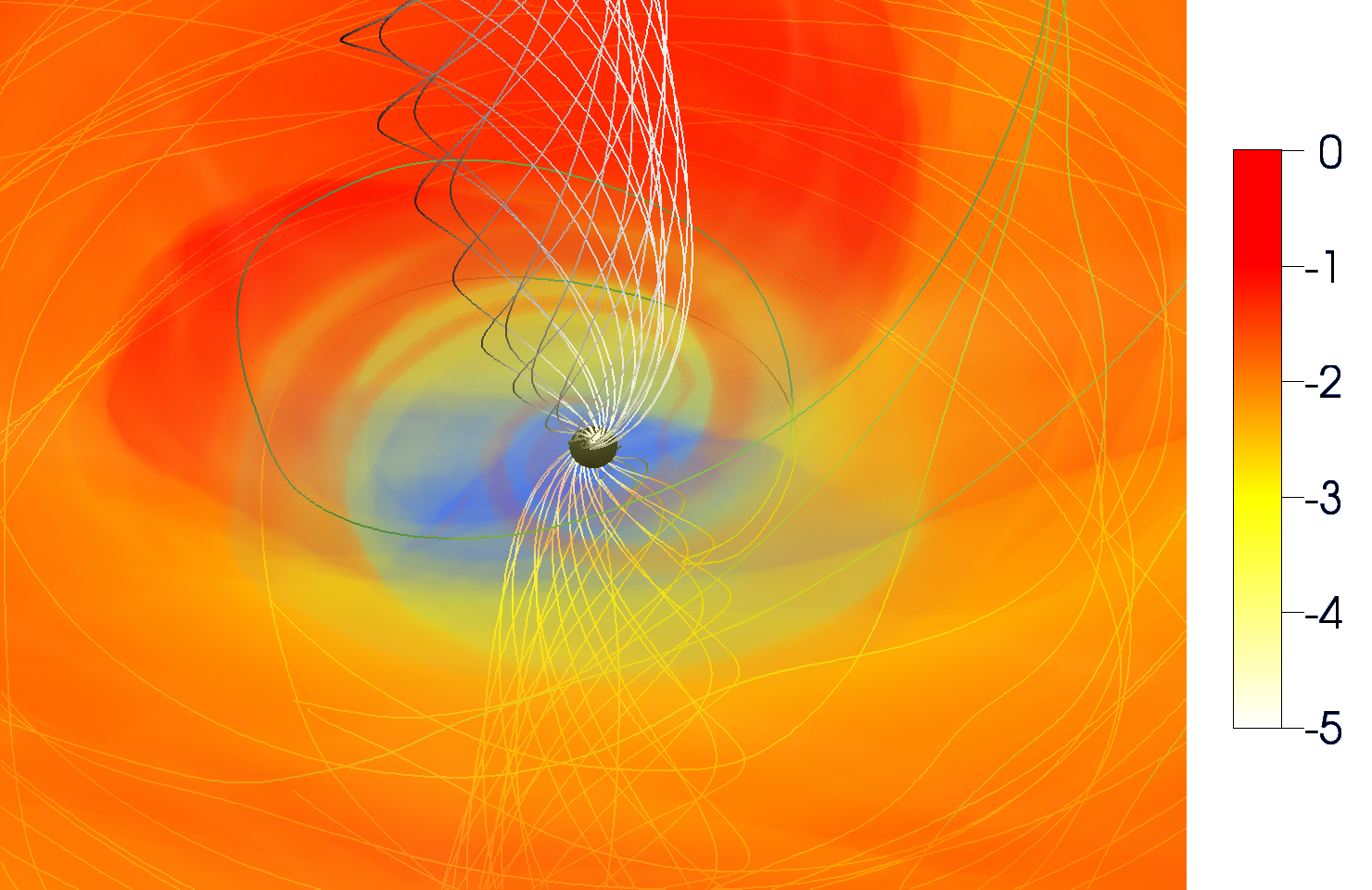}
        \caption{Volume rendering of rest-mass density, normalized to
          its initial maximum value $\rho_{0,\rm max}$ (see color
          coding), and magnetic field lines for the 1:1 case. Left
          panels: Halfway through the the inspiral $t-t_m =
          500M$. Right panels: Time $t-t_m\sim 100M$ after merger. Top
          panels: Global view out to $r/M\sim 150M$. Bottom panels:
          Closeup view within $r/M\sim 20M$. White field lines emanate
          from the BH apparent horizons. Green field lines emanate
          from the disk. The blue background indicates densities less
          than $10^{-5}\rho_{0,\rm max}$. Incipient jets are launched
          above each BH, merge at larger radii, and further collimate
          shortly after merger.
          \label{fig:jets}}
  \end{center}
\end{figure*}

In \cite{Gold:2013zma} we have identified collimated, magnetized
outflows in the predecoupling epoch. As expected, no collimated
outflows are observed for the non-spinning single BH case with the
same initial disk and numerical parameters. In all binary cases we
find that the incipient jets persist through merger and the immediate
post-merger evolution; see Figs.~\ref{fig:outflow} and \ref{fig:jets}
for all cases. The difference between single and binary cases as well
as visualizations of B-field lines (Fig.~\ref{fig:jets}) throughout
the evolution lead us to attribute the outflows to magnetic winding
and buildup of magnetic pressure above the poles of the orbiting black
holes. Through accretion, B-field is accreted onto the black holes
which can then tap the orbital kinetic energy as in a single spinning
BH magnetic fields can tap the rotational kinetic energy of the BH,
eventually giving rise to collimated, relativistic outflows, see
Fig.~\ref{fig:outflow}.

After merger all cases reveal an increase in the Lorentz factor $W$
(measured by normal observers) of the flow in the funnel accompanied
by an increase in $b^2/2\rho_0$ (where $b^2/2$ is the magnetic
pressure and $\rho_0$ the rest-mass density). Note that $b^2/2\rho_0$ not
only shows how dominant the B-field is over the inertia of the matter,
but also equals the terminal Lorentz factor achieved by a
steady-state, axisymmetric jet model
\cite{B2_over_2RHO_yields_target_Lorentz_factor}. Until merger we find
mildly relativistic $W \gtrsim 1.2$ outflows with maximum $b^2/2\rho_0
\sim 10$ at larger distance from the BHs. After merger $b^2/2\rho_0$ and
$W$ inside the funnel above the polar region increase to $W\sim 2.4$
($q=1$), $W\sim 2$ ($q=2$), $W\sim 1.6$ ($q=4$) and maximum $b^2/2\rho_0
\gtrsim 100$.

\subsection{Distinguishing pre-, postdecoupling and merger}
\label{sec:decoupling-merger}

In \cite{Gold:2013zma} we have presented an analysis of the
predecoupling phase and the relaxed disk properties.  Here we report
our results during the postdecoupling epoch, adopting the relaxed disk
as initial data for a realistic calculation of the postdecoupling
evolution.

\subsubsection{Postdecoupling}
\label{sec:postdecoupling}

For most cases $\Sigma(r)$ remains rather similar at merger and
decoupling, at least in the bulk of the disk (see
Fig.~\ref{fig:sigma}). This demonstrates that the response of the {\it
  bulk} of the disk material is slow compared to the merger time. Near
the inner edge all cases show a small drift of matter inward. This
behavior is expected, as the binary tidal torques decrease during
inspiral while the outward angular momentum transport due to MHD
turbulence persists. In all cases considered here, there is a
significant reduction in density relative to the single BH case near
the BHs.

For thin disks the postdecoupling evolution leads to a dimming of the
source, as the binary inspirals in the nearly empty cavity while
running away from the inner disk edge
\cite{MacFadyen:2006jx,Shapiro:2013qsa}, but see
\cite{Farris:2014qma}.
For thick disks the cavity contains a considerable amount of gas,
which leads to a smearing of the classic decoupling picture for thin
disks -- in particular for mass ratios different from unity. The
snapshots of the rest-mass density covering inspiral and merger, see
Fig.~\ref{fig:snapshots}, clearly demonstrate the persistence of two
dense accretion streams threading the horizons and a gaseous
environment, in which the BHs remain embedded in all the way through
merger.

In the 1:1 case the onset of the postdecoupling evolution is signaled
by a gradual {\it decrease} in the accretion rate (upper panel of
Fig.~\ref{fig:multimessenger}; compare to $\dot{M}/\dot{M}_{q=0}\sim
0.43$ in the predecoupling phase \cite{Gold:2013zma}), $L_{\rm cool}$
and $L_{\rm kin}$. In both the 1:2 and 1:4 cases there is no
luminosity decrease during the inspiral of the binary. In all cases,
the amplitude and frequency of the GW signal {\it increase}
substantially, see Fig.~\ref{fig:multimessenger}.

\subsubsection{Early merger aftermath}
\label{sec:aftermath}

A few hundred to $1000M$ after merger (depending on $q$), the Poynting
luminosity $L_{\rm EM}$ (see \cite{Gold:2013zma} for the definition)
undergoes a sudden increase.  For 1:1 $L_{\rm EM}$ increases by a
factor of $~5$ within $~1000M$ following merger. By contrast, the
cooling luminosity $L_{\rm cool}$ decreases by $30\%$ over a shorter
time interval of $~300M$ before reaching a plateau at half the
predecoupling value until the end of our simulation. Even so, the
cooling luminosity still dominates: $L_{\rm cool} \sim 8 L_{\rm EM}$
at the end of our simulation. We do not find a sudden, large increase
in $L_{\rm cool}$ at merger as in \cite{Farris:2012ux} (or as for thin
disks \cite{Shapiro:2013qsa}). This could be due to the different
cooling emissivity or differences in the disk (adiabatic index) we
employ here.

We confirm our findings in \cite{Farris:2012ux} regarding a rapid
change in the properties of the collimated, magnetized outflows in the
polar regions during and shortly after merger in all cases. This can
be seen, e.g., by comparing $b^2/2\rho_0$ in the meridional plane at
merger and shortly after merger, see Fig.~\ref{fig:outflow}. In all
cases and for all epochs the magnetic pressure is sub-dominant
relative to the rest-mass density in the bulk of the disk. The polar
regions are dominated by magnetic pressure. Prior to merger we find
relatively (compared to other single BH GRMHD accretion studies) small
values $b^2/2\rho_0 \gtrsim 10$ at large distances from the BHs, but
shortly after merger $b^2/2\rho_0 \gtrsim 100$ in a collimated cone,
which quickly expands into the polar directions. This trend is shown
in Fig.~\ref{fig:outflow}. The same collimation effect just after
merger is obvious in 3D visualizations including the magnetic field
lines, see Fig.~\ref{fig:jets}.  The simultaneous onset of the rapid
change in the outflows with the increase in $L_{\rm EM}$ strongly
suggests, that the increased magnetization and acceleration of the
outflow is the main cause for the brightening. The collimation just
after merger is observed in a similar way in all binary cases.

All currently existing ideal MHD schemes (either relativistic or
Newtonian) can accurately evolve regions only up to a certain critical
value of the plasma $\beta$ parameter $\beta \equiv 2P/b^2$. Once the
critical value is reached or exceeded (typically in the low-density
atmosphere) certain inequalities must be imposed to continue the
simulations, with their impact designed to be minimal. Based on
previous results and tests with our code, we are convinced that the
postmerger {\it increase} in the magnetization in the funnel is
robust, but terminal values of $b^2/2\rho_0 \gtrsim 100$ in those
regions may not be reliable.

The Blandford-Payne mechanism \cite{1982MNRAS.199..883B} is probably
not the cause for this transient behavior because of the special
conditions under which the mechanism operates.

However, it is natural to attribute part of the increase in $L_{\rm
  EM}$ to the Blandford-Znajek \cite{Blandford1977} effect (see also
\cite{Neilsen:2010ax,Palenzuela:2010nf}), because all of our BH
remnants are spinning with the funnel area above the remnant BH poles being 
nearly force-free. The BZ solution is known to describe the
force-free regions in the funnel of magnetized, geometrically thick
disks accreting onto single spinning BHs \cite{McKinney:2004ka}.

As in \cite{Farris:2012ux}, we can see the disk beginning to drift
inwards towards the remnant BH by comparing $\Sigma(r)$ at different
times, see Fig.~\ref{fig:sigma}. By the time the bulk of the material
will reach the remnant BH the system will likely undergo a {\it
  re-brightening} \cite{Milosavljevic:2004cg}.

Due to asymmetries in the gravitational radiation and momentum
conservation the remnant BHs in the 1:2 and 1:4 cases, experience a
recoil or ``kick'' (see \cite{Baker:2006vn,Megevand:2009} and
references therein). However, for initially non-spinning BHs the
recoil velocity has a maximum value of $\sim 175$km/s
\cite{Gonzalez:2006md}, and hence it is small compared to other
characteristic velocities in the system (see also
\cite{Bogdanovic:2007hp,Dotti11022010}) unlike
\cite{Ponce-2012}. Therefore, the remnant BH recoil in our simulations
does not have any significant impact on the accretion flow.

\section{Astrophysical implications}
\label{sec:implications}

Observational evidence for a SMBH binary near the postdecoupling
regime remains elusive. There are two possible explanations: (I) There
are too few sources. (II) Due to various reasons, identifying SMBH
binaries is difficult. Point (II) includes the possible
misinterpretation of a binary AGN as a single BH AGN. This confusion
can arise because the second BH may not be an ``active'' AGN or may
not be massive enough to alter the outgoing radiation at an observable
level.


In our models several diagnostics reveal differences between the
single and binary models. In the single BH more matter resides closer
to the BH, which is visible in $\Sigma(r)$ comparisons, see
Fig.~\ref{fig:sigma}. In the binary case $\dot{M}$ is reduced compared
to the single BH case for the same disk. The decrease over time in
$\dot{M}$ and $L_{\rm EM}$ in the binary system after decoupling
[$\sim 1 \bra{M/10^8M_\odot} \rm week$ prior to merger], which we report
here, is another signature which is absent in the single BH case. This
reduction is expected to last until re-brightening ($\sim 2
\bra{M/10^8M_\odot} \rm months$ after merger).

The magnetically-driven transient during merger and the resulting
acceleration of matter along the polar regions is characteristic of
the binary merger in all cases and is not observed (nor expected) in
the single BH case. Instead, in the single non-spinning BH we observe
little to no polar outflows. There are outflows from spinning BHs, but
these are not likely to exhibit a one-time, dramatic transient
behavior such as the one exhibited in the binary case.  Based on our
findings, the strongest evidence for the presence of a BH binary is
the transient in the magnetized outflows/jets during and shortly after
merger. The increased magnetic field strength and outflow velocity
will likely lead to enhancements of the radio emission from the jet
and perhaps the X-rays and additional brightening due to relativistic
beaming.

While there are known effects that can cause a single BH accretion
flow to flare (recurringly) -- such as hotspots -- the flare in the
Poynting luminosity near a binary merger is a one-time event.

Our results motivate a search for binary SMBH candidates based on jet
morphology. Even a merger event long in the past could be identified
by a change in the collimation from the foot of the jet towards its
head. In fact, observations in time of such jets could reveal that the
stronger emission is propagating outwards. This is similar to the
interpretation that X-shaped radio sources originate from a sudden
spin-flip following a past BHBH merger. The difference here is that
there is a transient feature in the jet even in the absence of a spin
flip.

The effective temperature, magnetic energy density, and characteristic
cyclotron frequencies during the inspiral phase remain similar to the
values we reported in \cite{Gold:2013zma} for the predecoupling phase:
\labeq{}{
T_{\rm eff}
\sim 10^5\bra{\frac{L_{\rm b}}{L_{\rm Edd}}}^{1/4}\bra{\frac{M}{10^8M_\odot}}^{-1/4}\rm K,
} 
\labeq{}{
\nu_{\rm cy} \sim
10^{6}\bra{\frac{L_{\rm b}}{L_{\rm
      Edd}}}^{1/2}\bra{\frac{M}{10^8M_\odot}}^{-1/2}\rm Hz.
}
However, a few hundred $M$ after merger we find an increase in the
magnetic energy density in the funnel region by a factor of $\sim
10$. This effect shifts the cutoff frequency of synchrotron emission,
which may arise in these systems from the presence of relativistic
electrons, towards higher frequencies; see e.g. Problem 4.2 in
\cite{rybicki86}. Therefore, a one-time frequency shift in the
synchrotron emission could be detected in radio surveys
\cite{Kaplan:2011mz,O'Shaughnessy:2011zz} and may reveal the presence of a
BHBH merger.

Also blazar systems, similar to the SMBH binary candidate OJ-287
\cite{Sillanpaa88,Valtonen:2011ny}, constitute a promising class of
systems where binaries might be identified through EM observations
\cite{Decarli:2013xwa} based on variability studies
\cite{Tanaka:2013cva}.

The dimming of total luminosity observed in 1:1 signals the onset of
the postdecoupling epoch and serves as a precursor for the upcoming
merger with a lead time of $500M \sim 3 \, \bra{M/10^8M_\odot}{\rm
  days}$.  Such a one-time dimming is unique to the equal-mass SMBH
binary and does not seem to occur either in a single BH accretion flow
or for mass ratios significantly different from unity. Thus, a
near-equal-mass binary AGN can potentially be distinguished from both
a single BH AGN {\it prior} to merger and binary AGNs with mass ratios
deviating from unity, even in the absence of a ``sudden'' EM feature
during merger.

Force-free simulations in full GR have suggested that dual jets from
BHBH systems may be detectable \cite{Palenzuela:2010nf}. If such dual
jets were detected they would be strong evidence for the presence of
an accretion disk onto a BHBH. However, the force-free simulations of
\cite{Moesta:2011bn} argue that the power in these dual jets is only a
small fraction of an otherwise more isotropic emission, suggesting
that dual jets are likely not detectable. Our GRMHD simulations show,
that the individual jets launched by each BH merge into one common jet
structure (at least during the late inspiral). Therefore, we conclude
that dual jets are unlikely to be detected from magnetized accretion
disk-BHBH systems in which the BHs are slowly spinning and the disk
orbital angular momentum is aligned with binary orbital angular
momentum, see also \cite{O'Shaughnessy:2011zz}.

The merger of the two BHs poses a major change to the whole
system. The results we report here are attributed to the effects
immediately ($t < 1000M \sim 6 (M/10^8M_\odot) \rm days$) following
the merger event. By ``immediate'' we mean over time scales which
involve the dynamics of material within a few M near the BH left over
from the merger. This is in contrast to the independent effects of
material from the bulk of the disk beginning to fall in on a (much
longer) viscous time scale ($t_{\rm vis} > 10,000M = 60
(M/10^8M_\odot) \rm days$) after the merger has occurred.

\section{Conclusions}
\label{sec:conclusions}

We have presented results from our follow-up study of
\cite{Gold:2013zma} by evolving {\it relaxed} GRMHD accretion flows
through the binary inspiral and merger phases in full GR.

The key differences between signatures arising from the BHBH-disk at
decoupling as compared to those near merger are:

\begin{enumerate}

\item The mildly relativistic dual jets observed near decoupling and
  prior to merger, coalesce and form one common jet. The common jet
  further collimates near merger (see Figs.~\ref{fig:outflow} and
  \ref{fig:jets}), while the outflow Lorentz factors are boosted
  following merger by $\sim 60\%$.

\item The Poynting luminosity increases shortly after merger by a
  factor of $\sim 1.5-2$, and its value is sustained until the end of
  our simulations (see Fig.~\ref{fig:multimessenger}).

\item The kinetic luminosity exhibits a large peak near merger (see
  Fig.~\ref{fig:multimessenger}) whose height is $\sim 1.5-2$ times
  larger than the values prior to merger.

\item The cooling luminosity is largely insensitive to the dynamics
  during postdecoupling and merger (see
  Figs.~\ref{fig:multimessenger} and \ref{fig:cavity}). We find a
  decrease during postdecoupling only in the 1:1 case.

\end{enumerate}

For decreasing mass ratio (1:1, vs. 1:2, vs. 1:4), the key trends of
the BHBH-disk systems are:
\begin{enumerate}

\item The increase in the Poynting and kinetic luminosities after
  merger begins earlier as $q$ decreases. For 1:4 the kinetic
  luminosity peaks almost simultaneously with the GW burst at merger.

\item The boost in the Poynting luminosity after merger decreases with
  decreasing $q$. This is most likely due to the fact that the spin of
  the remnant BH decreases with $q$.

\item There is significant variability in the accretion rate and the
  cooling luminosity in the 1:4 case, which is not observed for the
  other cases.

\item The non-axisymmetric ``lump'' feature becomes weaker as $q$
  decreases.

\end{enumerate}

We find little decrease in nearly all luminosity diagnostics after
decoupling, indicating that such sources may be bright.  Aftermath EM
signatures are more prominent than precursor EM signals. Generally,
the dependence of EM signatures (the increase in the Poynting
luminosity and its time lag after merger) on mass ratio is stronger
after merger than before merger or in the predecoupling epoch. A
robust acceleration and boost in magnetic energy density of the
outflowing material is observed, which is an excellent candidate for a
clear and pronounced, one-time EM signature for merging SMBH
binaries. This transient and the reported features in the light curves
are unlikely to occur in single BH disk systems.

In the future we intend to include realistic radiation processes and
radiative transport, and refine our results by calculating an EM
spectrum, in order to identify distinguishing features between single BH
and binary BH AGN.

\acknowledgments It is a pleasure to thank the Illinois Relativity
group REU team (Brian R.~Taylor, Lingyi Kong, Abid Khan, Sean
E. Connelly, Albert Kim and Francis J.~Walsh) for assistance in
creating Fig.~\ref{fig:jets}.  We thank Charles Gammie, Brian Farris,
Jonathan McKinney, and Alberto Sesana for useful discussions. This
paper was supported in part by NSF Grants No. PHY-0963136 and
No. PHY-1300903 as well as NASA Grants NNX11AE11G and NNX13AH44G at
the University of Illinois at Urbana-Champaign. V.P. gratefully
acknowledges support from a Fortner Fellowship at UIUC. H. P. P.
acknowledges support by NSERC of Canada, the Canada Chairs Program,
and the Canadian Institute for Advanced Research. The metric initial
data was computed on the GPC supercomputer at the SciNet HPC
Consortium~\cite{Scinet}. SciNet is funded by: the Canada Foundation
for Innovation under the auspices of Compute Canada; the Government of
Ontario; Ontario Research Fund--Research Excellence; and the
University of Toronto. This work used the Extreme Science and
Engineering Discovery Environment (XSEDE), which is supported by NSF
Grant number OCI-1053575. This research is part of the Blue Waters
sustained-petascale computing project, which is supported by the
National Science Foundation (Award No. OCI 07-25070) and the state
of Illinois. Blue Waters is a joint effort of the University of
Illinois at Urbana-Champaign and its National Center for
Supercomputing Applications.









\bibliography{paper}{}

\begin{thebibliography}{114}%
\makeatletter
\providecommand \@ifxundefined [1]{%
 \@ifx{#1\undefined}
}%
\providecommand \@ifnum [1]{%
 \ifnum #1\expandafter \@firstoftwo
 \else \expandafter \@secondoftwo
 \fi
}%
\providecommand \@ifx [1]{%
 \ifx #1\expandafter \@firstoftwo
 \else \expandafter \@secondoftwo
 \fi
}%
\providecommand \natexlab [1]{#1}%
\providecommand \enquote  [1]{``#1''}%
\providecommand \bibnamefont  [1]{#1}%
\providecommand \bibfnamefont [1]{#1}%
\providecommand \citenamefont [1]{#1}%
\providecommand \href@noop [0]{\@secondoftwo}%
\providecommand \href [0]{\begingroup \@sanitize@url \@href}%
\providecommand \@href[1]{\@@startlink{#1}\@@href}%
\providecommand \@@href[1]{\endgroup#1\@@endlink}%
\providecommand \@sanitize@url [0]{\catcode `\\12\catcode `\$12\catcode
  `\&12\catcode `\#12\catcode `\^12\catcode `\_12\catcode `\%12\relax}%
\providecommand \@@startlink[1]{}%
\providecommand \@@endlink[0]{}%
\providecommand \url  [0]{\begingroup\@sanitize@url \@url }%
\providecommand \@url [1]{\endgroup\@href {#1}{\urlprefix }}%
\providecommand \urlprefix  [0]{URL }%
\providecommand \Eprint [0]{\href }%
\providecommand \doibase [0]{http://dx.doi.org/}%
\providecommand \selectlanguage [0]{\@gobble}%
\providecommand \bibinfo  [0]{\@secondoftwo}%
\providecommand \bibfield  [0]{\@secondoftwo}%
\providecommand \translation [1]{[#1]}%
\providecommand \BibitemOpen [0]{}%
\providecommand \bibitemStop [0]{}%
\providecommand \bibitemNoStop [0]{.\EOS\space}%
\providecommand \EOS [0]{\spacefactor3000\relax}%
\providecommand \BibitemShut  [1]{\csname bibitem#1\endcsname}%
\let\auto@bib@innerbib\@empty
\bibitem [{\citenamefont {{Soltan}}(1982)}]{Soltan1982}%
  \BibitemOpen
  \bibfield  {author} {\bibinfo {author} {\bibfnamefont {A.}~\bibnamefont
  {{Soltan}}},\ }\href@noop {} {\bibfield  {journal} {\bibinfo  {journal}
  {\mnras}\ }\textbf {\bibinfo {volume} {200}},\ \bibinfo {pages} {115}
  (\bibinfo {year} {1982})}\BibitemShut {NoStop}%
\bibitem [{\citenamefont {Tanaka}\ and\ \citenamefont
  {Haiman}(2009)}]{Tanaka:2008bv}%
  \BibitemOpen
  \bibfield  {author} {\bibinfo {author} {\bibfnamefont {T.}~\bibnamefont
  {Tanaka}}\ and\ \bibinfo {author} {\bibfnamefont {Z.}~\bibnamefont
  {Haiman}},\ }\href {\doibase 10.1088/0004-637X/696/2/1798} {\bibfield
  {journal} {\bibinfo  {journal} {Astrophys.J.}\ }\textbf {\bibinfo {volume}
  {696}},\ \bibinfo {pages} {1798} (\bibinfo {year} {2009})},\ \Eprint
  {http://arxiv.org/abs/astro-ph/0807.4702} {arXiv:astro-ph/0807.4702
  [astro-ph]} \BibitemShut {NoStop}%
\bibitem [{\citenamefont {Volonteri}\ \emph {et~al.}(2003)\citenamefont
  {Volonteri}, \citenamefont {Haardt},\ and\ \citenamefont
  {Madau}}]{Volonteri:2002vz}%
  \BibitemOpen
  \bibfield  {author} {\bibinfo {author} {\bibfnamefont {M.}~\bibnamefont
  {Volonteri}}, \bibinfo {author} {\bibfnamefont {F.}~\bibnamefont {Haardt}}, \
  and\ \bibinfo {author} {\bibfnamefont {P.}~\bibnamefont {Madau}},\ }\href
  {\doibase 10.1086/344675} {\bibfield  {journal} {\bibinfo  {journal}
  {Astrophys.J.}\ }\textbf {\bibinfo {volume} {582}},\ \bibinfo {pages} {559}
  (\bibinfo {year} {2003})},\ \Eprint {http://arxiv.org/abs/astro-ph/0207276}
  {arXiv:astro-ph/0207276 [astro-ph]} \BibitemShut {NoStop}%
\bibitem [{\citenamefont {Begelman}\ \emph {et~al.}(1980)\citenamefont
  {Begelman}, \citenamefont {Blandford},\ and\ \citenamefont
  {Rees}}]{Begelman:1980vb}%
  \BibitemOpen
  \bibfield  {author} {\bibinfo {author} {\bibfnamefont {M.}~\bibnamefont
  {Begelman}}, \bibinfo {author} {\bibfnamefont {R.}~\bibnamefont {Blandford}},
  \ and\ \bibinfo {author} {\bibfnamefont {M.}~\bibnamefont {Rees}},\ }\href
  {\doibase 10.1038/287307a0} {\bibfield  {journal} {\bibinfo  {journal}
  {Nature}\ }\textbf {\bibinfo {volume} {287}},\ \bibinfo {pages} {307}
  (\bibinfo {year} {1980})}\BibitemShut {NoStop}%
\bibitem [{\citenamefont {Mayer}(2013)}]{Mayer:2013jja}%
  \BibitemOpen
  \bibfield  {author} {\bibinfo {author} {\bibfnamefont {L.}~\bibnamefont
  {Mayer}},\ }\href {\doibase 10.1088/0264-9381/30/24/244008} {\bibfield
  {journal} {\bibinfo  {journal} {Class.Quant.Grav.}\ }\textbf {\bibinfo
  {volume} {30}},\ \bibinfo {pages} {244008} (\bibinfo {year} {2013})},\
  \Eprint {http://arxiv.org/abs/arxiv:1308.0431} {arXiv:arxiv:1308.0431
  [astro-ph.CO]} \BibitemShut {NoStop}%
\bibitem [{\citenamefont {Gold}\ \emph {et~al.}(2014)\citenamefont {Gold},
  \citenamefont {Paschalidis}, \citenamefont {Etienne}, \citenamefont
  {Shapiro},\ and\ \citenamefont {Pfeiffer}}]{Gold:2013zma}%
  \BibitemOpen
  \bibfield  {author} {\bibinfo {author} {\bibfnamefont {R.}~\bibnamefont
  {Gold}}, \bibinfo {author} {\bibfnamefont {V.}~\bibnamefont {Paschalidis}},
  \bibinfo {author} {\bibfnamefont {Z.~B.}\ \bibnamefont {Etienne}}, \bibinfo
  {author} {\bibfnamefont {S.~L.}\ \bibnamefont {Shapiro}}, \ and\ \bibinfo
  {author} {\bibfnamefont {H.~P.}\ \bibnamefont {Pfeiffer}},\ }\href {\doibase
  10.1103/PhysRevD.89.064060} {\bibfield  {journal} {\bibinfo  {journal}
  {Phys.Rev.}\ }\textbf {\bibinfo {volume} {D89}},\ \bibinfo {pages} {064060}
  (\bibinfo {year} {2014})},\ \Eprint {http://arxiv.org/abs/astro-ph/1312.0600}
  {arXiv:astro-ph/1312.0600 [astro-ph.HE]} \BibitemShut {NoStop}%
\bibitem [{LIG()}]{LIGO_web}%
  \BibitemOpen
  \href@noop {} {}\bibinfo {note} {{LIGO} --
  http://www.ligo.caltech.edu/}\BibitemShut {NoStop}%
\bibitem [{\citenamefont {Hobbs}\ \emph {et~al.}(2010)\citenamefont {Hobbs},
  \citenamefont {Archibald}, \citenamefont {Arzoumanian}, \citenamefont
  {Backer}, \citenamefont {Bailes} \emph {et~al.}}]{Hobbs:2009yy}%
  \BibitemOpen
  \bibfield  {author} {\bibinfo {author} {\bibfnamefont {G.}~\bibnamefont
  {Hobbs}}, \bibinfo {author} {\bibfnamefont {A.}~\bibnamefont {Archibald}},
  \bibinfo {author} {\bibfnamefont {Z.}~\bibnamefont {Arzoumanian}}, \bibinfo
  {author} {\bibfnamefont {D.}~\bibnamefont {Backer}}, \bibinfo {author}
  {\bibfnamefont {M.}~\bibnamefont {Bailes}},  \emph {et~al.},\ }\href
  {\doibase 10.1088/0264-9381/27/8/084013} {\bibfield  {journal} {\bibinfo
  {journal} {Class.Quant.Grav.}\ }\textbf {\bibinfo {volume} {27}},\ \bibinfo
  {pages} {084013} (\bibinfo {year} {2010})},\ \Eprint
  {http://arxiv.org/abs/arXiv:0911.5206} {arXiv:arXiv:0911.5206 [astro-ph.SR]}
  \BibitemShut {NoStop}%
\bibitem [{\citenamefont {Sesana}\ \emph
  {et~al.}(2008{\natexlab{a}})\citenamefont {Sesana}, \citenamefont {Vecchio},\
  and\ \citenamefont {Volonteri}}]{Sesana:2008xk}%
  \BibitemOpen
  \bibfield  {author} {\bibinfo {author} {\bibfnamefont {A.}~\bibnamefont
  {Sesana}}, \bibinfo {author} {\bibfnamefont {A.}~\bibnamefont {Vecchio}}, \
  and\ \bibinfo {author} {\bibfnamefont {M.}~\bibnamefont {Volonteri}},\
  }\href@noop {} {\  (\bibinfo {year} {2008}{\natexlab{a}})},\ \Eprint
  {http://arxiv.org/abs/arXiv:0809.3412} {arXiv:arXiv:0809.3412 [astro-ph]}
  \BibitemShut {NoStop}%
\bibitem [{\citenamefont {Sesana}(2012)}]{Sesana:2012ak}%
  \BibitemOpen
  \bibfield  {author} {\bibinfo {author} {\bibfnamefont {A.}~\bibnamefont
  {Sesana}},\ }\href@noop {} {\  (\bibinfo {year} {2012})},\ \Eprint
  {http://arxiv.org/abs/arXiv:1211.5375} {arXiv:arXiv:1211.5375 [astro-ph.CO]}
  \BibitemShut {NoStop}%
\bibitem [{\citenamefont {Sesana}\ \emph
  {et~al.}(2008{\natexlab{b}})\citenamefont {Sesana}, \citenamefont {Vecchio},\
  and\ \citenamefont {Colacino}}]{Sesana:2008mz}%
  \BibitemOpen
  \bibfield  {author} {\bibinfo {author} {\bibfnamefont {A.}~\bibnamefont
  {Sesana}}, \bibinfo {author} {\bibfnamefont {A.}~\bibnamefont {Vecchio}}, \
  and\ \bibinfo {author} {\bibfnamefont {C.~N.}\ \bibnamefont {Colacino}},\
  }\href@noop {} {\  (\bibinfo {year} {2008}{\natexlab{b}})},\ \Eprint
  {http://arxiv.org/abs/arXiv:0804.4476} {arXiv:arXiv:0804.4476 [astro-ph]}
  \BibitemShut {NoStop}%
\bibitem [{\citenamefont {Taylor}\ \emph {et~al.}(2014)\citenamefont {Taylor},
  \citenamefont {Ellis},\ and\ \citenamefont {Gair}}]{Taylor:2014iua}%
  \BibitemOpen
  \bibfield  {author} {\bibinfo {author} {\bibfnamefont {S.}~\bibnamefont
  {Taylor}}, \bibinfo {author} {\bibfnamefont {J.}~\bibnamefont {Ellis}}, \
  and\ \bibinfo {author} {\bibfnamefont {J.}~\bibnamefont {Gair}},\ }\href@noop
  {} {\  (\bibinfo {year} {2014})},\ \Eprint
  {http://arxiv.org/abs/gr-qc/1406.5224} {arXiv:gr-qc/1406.5224 [gr-qc]}
  \BibitemShut {NoStop}%
\bibitem [{\citenamefont {Wang}\ \emph {et~al.}(2014)\citenamefont {Wang},
  \citenamefont {Mohanty},\ and\ \citenamefont {Jenet}}]{Wang:2014ava}%
  \BibitemOpen
  \bibfield  {author} {\bibinfo {author} {\bibfnamefont {Y.}~\bibnamefont
  {Wang}}, \bibinfo {author} {\bibfnamefont {S.~D.}\ \bibnamefont {Mohanty}}, \
  and\ \bibinfo {author} {\bibfnamefont {F.~A.}\ \bibnamefont {Jenet}},\ }\href
  {\doibase 10.1088/0004-637X/795/1/96} {\  (\bibinfo {year} {2014}),\
  10.1088/0004-637X/795/1/96},\ \Eprint {http://arxiv.org/abs/arXiv:1406.5496}
  {arXiv:arXiv:1406.5496 [astro-ph.IM]} \BibitemShut {NoStop}%
\bibitem [{\citenamefont {Sesana}\ \emph {et~al.}(2011)\citenamefont {Sesana},
  \citenamefont {Roedig}, \citenamefont {Reynolds},\ and\ \citenamefont
  {Dotti}}]{Sesana:2011zv}%
  \BibitemOpen
  \bibfield  {author} {\bibinfo {author} {\bibfnamefont {A.}~\bibnamefont
  {Sesana}}, \bibinfo {author} {\bibfnamefont {C.}~\bibnamefont {Roedig}},
  \bibinfo {author} {\bibfnamefont {M.}~\bibnamefont {Reynolds}}, \ and\
  \bibinfo {author} {\bibfnamefont {M.}~\bibnamefont {Dotti}},\ }\href@noop {}
  {\  (\bibinfo {year} {2011})},\ \Eprint
  {http://arxiv.org/abs/arXiv:1107.2927} {arXiv:arXiv:1107.2927 [astro-ph.CO]}
  \BibitemShut {NoStop}%
\bibitem [{\citenamefont {{Tanaka}}\ \emph {et~al.}(2012)\citenamefont
  {{Tanaka}}, \citenamefont {{Menou}},\ and\ \citenamefont
  {{Haiman}}}]{Tanaka:2011af}%
  \BibitemOpen
  \bibfield  {author} {\bibinfo {author} {\bibfnamefont {T.}~\bibnamefont
  {{Tanaka}}}, \bibinfo {author} {\bibfnamefont {K.}~\bibnamefont {{Menou}}}, \
  and\ \bibinfo {author} {\bibfnamefont {Z.}~\bibnamefont {{Haiman}}},\ }\href
  {\doibase 10.1111/j.1365-2966.2011.20083.x} {\bibfield  {journal} {\bibinfo
  {journal} {\mnras}\ }\textbf {\bibinfo {volume} {420}},\ \bibinfo {pages}
  {705} (\bibinfo {year} {2012})},\ \Eprint
  {http://arxiv.org/abs/arXiv:1107.2937} {arXiv:arXiv:1107.2937 [astro-ph.CO]}
  \BibitemShut {NoStop}%
\bibitem [{\citenamefont {{Tanaka}}\ and\ \citenamefont
  {{Haiman}}(2013)}]{Tanaka:2013oju}%
  \BibitemOpen
  \bibfield  {author} {\bibinfo {author} {\bibfnamefont {T.~L.}\ \bibnamefont
  {{Tanaka}}}\ and\ \bibinfo {author} {\bibfnamefont {Z.}~\bibnamefont
  {{Haiman}}},\ }\href {\doibase 10.1088/0264-9381/30/22/224012} {\bibfield
  {journal} {\bibinfo  {journal} {Classical and Quantum Gravity}\ }\textbf
  {\bibinfo {volume} {30}},\ \bibinfo {eid} {224012} (\bibinfo {year}
  {2013})},\ \Eprint {http://arxiv.org/abs/arXiv:1309.2302}
  {arXiv:arXiv:1309.2302 [astro-ph.CO]} \BibitemShut {NoStop}%
\bibitem [{\citenamefont {Volonteri}\ \emph {et~al.}(2009)\citenamefont
  {Volonteri}, \citenamefont {Miller},\ and\ \citenamefont
  {Dotti}}]{Volonteri:2009nh}%
  \BibitemOpen
  \bibfield  {author} {\bibinfo {author} {\bibfnamefont {M.}~\bibnamefont
  {Volonteri}}, \bibinfo {author} {\bibfnamefont {J.}~\bibnamefont {Miller}}, \
  and\ \bibinfo {author} {\bibfnamefont {M.}~\bibnamefont {Dotti}},\ }\href
  {\doibase 10.1088/0004-637X/703/1/L86} {\bibfield  {journal} {\bibinfo
  {journal} {Astrophys.J.}\ }\textbf {\bibinfo {volume} {703}},\ \bibinfo
  {pages} {L86} (\bibinfo {year} {2009})},\ \Eprint
  {http://arxiv.org/abs/arXiv:0903.3947} {arXiv:arXiv:0903.3947 [astro-ph.CO]}
  \BibitemShut {NoStop}%
\bibitem [{\citenamefont {Ju}\ \emph {et~al.}(2013)\citenamefont {Ju},
  \citenamefont {Greene}, \citenamefont {Rafikov}, \citenamefont {Bickerton},\
  and\ \citenamefont {Badenes}}]{Ju:2013hna}%
  \BibitemOpen
  \bibfield  {author} {\bibinfo {author} {\bibfnamefont {W.}~\bibnamefont
  {Ju}}, \bibinfo {author} {\bibfnamefont {J.~E.}\ \bibnamefont {Greene}},
  \bibinfo {author} {\bibfnamefont {R.~R.}\ \bibnamefont {Rafikov}}, \bibinfo
  {author} {\bibfnamefont {S.~J.}\ \bibnamefont {Bickerton}}, \ and\ \bibinfo
  {author} {\bibfnamefont {C.}~\bibnamefont {Badenes}},\ }\href {\doibase
  10.1088/0004-637X/777/1/44} {\bibfield  {journal} {\bibinfo  {journal}
  {Astrophys.J.}\ }\textbf {\bibinfo {volume} {777}},\ \bibinfo {pages} {44}
  (\bibinfo {year} {2013})},\ \Eprint {http://arxiv.org/abs/astro-ph/1306.4987}
  {arXiv:astro-ph/1306.4987 [astro-ph.CO]} \BibitemShut {NoStop}%
\bibitem [{\citenamefont {Kaiser}\ \emph {et~al.}(2002)\citenamefont {Kaiser},
  \citenamefont {Aussel}, \citenamefont {Boesgaard}, \citenamefont {Chambers},
  \citenamefont {Heasley} \emph {et~al.}}]{Kaiser:2002zz}%
  \BibitemOpen
  \bibfield  {author} {\bibinfo {author} {\bibfnamefont {N.}~\bibnamefont
  {Kaiser}}, \bibinfo {author} {\bibfnamefont {H.}~\bibnamefont {Aussel}},
  \bibinfo {author} {\bibfnamefont {H.}~\bibnamefont {Boesgaard}}, \bibinfo
  {author} {\bibfnamefont {K.}~\bibnamefont {Chambers}}, \bibinfo {author}
  {\bibfnamefont {J.~N.}\ \bibnamefont {Heasley}},  \emph {et~al.},\ }\href
  {\doibase 10.1117/12.457365} {\bibfield  {journal} {\bibinfo  {journal}
  {Proc.SPIE Int.Soc.Opt.Eng.}\ }\textbf {\bibinfo {volume} {4836}},\ \bibinfo
  {pages} {154} (\bibinfo {year} {2002})}\BibitemShut {NoStop}%
\bibitem [{\citenamefont {Abell}\ \emph {et~al.}(2009)\citenamefont {Abell}
  \emph {et~al.}}]{Abell:2009aa}%
  \BibitemOpen
  \bibfield  {author} {\bibinfo {author} {\bibfnamefont {P.~A.}\ \bibnamefont
  {Abell}} \emph {et~al.} (\bibinfo {collaboration} {LSST Science
  Collaborations, LSST Project}),\ }\href@noop {} {\  (\bibinfo {year}
  {2009})},\ \Eprint {http://arxiv.org/abs/astro-ph/0912.0201}
  {arXiv:astro-ph/0912.0201 [astro-ph.IM]} \BibitemShut {NoStop}%
\bibitem [{\citenamefont {Green}\ \emph {et~al.}(2012)\citenamefont {Green},
  \citenamefont {Schechter}, \citenamefont {Baltay}, \citenamefont {Bean},
  \citenamefont {Bennett} \emph {et~al.}}]{Green:2012mj}%
  \BibitemOpen
  \bibfield  {author} {\bibinfo {author} {\bibfnamefont {J.}~\bibnamefont
  {Green}}, \bibinfo {author} {\bibfnamefont {P.}~\bibnamefont {Schechter}},
  \bibinfo {author} {\bibfnamefont {C.}~\bibnamefont {Baltay}}, \bibinfo
  {author} {\bibfnamefont {R.}~\bibnamefont {Bean}}, \bibinfo {author}
  {\bibfnamefont {D.}~\bibnamefont {Bennett}},  \emph {et~al.},\ }\href@noop {}
  {\  (\bibinfo {year} {2012})},\ \Eprint
  {http://arxiv.org/abs/astro-ph/1208.4012} {arXiv:astro-ph/1208.4012
  [astro-ph.IM]} \BibitemShut {NoStop}%
\bibitem [{\citenamefont {Haiman}\ \emph {et~al.}(2009)\citenamefont {Haiman},
  \citenamefont {Kocsis},\ and\ \citenamefont {Menou}}]{Haiman:2009te}%
  \BibitemOpen
  \bibfield  {author} {\bibinfo {author} {\bibfnamefont {Z.}~\bibnamefont
  {Haiman}}, \bibinfo {author} {\bibfnamefont {B.}~\bibnamefont {Kocsis}}, \
  and\ \bibinfo {author} {\bibfnamefont {K.}~\bibnamefont {Menou}},\ }\href
  {\doibase 10.1088/0004-637X/700/2/1952} {\bibfield  {journal} {\bibinfo
  {journal} {Astrophys.J.}\ }\textbf {\bibinfo {volume} {700}},\ \bibinfo
  {pages} {1952} (\bibinfo {year} {2009})},\ \Eprint
  {http://arxiv.org/abs/astro-ph/0904.1383} {arXiv:astro-ph/0904.1383
  [astro-ph.CO]} \BibitemShut {NoStop}%
\bibitem [{\citenamefont {Tanaka}\ and\ \citenamefont
  {Menou}(2010)}]{Tanaka:2009iy}%
  \BibitemOpen
  \bibfield  {author} {\bibinfo {author} {\bibfnamefont {T.}~\bibnamefont
  {Tanaka}}\ and\ \bibinfo {author} {\bibfnamefont {K.}~\bibnamefont {Menou}},\
  }\href {\doibase 10.1088/0004-637X/714/1/404} {\bibfield  {journal} {\bibinfo
   {journal} {Astrophys.J.}\ }\textbf {\bibinfo {volume} {714}},\ \bibinfo
  {pages} {404} (\bibinfo {year} {2010})},\ \Eprint
  {http://arxiv.org/abs/astro-ph/0912.2054} {arXiv:astro-ph/0912.2054
  [astro-ph.CO]} \BibitemShut {NoStop}%
\bibitem [{\citenamefont {Shapiro}(2010)}]{Shapiro:2009uy}%
  \BibitemOpen
  \bibfield  {author} {\bibinfo {author} {\bibfnamefont {S.~L.}\ \bibnamefont
  {Shapiro}},\ }\href {\doibase 10.1103/PhysRevD.81.024019} {\bibfield
  {journal} {\bibinfo  {journal} {Phys.Rev.}\ }\textbf {\bibinfo {volume}
  {D81}},\ \bibinfo {pages} {024019} (\bibinfo {year} {2010})},\ \Eprint
  {http://arxiv.org/abs/astro-ph/0912.2345} {arXiv:astro-ph/0912.2345
  [astro-ph.HE]} \BibitemShut {NoStop}%
\bibitem [{\citenamefont {Liu}\ and\ \citenamefont
  {Shapiro}(2010)}]{Liu:2010mh}%
  \BibitemOpen
  \bibfield  {author} {\bibinfo {author} {\bibfnamefont {Y.~T.}\ \bibnamefont
  {Liu}}\ and\ \bibinfo {author} {\bibfnamefont {S.~L.}\ \bibnamefont
  {Shapiro}},\ }\href {\doibase 10.1103/PhysRevD.82.123011} {\bibfield
  {journal} {\bibinfo  {journal} {Phys.Rev.}\ }\textbf {\bibinfo {volume}
  {D82}},\ \bibinfo {pages} {123011} (\bibinfo {year} {2010})},\ \Eprint
  {http://arxiv.org/abs/astro-ph/1011.0002} {arXiv:astro-ph/1011.0002
  [astro-ph.HE]} \BibitemShut {NoStop}%
\bibitem [{\citenamefont {Kocsis}\ \emph {et~al.}(2012)\citenamefont {Kocsis},
  \citenamefont {Haiman},\ and\ \citenamefont {Loeb}}]{Kocsis:2012ui}%
  \BibitemOpen
  \bibfield  {author} {\bibinfo {author} {\bibfnamefont {B.}~\bibnamefont
  {Kocsis}}, \bibinfo {author} {\bibfnamefont {Z.}~\bibnamefont {Haiman}}, \
  and\ \bibinfo {author} {\bibfnamefont {A.}~\bibnamefont {Loeb}},\ }\href@noop
  {} {\bibfield  {journal} {\bibinfo  {journal} {Mon. Not. Roy. Astron. Soc.}\
  }\textbf {\bibinfo {volume} {427}},\ \bibinfo {pages} {2680} (\bibinfo {year}
  {2012})},\ \Eprint {http://arxiv.org/abs/astro-ph/1205.5268}
  {arXiv:astro-ph/1205.5268 [astro-ph.HE]} \BibitemShut {NoStop}%
\bibitem [{\citenamefont {Shapiro}(2013)}]{Shapiro:2013qsa}%
  \BibitemOpen
  \bibfield  {author} {\bibinfo {author} {\bibfnamefont {S.~L.}\ \bibnamefont
  {Shapiro}},\ }\href {\doibase 10.1103/PhysRevD.87.103009} {\bibfield
  {journal} {\bibinfo  {journal} {Phys.Rev.}\ }\textbf {\bibinfo {volume}
  {D87}},\ \bibinfo {pages} {103009} (\bibinfo {year} {2013})},\ \Eprint
  {http://arxiv.org/abs/astro-ph/1304.6090} {arXiv:astro-ph/1304.6090
  [astro-ph.HE]} \BibitemShut {NoStop}%
\bibitem [{\citenamefont {Barausse}\ \emph {et~al.}(2014)\citenamefont
  {Barausse}, \citenamefont {Bellovary}, \citenamefont {Berti}, \citenamefont
  {Holley-Bockelmann}, \citenamefont {Farris} \emph
  {et~al.}}]{Barausse:2014oca}%
  \BibitemOpen
  \bibfield  {author} {\bibinfo {author} {\bibfnamefont {E.}~\bibnamefont
  {Barausse}}, \bibinfo {author} {\bibfnamefont {J.}~\bibnamefont {Bellovary}},
  \bibinfo {author} {\bibfnamefont {E.}~\bibnamefont {Berti}}, \bibinfo
  {author} {\bibfnamefont {K.}~\bibnamefont {Holley-Bockelmann}}, \bibinfo
  {author} {\bibfnamefont {B.}~\bibnamefont {Farris}},  \emph {et~al.},\
  }\href@noop {} {\  (\bibinfo {year} {2014})},\ \Eprint
  {http://arxiv.org/abs/1410.2907} {arXiv:1410.2907 [astro-ph.HE]} \BibitemShut
  {NoStop}%
\bibitem [{\citenamefont {Artymowicz}\ and\ \citenamefont
  {Lubow}(1994)}]{Artymowicz:1994bw}%
  \BibitemOpen
  \bibfield  {author} {\bibinfo {author} {\bibfnamefont {P.}~\bibnamefont
  {Artymowicz}}\ and\ \bibinfo {author} {\bibfnamefont {S.~H.}\ \bibnamefont
  {Lubow}},\ }\href {\doibase 10.1086/173679} {\bibfield  {journal} {\bibinfo
  {journal} {Astrophys.J.}\ }\textbf {\bibinfo {volume} {421}},\ \bibinfo
  {pages} {651} (\bibinfo {year} {1994})}\BibitemShut {NoStop}%
\bibitem [{\citenamefont {Cuadra}\ \emph {et~al.}(2008)\citenamefont {Cuadra},
  \citenamefont {Armitage}, \citenamefont {Alexander},\ and\ \citenamefont
  {Begelman}}]{Cuadra:2008xn}%
  \BibitemOpen
  \bibfield  {author} {\bibinfo {author} {\bibfnamefont {J.}~\bibnamefont
  {Cuadra}}, \bibinfo {author} {\bibfnamefont {P.}~\bibnamefont {Armitage}},
  \bibinfo {author} {\bibfnamefont {R.}~\bibnamefont {Alexander}}, \ and\
  \bibinfo {author} {\bibfnamefont {M.}~\bibnamefont {Begelman}},\ }\href@noop
  {} {\  (\bibinfo {year} {2008})},\ \Eprint
  {http://arxiv.org/abs/arXiv:0809.0311} {arXiv:arXiv:0809.0311 [astro-ph]}
  \BibitemShut {NoStop}%
\bibitem [{\citenamefont {{Roedig}}\ \emph {et~al.}(2011)\citenamefont
  {{Roedig}}, \citenamefont {{Dotti}}, \citenamefont {{Sesana}}, \citenamefont
  {{Cuadra}},\ and\ \citenamefont {{Colpi}}}]{Rodig:2011jz}%
  \BibitemOpen
  \bibfield  {author} {\bibinfo {author} {\bibfnamefont {C.}~\bibnamefont
  {{Roedig}}}, \bibinfo {author} {\bibfnamefont {M.}~\bibnamefont {{Dotti}}},
  \bibinfo {author} {\bibfnamefont {A.}~\bibnamefont {{Sesana}}}, \bibinfo
  {author} {\bibfnamefont {J.}~\bibnamefont {{Cuadra}}}, \ and\ \bibinfo
  {author} {\bibfnamefont {M.}~\bibnamefont {{Colpi}}},\ }\href {\doibase
  10.1111/j.1365-2966.2011.18927.x} {\bibfield  {journal} {\bibinfo  {journal}
  {\mnras}\ }\textbf {\bibinfo {volume} {415}},\ \bibinfo {pages} {3033}
  (\bibinfo {year} {2011})},\ \Eprint {http://arxiv.org/abs/arXiv:1104.3868}
  {arXiv:arXiv:1104.3868 [astro-ph.CO]} \BibitemShut {NoStop}%
\bibitem [{\citenamefont {Roedig}\ \emph {et~al.}(2012)\citenamefont {Roedig},
  \citenamefont {Sesana}, \citenamefont {Dotti}, \citenamefont {Cuadra},
  \citenamefont {Amaro-Seoane} \emph {et~al.}}]{Roedig:2012nc}%
  \BibitemOpen
  \bibfield  {author} {\bibinfo {author} {\bibfnamefont {C.}~\bibnamefont
  {Roedig}}, \bibinfo {author} {\bibfnamefont {A.}~\bibnamefont {Sesana}},
  \bibinfo {author} {\bibfnamefont {M.}~\bibnamefont {Dotti}}, \bibinfo
  {author} {\bibfnamefont {J.}~\bibnamefont {Cuadra}}, \bibinfo {author}
  {\bibfnamefont {P.}~\bibnamefont {Amaro-Seoane}},  \emph {et~al.},\
  }\href@noop {} {\  (\bibinfo {year} {2012})},\ \Eprint
  {http://arxiv.org/abs/arXiv:1202.6063} {arXiv:arXiv:1202.6063 [astro-ph.CO]}
  \BibitemShut {NoStop}%
\bibitem [{\citenamefont {MacFadyen}\ and\ \citenamefont
  {Milosavljevic}(2008)}]{MacFadyen:2006jx}%
  \BibitemOpen
  \bibfield  {author} {\bibinfo {author} {\bibfnamefont {A.~I.}\ \bibnamefont
  {MacFadyen}}\ and\ \bibinfo {author} {\bibfnamefont {M.}~\bibnamefont
  {Milosavljevic}},\ }\href {\doibase 10.1086/523869} {\bibfield  {journal}
  {\bibinfo  {journal} {Astrophys.J.}\ }\textbf {\bibinfo {volume} {672}},\
  \bibinfo {pages} {83} (\bibinfo {year} {2008})},\ \Eprint
  {http://arxiv.org/abs/astro-ph/0607467} {arXiv:astro-ph/0607467 [astro-ph]}
  \BibitemShut {NoStop}%
\bibitem [{\citenamefont {D'Orazio}\ \emph {et~al.}(2012)\citenamefont
  {D'Orazio}, \citenamefont {Haiman},\ and\ \citenamefont
  {MacFadyen}}]{D'Orazio:2012nz}%
  \BibitemOpen
  \bibfield  {author} {\bibinfo {author} {\bibfnamefont {D.~J.}\ \bibnamefont
  {D'Orazio}}, \bibinfo {author} {\bibfnamefont {Z.}~\bibnamefont {Haiman}}, \
  and\ \bibinfo {author} {\bibfnamefont {A.~I.}\ \bibnamefont {MacFadyen}},\
  }\href@noop {} {\  (\bibinfo {year} {2012})},\ \Eprint
  {http://arxiv.org/abs/astro-ph/1210.0536} {arXiv:astro-ph/1210.0536
  [astro-ph.GA]} \BibitemShut {NoStop}%
\bibitem [{\citenamefont {Farris}\ \emph
  {et~al.}(2014{\natexlab{a}})\citenamefont {Farris}, \citenamefont {Duffell},
  \citenamefont {MacFadyen},\ and\ \citenamefont {Haiman}}]{Farris:2014qma}%
  \BibitemOpen
  \bibfield  {author} {\bibinfo {author} {\bibfnamefont {B.~D.}\ \bibnamefont
  {Farris}}, \bibinfo {author} {\bibfnamefont {P.}~\bibnamefont {Duffell}},
  \bibinfo {author} {\bibfnamefont {A.~I.}\ \bibnamefont {MacFadyen}}, \ and\
  \bibinfo {author} {\bibfnamefont {Z.}~\bibnamefont {Haiman}},\ }\href@noop {}
  {\  (\bibinfo {year} {2014}{\natexlab{a}})},\ \Eprint
  {http://arxiv.org/abs/astro-ph/1409.5124} {arXiv:astro-ph/1409.5124
  [astro-ph.HE]} \BibitemShut {NoStop}%
\bibitem [{\citenamefont {Shi}\ \emph {et~al.}(2012)\citenamefont {Shi},
  \citenamefont {Krolik}, \citenamefont {Lubow},\ and\ \citenamefont
  {Hawley}}]{Shi:2011us}%
  \BibitemOpen
  \bibfield  {author} {\bibinfo {author} {\bibfnamefont {J.-M.}\ \bibnamefont
  {Shi}}, \bibinfo {author} {\bibfnamefont {J.~H.}\ \bibnamefont {Krolik}},
  \bibinfo {author} {\bibfnamefont {S.~H.}\ \bibnamefont {Lubow}}, \ and\
  \bibinfo {author} {\bibfnamefont {J.~F.}\ \bibnamefont {Hawley}},\ }\href
  {\doibase 10.1088/0004-637X/749/2/118} {\bibfield  {journal} {\bibinfo
  {journal} {Astrophys.J.}\ }\textbf {\bibinfo {volume} {749}},\ \bibinfo
  {pages} {118} (\bibinfo {year} {2012})},\ \Eprint
  {http://arxiv.org/abs/astro-ph/1110.4866} {arXiv:astro-ph/1110.4866
  [astro-ph.HE]} \BibitemShut {NoStop}%
\bibitem [{\citenamefont {Noble}\ \emph {et~al.}(2012)\citenamefont {Noble},
  \citenamefont {Mundim}, \citenamefont {Nakano}, \citenamefont {Krolik},
  \citenamefont {Campanelli} \emph {et~al.}}]{Noble:2012xz}%
  \BibitemOpen
  \bibfield  {author} {\bibinfo {author} {\bibfnamefont {S.~C.}\ \bibnamefont
  {Noble}}, \bibinfo {author} {\bibfnamefont {B.~C.}\ \bibnamefont {Mundim}},
  \bibinfo {author} {\bibfnamefont {H.}~\bibnamefont {Nakano}}, \bibinfo
  {author} {\bibfnamefont {J.~H.}\ \bibnamefont {Krolik}}, \bibinfo {author}
  {\bibfnamefont {M.}~\bibnamefont {Campanelli}},  \emph {et~al.},\ }\href
  {\doibase 10.1088/0004-637X/755/1/51} {\bibfield  {journal} {\bibinfo
  {journal} {Astrophys.J.}\ }\textbf {\bibinfo {volume} {755}},\ \bibinfo
  {pages} {51} (\bibinfo {year} {2012})},\ \Eprint
  {http://arxiv.org/abs/astro-ph/1204.1073} {arXiv:astro-ph/1204.1073
  [astro-ph.HE]} \BibitemShut {NoStop}%
\bibitem [{\citenamefont {Zilhão}\ \emph {et~al.}(2014)\citenamefont
  {Zilhão}, \citenamefont {Noble}, \citenamefont {Campanelli},\ and\
  \citenamefont {Zlochower}}]{Zilhao:2014ida}%
  \BibitemOpen
  \bibfield  {author} {\bibinfo {author} {\bibfnamefont {M.}~\bibnamefont
  {Zilhão}}, \bibinfo {author} {\bibfnamefont {S.~C.}\ \bibnamefont {Noble}},
  \bibinfo {author} {\bibfnamefont {M.}~\bibnamefont {Campanelli}}, \ and\
  \bibinfo {author} {\bibfnamefont {Y.}~\bibnamefont {Zlochower}},\ }\href@noop
  {} {\  (\bibinfo {year} {2014})},\ \Eprint
  {http://arxiv.org/abs/astro-ph/1409.4787} {arXiv:astro-ph/1409.4787 [gr-qc]}
  \BibitemShut {NoStop}%
\bibitem [{\citenamefont {Dunhill}\ \emph {et~al.}(2014)\citenamefont
  {Dunhill}, \citenamefont {Alexander}, \citenamefont {Nixon},\ and\
  \citenamefont {King}}]{Dunhill:2014oka}%
  \BibitemOpen
  \bibfield  {author} {\bibinfo {author} {\bibfnamefont {A.}~\bibnamefont
  {Dunhill}}, \bibinfo {author} {\bibfnamefont {R.}~\bibnamefont {Alexander}},
  \bibinfo {author} {\bibfnamefont {C.}~\bibnamefont {Nixon}}, \ and\ \bibinfo
  {author} {\bibfnamefont {A.}~\bibnamefont {King}},\ }\href@noop {} {\
  (\bibinfo {year} {2014})},\ \Eprint {http://arxiv.org/abs/astro-ph/1409.3842}
  {arXiv:astro-ph/1409.3842 [astro-ph.HE]} \BibitemShut {NoStop}%
\bibitem [{\citenamefont {Mosta}\ \emph {et~al.}(2010)\citenamefont {Mosta},
  \citenamefont {Palenzuela}, \citenamefont {Rezzolla}, \citenamefont {Lehner},
  \citenamefont {Yoshida} \emph {et~al.}}]{Mosta:2009rr}%
  \BibitemOpen
  \bibfield  {author} {\bibinfo {author} {\bibfnamefont {P.}~\bibnamefont
  {Mosta}}, \bibinfo {author} {\bibfnamefont {C.}~\bibnamefont {Palenzuela}},
  \bibinfo {author} {\bibfnamefont {L.}~\bibnamefont {Rezzolla}}, \bibinfo
  {author} {\bibfnamefont {L.}~\bibnamefont {Lehner}}, \bibinfo {author}
  {\bibfnamefont {S.}~\bibnamefont {Yoshida}},  \emph {et~al.},\ }\href
  {\doibase 10.1103/PhysRevD.81.064017} {\bibfield  {journal} {\bibinfo
  {journal} {Phys.Rev.}\ }\textbf {\bibinfo {volume} {D81}},\ \bibinfo {pages}
  {064017} (\bibinfo {year} {2010})},\ \Eprint
  {http://arxiv.org/abs/arXiv:0912.2330} {arXiv:arXiv:0912.2330 [gr-qc]}
  \BibitemShut {NoStop}%
\bibitem [{\citenamefont {Neilsen}\ \emph {et~al.}(2011)\citenamefont
  {Neilsen}, \citenamefont {Lehner}, \citenamefont {Palenzuela}, \citenamefont
  {Hirschmann}, \citenamefont {Liebling} \emph {et~al.}}]{Neilsen:2010ax}%
  \BibitemOpen
  \bibfield  {author} {\bibinfo {author} {\bibfnamefont {D.}~\bibnamefont
  {Neilsen}}, \bibinfo {author} {\bibfnamefont {L.}~\bibnamefont {Lehner}},
  \bibinfo {author} {\bibfnamefont {C.}~\bibnamefont {Palenzuela}}, \bibinfo
  {author} {\bibfnamefont {E.~W.}\ \bibnamefont {Hirschmann}}, \bibinfo
  {author} {\bibfnamefont {S.~L.}\ \bibnamefont {Liebling}},  \emph {et~al.},\
  }\href {\doibase 10.1073/pnas.1019618108} {\bibfield  {journal} {\bibinfo
  {journal} {Proc.Nat.Acad.Sci.}\ }\textbf {\bibinfo {volume} {108}},\ \bibinfo
  {pages} {12641} (\bibinfo {year} {2011})},\ \Eprint
  {http://arxiv.org/abs/astro-ph/1012.5661} {arXiv:astro-ph/1012.5661
  [astro-ph.HE]} \BibitemShut {NoStop}%
\bibitem [{\citenamefont {Palenzuela}\ \emph
  {et~al.}(2010{\natexlab{a}})\citenamefont {Palenzuela}, \citenamefont
  {Garrett}, \citenamefont {Lehner},\ and\ \citenamefont
  {Liebling}}]{Palenzuela:2010xn}%
  \BibitemOpen
  \bibfield  {author} {\bibinfo {author} {\bibfnamefont {C.}~\bibnamefont
  {Palenzuela}}, \bibinfo {author} {\bibfnamefont {T.}~\bibnamefont {Garrett}},
  \bibinfo {author} {\bibfnamefont {L.}~\bibnamefont {Lehner}}, \ and\ \bibinfo
  {author} {\bibfnamefont {S.~L.}\ \bibnamefont {Liebling}},\ }\href {\doibase
  10.1103/PhysRevD.82.044045} {\bibfield  {journal} {\bibinfo  {journal}
  {Phys.Rev.}\ }\textbf {\bibinfo {volume} {D82}},\ \bibinfo {pages} {044045}
  (\bibinfo {year} {2010}{\natexlab{a}})},\ \Eprint
  {http://arxiv.org/abs/gr-qc/1007.1198} {arXiv:gr-qc/1007.1198 [gr-qc]}
  \BibitemShut {NoStop}%
\bibitem [{\citenamefont {Palenzuela}\ \emph
  {et~al.}(2010{\natexlab{b}})\citenamefont {Palenzuela}, \citenamefont
  {Lehner},\ and\ \citenamefont {Liebling}}]{Palenzuela:2010nf}%
  \BibitemOpen
  \bibfield  {author} {\bibinfo {author} {\bibfnamefont {C.}~\bibnamefont
  {Palenzuela}}, \bibinfo {author} {\bibfnamefont {L.}~\bibnamefont {Lehner}},
  \ and\ \bibinfo {author} {\bibfnamefont {S.~L.}\ \bibnamefont {Liebling}},\
  }\href {\doibase 10.1126/science.1191766} {\bibfield  {journal} {\bibinfo
  {journal} {Science}\ }\textbf {\bibinfo {volume} {329}},\ \bibinfo {pages}
  {927} (\bibinfo {year} {2010}{\natexlab{b}})},\ \Eprint
  {http://arxiv.org/abs/astro-ph/1005.1067} {arXiv:astro-ph/1005.1067
  [astro-ph.HE]} \BibitemShut {NoStop}%
\bibitem [{\citenamefont {Alic}\ \emph {et~al.}(2012)\citenamefont {Alic},
  \citenamefont {Mosta}, \citenamefont {Rezzolla}, \citenamefont {Zanotti},\
  and\ \citenamefont {Jaramillo}}]{Alic:2012df}%
  \BibitemOpen
  \bibfield  {author} {\bibinfo {author} {\bibfnamefont {D.}~\bibnamefont
  {Alic}}, \bibinfo {author} {\bibfnamefont {P.}~\bibnamefont {Mosta}},
  \bibinfo {author} {\bibfnamefont {L.}~\bibnamefont {Rezzolla}}, \bibinfo
  {author} {\bibfnamefont {O.}~\bibnamefont {Zanotti}}, \ and\ \bibinfo
  {author} {\bibfnamefont {J.~L.}\ \bibnamefont {Jaramillo}},\ }\href {\doibase
  10.1088/0004-637X/754/1/36} {\bibfield  {journal} {\bibinfo  {journal}
  {Astrophys.J.}\ }\textbf {\bibinfo {volume} {754}},\ \bibinfo {pages} {36}
  (\bibinfo {year} {2012})},\ \Eprint {http://arxiv.org/abs/arXiv:1204.2226}
  {arXiv:arXiv:1204.2226 [gr-qc]} \BibitemShut {NoStop}%
\bibitem [{\citenamefont {Bode}\ \emph {et~al.}(2010)\citenamefont {Bode},
  \citenamefont {Haas}, \citenamefont {Bogdanovic}, \citenamefont {Laguna},\
  and\ \citenamefont {Shoemaker}}]{Bode:2009mt}%
  \BibitemOpen
  \bibfield  {author} {\bibinfo {author} {\bibfnamefont {T.}~\bibnamefont
  {Bode}}, \bibinfo {author} {\bibfnamefont {R.}~\bibnamefont {Haas}}, \bibinfo
  {author} {\bibfnamefont {T.}~\bibnamefont {Bogdanovic}}, \bibinfo {author}
  {\bibfnamefont {P.}~\bibnamefont {Laguna}}, \ and\ \bibinfo {author}
  {\bibfnamefont {D.}~\bibnamefont {Shoemaker}},\ }\href {\doibase
  10.1088/0004-637X/715/2/1117} {\bibfield  {journal} {\bibinfo  {journal}
  {Astrophys.J.}\ }\textbf {\bibinfo {volume} {715}},\ \bibinfo {pages} {1117}
  (\bibinfo {year} {2010})},\ \Eprint {http://arxiv.org/abs/gr-qc/0912.0087}
  {arXiv:gr-qc/0912.0087 [gr-qc]} \BibitemShut {NoStop}%
\bibitem [{\citenamefont {Bogdanovic}\ \emph {et~al.}(2011)\citenamefont
  {Bogdanovic}, \citenamefont {Bode}, \citenamefont {Haas}, \citenamefont
  {Laguna},\ and\ \citenamefont {Shoemaker}}]{Bogdanovic:2010he}%
  \BibitemOpen
  \bibfield  {author} {\bibinfo {author} {\bibfnamefont {T.}~\bibnamefont
  {Bogdanovic}}, \bibinfo {author} {\bibfnamefont {T.}~\bibnamefont {Bode}},
  \bibinfo {author} {\bibfnamefont {R.}~\bibnamefont {Haas}}, \bibinfo {author}
  {\bibfnamefont {P.}~\bibnamefont {Laguna}}, \ and\ \bibinfo {author}
  {\bibfnamefont {D.}~\bibnamefont {Shoemaker}},\ }\href {\doibase
  10.1088/0264-9381/28/9/094020} {\bibfield  {journal} {\bibinfo  {journal}
  {Class.Quant.Grav.}\ }\textbf {\bibinfo {volume} {28}},\ \bibinfo {pages}
  {094020} (\bibinfo {year} {2011})},\ \Eprint
  {http://arxiv.org/abs/astro-ph/1010.2496} {arXiv:astro-ph/1010.2496
  [astro-ph.CO]} \BibitemShut {NoStop}%
\bibitem [{\citenamefont {Bode}\ \emph {et~al.}(2012)\citenamefont {Bode},
  \citenamefont {Bogdanovic}, \citenamefont {Haas}, \citenamefont {Healy},
  \citenamefont {Laguna} \emph {et~al.}}]{Bode:2011tq}%
  \BibitemOpen
  \bibfield  {author} {\bibinfo {author} {\bibfnamefont {T.}~\bibnamefont
  {Bode}}, \bibinfo {author} {\bibfnamefont {T.}~\bibnamefont {Bogdanovic}},
  \bibinfo {author} {\bibfnamefont {R.}~\bibnamefont {Haas}}, \bibinfo {author}
  {\bibfnamefont {J.}~\bibnamefont {Healy}}, \bibinfo {author} {\bibfnamefont
  {P.}~\bibnamefont {Laguna}},  \emph {et~al.},\ }\href {\doibase
  10.1088/0004-637X/744/1/45} {\bibfield  {journal} {\bibinfo  {journal}
  {Astrophys.J.}\ }\textbf {\bibinfo {volume} {744}},\ \bibinfo {pages} {45}
  (\bibinfo {year} {2012})},\ \Eprint {http://arxiv.org/abs/gr-qc/1101.4684}
  {arXiv:gr-qc/1101.4684 [gr-qc]} \BibitemShut {NoStop}%
\bibitem [{\citenamefont {Farris}\ \emph {et~al.}(2011)\citenamefont {Farris},
  \citenamefont {Liu},\ and\ \citenamefont {Shapiro}}]{Farris:2011vx}%
  \BibitemOpen
  \bibfield  {author} {\bibinfo {author} {\bibfnamefont {B.~D.}\ \bibnamefont
  {Farris}}, \bibinfo {author} {\bibfnamefont {Y.~T.}\ \bibnamefont {Liu}}, \
  and\ \bibinfo {author} {\bibfnamefont {S.~L.}\ \bibnamefont {Shapiro}},\
  }\href {\doibase 10.1103/PhysRevD.84.024024} {\bibfield  {journal} {\bibinfo
  {journal} {Phys.Rev.}\ }\textbf {\bibinfo {volume} {D84}},\ \bibinfo {pages}
  {024024} (\bibinfo {year} {2011})},\ \Eprint
  {http://arxiv.org/abs/astro-ph/1105.2821} {arXiv:astro-ph/1105.2821
  [astro-ph.HE]} \BibitemShut {NoStop}%
\bibitem [{\citenamefont {Farris}\ \emph {et~al.}(2012)\citenamefont {Farris},
  \citenamefont {Gold}, \citenamefont {Paschalidis}, \citenamefont {Etienne},\
  and\ \citenamefont {Shapiro}}]{Farris:2012ux}%
  \BibitemOpen
  \bibfield  {author} {\bibinfo {author} {\bibfnamefont {B.~D.}\ \bibnamefont
  {Farris}}, \bibinfo {author} {\bibfnamefont {R.}~\bibnamefont {Gold}},
  \bibinfo {author} {\bibfnamefont {V.}~\bibnamefont {Paschalidis}}, \bibinfo
  {author} {\bibfnamefont {Z.~B.}\ \bibnamefont {Etienne}}, \ and\ \bibinfo
  {author} {\bibfnamefont {S.~L.}\ \bibnamefont {Shapiro}},\ }\href {\doibase
  10.1103/PhysRevLett.109.221102} {\bibfield  {journal} {\bibinfo  {journal}
  {Phys.Rev.Lett.}\ }\textbf {\bibinfo {volume} {109}},\ \bibinfo {pages}
  {221102} (\bibinfo {year} {2012})},\ \Eprint
  {http://arxiv.org/abs/astro-ph/1207.3354} {arXiv:astro-ph/1207.3354
  [astro-ph.HE]} \BibitemShut {NoStop}%
\bibitem [{\citenamefont {Giacomazzo}\ \emph {et~al.}(2012)\citenamefont
  {Giacomazzo}, \citenamefont {Baker}, \citenamefont {Miller}, \citenamefont
  {Reynolds},\ and\ \citenamefont {van Meter}}]{Giacomazzo:2012iv}%
  \BibitemOpen
  \bibfield  {author} {\bibinfo {author} {\bibfnamefont {B.}~\bibnamefont
  {Giacomazzo}}, \bibinfo {author} {\bibfnamefont {J.~G.}\ \bibnamefont
  {Baker}}, \bibinfo {author} {\bibfnamefont {M.~C.}\ \bibnamefont {Miller}},
  \bibinfo {author} {\bibfnamefont {C.~S.}\ \bibnamefont {Reynolds}}, \ and\
  \bibinfo {author} {\bibfnamefont {J.~R.}\ \bibnamefont {van Meter}},\
  }\href@noop {} {\bibfield  {journal} {\bibinfo  {journal} {Astrophys.J.}\
  }\textbf {\bibinfo {volume} {752}},\ \bibinfo {pages} {L15} (\bibinfo {year}
  {2012})},\ \Eprint {http://arxiv.org/abs/astro-ph/1203.6108}
  {arXiv:astro-ph/1203.6108 [astro-ph.HE]} \BibitemShut {NoStop}%
\bibitem [{\citenamefont {Corrales}\ \emph {et~al.}(2009)\citenamefont
  {Corrales}, \citenamefont {Haiman},\ and\ \citenamefont
  {MacFadyen}}]{Corrales:2009nv}%
  \BibitemOpen
  \bibfield  {author} {\bibinfo {author} {\bibfnamefont {L.~R.}\ \bibnamefont
  {Corrales}}, \bibinfo {author} {\bibfnamefont {Z.}~\bibnamefont {Haiman}}, \
  and\ \bibinfo {author} {\bibfnamefont {A.~I.}\ \bibnamefont {MacFadyen}},\
  }\href@noop {} {\  (\bibinfo {year} {2009})},\ \Eprint
  {http://arxiv.org/abs/astro-ph/0910.0014} {arXiv:astro-ph/0910.0014
  [astro-ph.HE]} \BibitemShut {NoStop}%
\bibitem [{\citenamefont {{Rossi}}\ \emph {et~al.}(2010)\citenamefont
  {{Rossi}}, \citenamefont {{Lodato}}, \citenamefont {{Armitage}},
  \citenamefont {{Pringle}},\ and\ \citenamefont {{King}}}]{Rossi:2009nk}%
  \BibitemOpen
  \bibfield  {author} {\bibinfo {author} {\bibfnamefont {E.~M.}\ \bibnamefont
  {{Rossi}}}, \bibinfo {author} {\bibfnamefont {G.}~\bibnamefont {{Lodato}}},
  \bibinfo {author} {\bibfnamefont {P.~J.}\ \bibnamefont {{Armitage}}},
  \bibinfo {author} {\bibfnamefont {J.~E.}\ \bibnamefont {{Pringle}}}, \ and\
  \bibinfo {author} {\bibfnamefont {A.~R.}\ \bibnamefont {{King}}},\ }\href
  {\doibase 10.1111/j.1365-2966.2009.15802.x} {\bibfield  {journal} {\bibinfo
  {journal} {\mnras}\ }\textbf {\bibinfo {volume} {401}},\ \bibinfo {pages}
  {2021} (\bibinfo {year} {2010})},\ \Eprint {http://arxiv.org/abs/0910.0002}
  {arXiv:0910.0002 [astro-ph.HE]} \BibitemShut {NoStop}%
\bibitem [{\citenamefont {Anderson}\ \emph {et~al.}(2010)\citenamefont
  {Anderson}, \citenamefont {Lehner}, \citenamefont {Megevand},\ and\
  \citenamefont {Neilsen}}]{Anderson:2009fa}%
  \BibitemOpen
  \bibfield  {author} {\bibinfo {author} {\bibfnamefont {M.}~\bibnamefont
  {Anderson}}, \bibinfo {author} {\bibfnamefont {L.}~\bibnamefont {Lehner}},
  \bibinfo {author} {\bibfnamefont {M.}~\bibnamefont {Megevand}}, \ and\
  \bibinfo {author} {\bibfnamefont {D.}~\bibnamefont {Neilsen}},\ }\href
  {\doibase 10.1103/PhysRevD.81.044004} {\bibfield  {journal} {\bibinfo
  {journal} {Phys.Rev.}\ }\textbf {\bibinfo {volume} {D81}},\ \bibinfo {pages}
  {044004} (\bibinfo {year} {2010})},\ \Eprint
  {http://arxiv.org/abs/astro-ph/0910.4969} {arXiv:astro-ph/0910.4969
  [astro-ph.HE]} \BibitemShut {NoStop}%
\bibitem [{\citenamefont {Megevand}\ \emph
  {et~al.}(2009{\natexlab{a}})\citenamefont {Megevand}, \citenamefont
  {Anderson}, \citenamefont {Frank}, \citenamefont {Hirschmann}, \citenamefont
  {Lehner} \emph {et~al.}}]{Megevand:2009yx}%
  \BibitemOpen
  \bibfield  {author} {\bibinfo {author} {\bibfnamefont {M.}~\bibnamefont
  {Megevand}}, \bibinfo {author} {\bibfnamefont {M.}~\bibnamefont {Anderson}},
  \bibinfo {author} {\bibfnamefont {J.}~\bibnamefont {Frank}}, \bibinfo
  {author} {\bibfnamefont {E.~W.}\ \bibnamefont {Hirschmann}}, \bibinfo
  {author} {\bibfnamefont {L.}~\bibnamefont {Lehner}},  \emph {et~al.},\ }\href
  {\doibase 10.1103/PhysRevD.80.024012} {\bibfield  {journal} {\bibinfo
  {journal} {Phys.Rev.}\ }\textbf {\bibinfo {volume} {D80}},\ \bibinfo {pages}
  {024012} (\bibinfo {year} {2009}{\natexlab{a}})},\ \Eprint
  {http://arxiv.org/abs/astro-ph/0905.3390} {arXiv:astro-ph/0905.3390
  [astro-ph.HE]} \BibitemShut {NoStop}%
\bibitem [{\citenamefont {Zanotti}\ \emph {et~al.}(2010)\citenamefont
  {Zanotti}, \citenamefont {Rezzolla}, \citenamefont {Del~Zanna},\ and\
  \citenamefont {Palenzuela}}]{Zanotti:2010xs}%
  \BibitemOpen
  \bibfield  {author} {\bibinfo {author} {\bibfnamefont {O.}~\bibnamefont
  {Zanotti}}, \bibinfo {author} {\bibfnamefont {L.}~\bibnamefont {Rezzolla}},
  \bibinfo {author} {\bibfnamefont {L.}~\bibnamefont {Del~Zanna}}, \ and\
  \bibinfo {author} {\bibfnamefont {C.}~\bibnamefont {Palenzuela}},\ }\href
  {\doibase 10.1051/0004-6361/201014969} {\bibfield  {journal} {\bibinfo
  {journal} {Astron.Astrophys.}\ }\textbf {\bibinfo {volume} {523}},\ \bibinfo
  {pages} {A8} (\bibinfo {year} {2010})},\ \Eprint
  {http://arxiv.org/abs/1002.4185} {arXiv:1002.4185 [astro-ph.HE]} \BibitemShut
  {NoStop}%
\bibitem [{\citenamefont {Ponce}\ \emph
  {et~al.}(2012{\natexlab{a}})\citenamefont {Ponce}, \citenamefont {Faber},\
  and\ \citenamefont {Lombardi}}]{Ponce:2011kv}%
  \BibitemOpen
  \bibfield  {author} {\bibinfo {author} {\bibfnamefont {M.}~\bibnamefont
  {Ponce}}, \bibinfo {author} {\bibfnamefont {J.~A.}\ \bibnamefont {Faber}}, \
  and\ \bibinfo {author} {\bibfnamefont {J.}~\bibnamefont {Lombardi},
  \bibfnamefont {James~C.}},\ }\href {\doibase 10.1088/0004-637X/745/1/71}
  {\bibfield  {journal} {\bibinfo  {journal} {Astrophys.J.}\ }\textbf {\bibinfo
  {volume} {745}},\ \bibinfo {pages} {71} (\bibinfo {year}
  {2012}{\natexlab{a}})},\ \Eprint {http://arxiv.org/abs/astro-ph/1107.1711}
  {arXiv:astro-ph/1107.1711 [astro-ph.CO]} \BibitemShut {NoStop}%
\bibitem [{\citenamefont {{Abramowicz}}\ \emph {et~al.}(1988)\citenamefont
  {{Abramowicz}}, \citenamefont {{Czerny}}, \citenamefont {{Lasota}},\ and\
  \citenamefont {{Szuszkiewicz}}}]{Abramowicz1988}%
  \BibitemOpen
  \bibfield  {author} {\bibinfo {author} {\bibfnamefont {M.~A.}\ \bibnamefont
  {{Abramowicz}}}, \bibinfo {author} {\bibfnamefont {B.}~\bibnamefont
  {{Czerny}}}, \bibinfo {author} {\bibfnamefont {J.~P.}\ \bibnamefont
  {{Lasota}}}, \ and\ \bibinfo {author} {\bibfnamefont {E.}~\bibnamefont
  {{Szuszkiewicz}}},\ }\href {\doibase 10.1086/166683} {\bibfield  {journal}
  {\bibinfo  {journal} {\apj}\ }\textbf {\bibinfo {volume} {332}},\ \bibinfo
  {pages} {646} (\bibinfo {year} {1988})}\BibitemShut {NoStop}%
\bibitem [{\citenamefont {Abramowicz}\ and\ \citenamefont
  {Fragile}(2013)}]{lrr-2013-1}%
  \BibitemOpen
  \bibfield  {author} {\bibinfo {author} {\bibfnamefont {M.~A.}\ \bibnamefont
  {Abramowicz}}\ and\ \bibinfo {author} {\bibfnamefont {P.~C.}\ \bibnamefont
  {Fragile}},\ }\href {\doibase 10.12942/lrr-2013-1} {\bibfield  {journal}
  {\bibinfo  {journal} {Living Reviews in Relativity}\ }\textbf {\bibinfo
  {volume} {16}} (\bibinfo {year} {2013}),\ 10.12942/lrr-2013-1}\BibitemShut
  {NoStop}%
\bibitem [{\citenamefont {Paschalidis}\ \emph {et~al.}(2011)\citenamefont
  {Paschalidis}, \citenamefont {Liu}, \citenamefont {Etienne},\ and\
  \citenamefont {Shapiro}}]{Paschalidis:2011ez}%
  \BibitemOpen
  \bibfield  {author} {\bibinfo {author} {\bibfnamefont {V.}~\bibnamefont
  {Paschalidis}}, \bibinfo {author} {\bibfnamefont {Y.~T.}\ \bibnamefont
  {Liu}}, \bibinfo {author} {\bibfnamefont {Z.}~\bibnamefont {Etienne}}, \ and\
  \bibinfo {author} {\bibfnamefont {S.~L.}\ \bibnamefont {Shapiro}},\ }\href
  {\doibase 10.1103/PhysRevD.84.104032} {\bibfield  {journal} {\bibinfo
  {journal} {Phys.Rev.}\ }\textbf {\bibinfo {volume} {D84}},\ \bibinfo {pages}
  {104032} (\bibinfo {year} {2011})},\ \Eprint
  {http://arxiv.org/abs/astro-ph/1109.5177} {arXiv:astro-ph/1109.5177
  [astro-ph.HE]} \BibitemShut {NoStop}%
\bibitem [{\citenamefont {Pfeiffer}\ and\ \citenamefont
  {York}(2003)}]{Pfeiffer:2002iy}%
  \BibitemOpen
  \bibfield  {author} {\bibinfo {author} {\bibfnamefont {H.~P.}\ \bibnamefont
  {Pfeiffer}}\ and\ \bibinfo {author} {\bibfnamefont {J.}~\bibnamefont {York},
  \bibfnamefont {James~W.}},\ }\href {\doibase 10.1103/PhysRevD.67.044022}
  {\bibfield  {journal} {\bibinfo  {journal} {Phys.Rev.}\ }\textbf {\bibinfo
  {volume} {D67}},\ \bibinfo {pages} {044022} (\bibinfo {year} {2003})},\
  \Eprint {http://arxiv.org/abs/gr-qc/0207095} {arXiv:gr-qc/0207095 [gr-qc]}
  \BibitemShut {NoStop}%
\bibitem [{\citenamefont {Cook}\ and\ \citenamefont
  {Pfeiffer}(2004)}]{Cook:2004kt}%
  \BibitemOpen
  \bibfield  {author} {\bibinfo {author} {\bibfnamefont {G.~B.}\ \bibnamefont
  {Cook}}\ and\ \bibinfo {author} {\bibfnamefont {H.~P.}\ \bibnamefont
  {Pfeiffer}},\ }\href {\doibase 10.1103/PhysRevD.70.104016} {\bibfield
  {journal} {\bibinfo  {journal} {Phys.Rev.}\ }\textbf {\bibinfo {volume}
  {D70}},\ \bibinfo {pages} {104016} (\bibinfo {year} {2004})},\ \Eprint
  {http://arxiv.org/abs/gr-qc/0407078} {arXiv:gr-qc/0407078 [gr-qc]}
  \BibitemShut {NoStop}%
\bibitem [{\citenamefont {Caudill}\ \emph {et~al.}(2006)\citenamefont
  {Caudill}, \citenamefont {Cook}, \citenamefont {Grigsby},\ and\ \citenamefont
  {Pfeiffer}}]{Caudill:2006hw}%
  \BibitemOpen
  \bibfield  {author} {\bibinfo {author} {\bibfnamefont {M.}~\bibnamefont
  {Caudill}}, \bibinfo {author} {\bibfnamefont {G.~B.}\ \bibnamefont {Cook}},
  \bibinfo {author} {\bibfnamefont {J.~D.}\ \bibnamefont {Grigsby}}, \ and\
  \bibinfo {author} {\bibfnamefont {H.~P.}\ \bibnamefont {Pfeiffer}},\ }\href
  {\doibase 10.1103/PhysRevD.74.064011} {\bibfield  {journal} {\bibinfo
  {journal} {Phys.Rev.}\ }\textbf {\bibinfo {volume} {D74}},\ \bibinfo {pages}
  {064011} (\bibinfo {year} {2006})},\ \Eprint
  {http://arxiv.org/abs/gr-qc/0605053} {arXiv:gr-qc/0605053 [gr-qc]}
  \BibitemShut {NoStop}%
\bibitem [{\citenamefont {Baumgarte}\ and\ \citenamefont
  {Shapiro}(2010)}]{BSBook}%
  \BibitemOpen
  \bibfield  {author} {\bibinfo {author} {\bibfnamefont {T.}~\bibnamefont
  {Baumgarte}}\ and\ \bibinfo {author} {\bibfnamefont {S.}~\bibnamefont
  {Shapiro}},\ }\href@noop {} {\emph {\bibinfo {title} {{Numerical Relativity:
  Solving Einstein’s Equations on the Computer}}}}\ (\bibinfo  {publisher}
  {Cambridge University Press},\ \bibinfo {address} {Cambridge},\ \bibinfo
  {year} {2010})\BibitemShut {NoStop}%
\bibitem [{\citenamefont {Pfeiffer}\ \emph {et~al.}(2003)\citenamefont
  {Pfeiffer}, \citenamefont {Kidder}, \citenamefont {Scheel},\ and\
  \citenamefont {Teukolsky}}]{Pfeiffer:2002wt}%
  \BibitemOpen
  \bibfield  {author} {\bibinfo {author} {\bibfnamefont {H.~P.}\ \bibnamefont
  {Pfeiffer}}, \bibinfo {author} {\bibfnamefont {L.~E.}\ \bibnamefont
  {Kidder}}, \bibinfo {author} {\bibfnamefont {M.~A.}\ \bibnamefont {Scheel}},
  \ and\ \bibinfo {author} {\bibfnamefont {S.~A.}\ \bibnamefont {Teukolsky}},\
  }\href {\doibase 10.1016/S0010-4655(02)00847-0} {\bibfield  {journal}
  {\bibinfo  {journal} {Comput.Phys.Commun.}\ }\textbf {\bibinfo {volume}
  {152}},\ \bibinfo {pages} {253} (\bibinfo {year} {2003})},\ \Eprint
  {http://arxiv.org/abs/gr-qc/0202096} {arXiv:gr-qc/0202096 [gr-qc]}
  \BibitemShut {NoStop}%
\bibitem [{SpE()}]{SpECwebsite}%
  \BibitemOpen
  \href@noop {} {}\bibinfo {howpublished}
  {\url{http://www.black-holes.org/SpEC.html}}\BibitemShut {NoStop}%
\bibitem [{\citenamefont {Szilagyi}(2014)}]{Szilagyi:2014fna}%
  \BibitemOpen
  \bibfield  {author} {\bibinfo {author} {\bibfnamefont {B.}~\bibnamefont
  {Szilagyi}},\ }\href@noop {} {\  (\bibinfo {year} {2014})},\ \Eprint
  {http://arxiv.org/abs/astro-ph/1405.3693} {arXiv:astro-ph/1405.3693 [gr-qc]}
  \BibitemShut {NoStop}%
\bibitem [{\citenamefont {Mroue}\ \emph {et~al.}(2013)\citenamefont {Mroue},
  \citenamefont {Scheel}, \citenamefont {Szilagyi}, \citenamefont {Pfeiffer},
  \citenamefont {Boyle} \emph {et~al.}}]{Mroue:2013xna}%
  \BibitemOpen
  \bibfield  {author} {\bibinfo {author} {\bibfnamefont {A.~H.}\ \bibnamefont
  {Mroue}}, \bibinfo {author} {\bibfnamefont {M.~A.}\ \bibnamefont {Scheel}},
  \bibinfo {author} {\bibfnamefont {B.}~\bibnamefont {Szilagyi}}, \bibinfo
  {author} {\bibfnamefont {H.~P.}\ \bibnamefont {Pfeiffer}}, \bibinfo {author}
  {\bibfnamefont {M.}~\bibnamefont {Boyle}},  \emph {et~al.},\ }\href {\doibase
  10.1103/PhysRevLett.111.241104} {\bibfield  {journal} {\bibinfo  {journal}
  {Phys.Rev.Lett.}\ }\textbf {\bibinfo {volume} {111}},\ \bibinfo {pages}
  {241104} (\bibinfo {year} {2013})},\ \Eprint
  {http://arxiv.org/abs/gr-qc/1304.6077} {arXiv:gr-qc/1304.6077 [gr-qc]}
  \BibitemShut {NoStop}%
\bibitem [{\citenamefont {{Chakrabarti}}(1985)}]{Chakrabarti:1985}%
  \BibitemOpen
  \bibfield  {author} {\bibinfo {author} {\bibfnamefont {S.~K.}\ \bibnamefont
  {{Chakrabarti}}},\ }\href {\doibase 10.1086/162755} {\bibfield  {journal}
  {\bibinfo  {journal} {\apj}\ }\textbf {\bibinfo {volume} {288}},\ \bibinfo
  {pages} {1} (\bibinfo {year} {1985})}\BibitemShut {NoStop}%
\bibitem [{\citenamefont {De~Villiers}\ \emph {et~al.}(2003)\citenamefont
  {De~Villiers}, \citenamefont {Hawley},\ and\ \citenamefont
  {Krolik}}]{DeVilliers:2003gr}%
  \BibitemOpen
  \bibfield  {author} {\bibinfo {author} {\bibfnamefont {J.-P.}\ \bibnamefont
  {De~Villiers}}, \bibinfo {author} {\bibfnamefont {J.~F.}\ \bibnamefont
  {Hawley}}, \ and\ \bibinfo {author} {\bibfnamefont {J.~H.}\ \bibnamefont
  {Krolik}},\ }\href@noop {} {\  (\bibinfo {year} {2003})},\ \Eprint
  {http://arxiv.org/abs/astro-ph/0307260} {arXiv:astro-ph/0307260 [astro-ph]}
  \BibitemShut {NoStop}%
\bibitem [{\citenamefont {Balbus}\ and\ \citenamefont
  {Hawley}(1991)}]{Balbus:1991ay}%
  \BibitemOpen
  \bibfield  {author} {\bibinfo {author} {\bibfnamefont {S.~A.}\ \bibnamefont
  {Balbus}}\ and\ \bibinfo {author} {\bibfnamefont {J.~F.}\ \bibnamefont
  {Hawley}},\ }\href {\doibase 10.1086/170270} {\bibfield  {journal} {\bibinfo
  {journal} {Astrophys.J.}\ }\textbf {\bibinfo {volume} {376}},\ \bibinfo
  {pages} {214} (\bibinfo {year} {1991})}\BibitemShut {NoStop}%
\bibitem [{\citenamefont {Duez}\ \emph {et~al.}(2005)\citenamefont {Duez},
  \citenamefont {Liu}, \citenamefont {Shapiro},\ and\ \citenamefont
  {Stephens}}]{Duez:2005sf}%
  \BibitemOpen
  \bibfield  {author} {\bibinfo {author} {\bibfnamefont {M.~D.}\ \bibnamefont
  {Duez}}, \bibinfo {author} {\bibfnamefont {Y.~T.}\ \bibnamefont {Liu}},
  \bibinfo {author} {\bibfnamefont {S.~L.}\ \bibnamefont {Shapiro}}, \ and\
  \bibinfo {author} {\bibfnamefont {B.~C.}\ \bibnamefont {Stephens}},\ }\href
  {\doibase 10.1103/PhysRevD.72.024028} {\bibfield  {journal} {\bibinfo
  {journal} {Phys.Rev.}\ }\textbf {\bibinfo {volume} {D72}},\ \bibinfo {pages}
  {024028} (\bibinfo {year} {2005})},\ \Eprint
  {http://arxiv.org/abs/astro-ph/0503420} {arXiv:astro-ph/0503420 [astro-ph]}
  \BibitemShut {NoStop}%
\bibitem [{\citenamefont {Etienne}\ \emph {et~al.}(2010)\citenamefont
  {Etienne}, \citenamefont {Liu},\ and\ \citenamefont
  {Shapiro}}]{Etienne:2010ui}%
  \BibitemOpen
  \bibfield  {author} {\bibinfo {author} {\bibfnamefont {Z.~B.}\ \bibnamefont
  {Etienne}}, \bibinfo {author} {\bibfnamefont {Y.~T.}\ \bibnamefont {Liu}}, \
  and\ \bibinfo {author} {\bibfnamefont {S.~L.}\ \bibnamefont {Shapiro}},\
  }\href {\doibase 10.1103/PhysRevD.82.084031} {\bibfield  {journal} {\bibinfo
  {journal} {Phys.Rev.}\ }\textbf {\bibinfo {volume} {D82}},\ \bibinfo {pages}
  {084031} (\bibinfo {year} {2010})},\ \Eprint
  {http://arxiv.org/abs/astro-ph/1007.2848} {arXiv:astro-ph/1007.2848
  [astro-ph.HE]} \BibitemShut {NoStop}%
\bibitem [{\citenamefont {Etienne}\ \emph
  {et~al.}(2012{\natexlab{a}})\citenamefont {Etienne}, \citenamefont
  {Paschalidis}, \citenamefont {Liu},\ and\ \citenamefont
  {Shapiro}}]{Etienne:2011re}%
  \BibitemOpen
  \bibfield  {author} {\bibinfo {author} {\bibfnamefont {Z.~B.}\ \bibnamefont
  {Etienne}}, \bibinfo {author} {\bibfnamefont {V.}~\bibnamefont
  {Paschalidis}}, \bibinfo {author} {\bibfnamefont {Y.~T.}\ \bibnamefont
  {Liu}}, \ and\ \bibinfo {author} {\bibfnamefont {S.~L.}\ \bibnamefont
  {Shapiro}},\ }\href {\doibase 10.1103/PhysRevD.85.024013} {\bibfield
  {journal} {\bibinfo  {journal} {Phys.Rev.}\ }\textbf {\bibinfo {volume}
  {D85}},\ \bibinfo {pages} {024013} (\bibinfo {year} {2012}{\natexlab{a}})},\
  \Eprint {http://arxiv.org/abs/astro-ph/1110.4633} {arXiv:astro-ph/1110.4633
  [astro-ph.HE]} \BibitemShut {NoStop}%
\bibitem [{\citenamefont {Goodale}\ \emph {et~al.}(2003)\citenamefont
  {Goodale}, \citenamefont {Allen}, \citenamefont {Lanfermann}, \citenamefont
  {Mass{\'o}}, \citenamefont {Radke}, \citenamefont {Seidel},\ and\
  \citenamefont {Shalf}}]{Goodale2002a}%
  \BibitemOpen
  \bibfield  {author} {\bibinfo {author} {\bibfnamefont {T.}~\bibnamefont
  {Goodale}}, \bibinfo {author} {\bibfnamefont {G.}~\bibnamefont {Allen}},
  \bibinfo {author} {\bibfnamefont {G.}~\bibnamefont {Lanfermann}}, \bibinfo
  {author} {\bibfnamefont {J.}~\bibnamefont {Mass{\'o}}}, \bibinfo {author}
  {\bibfnamefont {T.}~\bibnamefont {Radke}}, \bibinfo {author} {\bibfnamefont
  {E.}~\bibnamefont {Seidel}}, \ and\ \bibinfo {author} {\bibfnamefont
  {J.}~\bibnamefont {Shalf}},\ }in\ \href {http://edoc.mpg.de/3341} {\emph
  {\bibinfo {booktitle} {5th Int. Conference, Lecture Notes in Comp.
  Science}}}\ (\bibinfo  {publisher} {Springer},\ \bibinfo {address} {Berlin},\
  \bibinfo {year} {2003})\BibitemShut {NoStop}%
\bibitem [{Car({\natexlab{a}})}]{Carpet-cactusweb}%
  \BibitemOpen
  \href {http://www.cactuscode.org/} {\enquote {\bibinfo {title} {{Cactus}
  {Computational} {Toolkit}},}\ } ({\natexlab{a}})\BibitemShut {NoStop}%
\bibitem [{Car({\natexlab{b}})}]{Carpet-carpetweb}%
  \BibitemOpen
  \href {http://www.carpetcode.org/} {\enquote {\bibinfo {title} {Mesh
  refinement with {Carpet}},}\ } ({\natexlab{b}})\BibitemShut {NoStop}%
\bibitem [{\citenamefont {{Paschalidis}}\ \emph {et~al.}(2011)\citenamefont
  {{Paschalidis}}, \citenamefont {{Etienne}}, \citenamefont {{Liu}},\ and\
  \citenamefont {{Shapiro}}}]{Paschalidis2011a}%
  \BibitemOpen
  \bibfield  {author} {\bibinfo {author} {\bibfnamefont {V.}~\bibnamefont
  {{Paschalidis}}}, \bibinfo {author} {\bibfnamefont {Z.}~\bibnamefont
  {{Etienne}}}, \bibinfo {author} {\bibfnamefont {Y.~T.}\ \bibnamefont
  {{Liu}}}, \ and\ \bibinfo {author} {\bibfnamefont {S.~L.}\ \bibnamefont
  {{Shapiro}}},\ }\href {\doibase 10.1103/PhysRevD.83.064002} {\bibfield
  {journal} {\bibinfo  {journal} {\prd}\ }\textbf {\bibinfo {volume} {83}},\
  \bibinfo {eid} {064002} (\bibinfo {year} {2011})},\ \Eprint
  {http://arxiv.org/abs/arXiv:1009.4932} {arXiv:arXiv:1009.4932 [astro-ph.HE]}
  \BibitemShut {NoStop}%
\bibitem [{\citenamefont {Etienne}\ \emph
  {et~al.}(2012{\natexlab{b}})\citenamefont {Etienne}, \citenamefont {Liu},
  \citenamefont {Paschalidis},\ and\ \citenamefont {Shapiro}}]{Etienne:2011ea}%
  \BibitemOpen
  \bibfield  {author} {\bibinfo {author} {\bibfnamefont {Z.~B.}\ \bibnamefont
  {Etienne}}, \bibinfo {author} {\bibfnamefont {Y.~T.}\ \bibnamefont {Liu}},
  \bibinfo {author} {\bibfnamefont {V.}~\bibnamefont {Paschalidis}}, \ and\
  \bibinfo {author} {\bibfnamefont {S.~L.}\ \bibnamefont {Shapiro}},\ }\href
  {\doibase 10.1103/PhysRevD.85.064029} {\bibfield  {journal} {\bibinfo
  {journal} {Phys.Rev.}\ }\textbf {\bibinfo {volume} {D85}},\ \bibinfo {pages}
  {064029} (\bibinfo {year} {2012}{\natexlab{b}})},\ \Eprint
  {http://arxiv.org/abs/astro-ph/1112.0568} {arXiv:astro-ph/1112.0568
  [astro-ph.HE]} \BibitemShut {NoStop}%
\bibitem [{\citenamefont {Etienne}\ \emph
  {et~al.}(2012{\natexlab{c}})\citenamefont {Etienne}, \citenamefont
  {Paschalidis},\ and\ \citenamefont {Shapiro}}]{Etienne:2012te}%
  \BibitemOpen
  \bibfield  {author} {\bibinfo {author} {\bibfnamefont {Z.~B.}\ \bibnamefont
  {Etienne}}, \bibinfo {author} {\bibfnamefont {V.}~\bibnamefont
  {Paschalidis}}, \ and\ \bibinfo {author} {\bibfnamefont {S.~L.}\ \bibnamefont
  {Shapiro}},\ }\href {\doibase 10.1103/PhysRevD.86.084026} {\bibfield
  {journal} {\bibinfo  {journal} {Phys.Rev.}\ }\textbf {\bibinfo {volume}
  {D86}},\ \bibinfo {pages} {084026} (\bibinfo {year} {2012}{\natexlab{c}})},\
  \Eprint {http://arxiv.org/abs/astro-ph/1209.1632} {arXiv:astro-ph/1209.1632
  [astro-ph.HE]} \BibitemShut {NoStop}%
\bibitem [{\citenamefont {Paschalidis}\ \emph {et~al.}(2012)\citenamefont
  {Paschalidis}, \citenamefont {Etienne},\ and\ \citenamefont
  {Shapiro}}]{Paschalidis:2012ff}%
  \BibitemOpen
  \bibfield  {author} {\bibinfo {author} {\bibfnamefont {V.}~\bibnamefont
  {Paschalidis}}, \bibinfo {author} {\bibfnamefont {Z.~B.}\ \bibnamefont
  {Etienne}}, \ and\ \bibinfo {author} {\bibfnamefont {S.~L.}\ \bibnamefont
  {Shapiro}},\ }\href {\doibase 10.1103/PhysRevD.86.064032} {\bibfield
  {journal} {\bibinfo  {journal} {Phys.Rev.}\ }\textbf {\bibinfo {volume}
  {D86}},\ \bibinfo {pages} {064032} (\bibinfo {year} {2012})},\ \Eprint
  {http://arxiv.org/abs/astro-ph/1208.5487} {arXiv:astro-ph/1208.5487
  [astro-ph.HE]} \BibitemShut {NoStop}%
\bibitem [{\citenamefont {Etienne}\ \emph {et~al.}(2013)\citenamefont
  {Etienne}, \citenamefont {Liu}, \citenamefont {Paschalidis},\ and\
  \citenamefont {Shapiro}}]{Etienne:2013qia}%
  \BibitemOpen
  \bibfield  {author} {\bibinfo {author} {\bibfnamefont {Z.~B.}\ \bibnamefont
  {Etienne}}, \bibinfo {author} {\bibfnamefont {Y.~T.}\ \bibnamefont {Liu}},
  \bibinfo {author} {\bibfnamefont {V.}~\bibnamefont {Paschalidis}}, \ and\
  \bibinfo {author} {\bibfnamefont {S.~L.}\ \bibnamefont {Shapiro}},\
  }\href@noop {} {\  (\bibinfo {year} {2013})},\ \Eprint
  {http://arxiv.org/abs/astro-ph/1303.0837} {arXiv:astro-ph/1303.0837
  [astro-ph.HE]} \BibitemShut {NoStop}%
\bibitem [{\citenamefont {Paschalidis}\ \emph {et~al.}(2013)\citenamefont
  {Paschalidis}, \citenamefont {Etienne},\ and\ \citenamefont
  {Shapiro}}]{Paschalidis:2013jsa}%
  \BibitemOpen
  \bibfield  {author} {\bibinfo {author} {\bibfnamefont {V.}~\bibnamefont
  {Paschalidis}}, \bibinfo {author} {\bibfnamefont {Z.~B.}\ \bibnamefont
  {Etienne}}, \ and\ \bibinfo {author} {\bibfnamefont {S.~L.}\ \bibnamefont
  {Shapiro}},\ }\href@noop {} {\  (\bibinfo {year} {2013})},\ \Eprint
  {http://arxiv.org/abs/astro-ph/1304.1805} {arXiv:astro-ph/1304.1805
  [astro-ph.HE]} \BibitemShut {NoStop}%
\bibitem [{\citenamefont {Farris}\ \emph {et~al.}(2010)\citenamefont {Farris},
  \citenamefont {Liu},\ and\ \citenamefont {Shapiro}}]{Farris:2009mt}%
  \BibitemOpen
  \bibfield  {author} {\bibinfo {author} {\bibfnamefont {B.~D.}\ \bibnamefont
  {Farris}}, \bibinfo {author} {\bibfnamefont {Y.~T.}\ \bibnamefont {Liu}}, \
  and\ \bibinfo {author} {\bibfnamefont {S.~L.}\ \bibnamefont {Shapiro}},\
  }\href {\doibase 10.1103/PhysRevD.81.084008} {\bibfield  {journal} {\bibinfo
  {journal} {Phys.Rev.}\ }\textbf {\bibinfo {volume} {D81}},\ \bibinfo {pages}
  {084008} (\bibinfo {year} {2010})},\ \Eprint
  {http://arxiv.org/abs/astro-ph/0912.2096} {arXiv:astro-ph/0912.2096
  [astro-ph.HE]} \BibitemShut {NoStop}%
\bibitem [{\citenamefont {Shibata}\ and\ \citenamefont
  {Nakamura}(1995)}]{Shibata:1995}%
  \BibitemOpen
  \bibfield  {author} {\bibinfo {author} {\bibfnamefont {M.}~\bibnamefont
  {Shibata}}\ and\ \bibinfo {author} {\bibfnamefont {T.}~\bibnamefont
  {Nakamura}},\ }\href {\doibase 10.1103/PhysRevD.52.5428} {\bibfield
  {journal} {\bibinfo  {journal} {Phys. Rev. D}\ }\textbf {\bibinfo {volume}
  {52}},\ \bibinfo {pages} {5428} (\bibinfo {year} {1995})}\BibitemShut
  {NoStop}%
\bibitem [{\citenamefont {Baumgarte}\ and\ \citenamefont
  {Shapiro}(1999)}]{Baumgarte:1998te}%
  \BibitemOpen
  \bibfield  {author} {\bibinfo {author} {\bibfnamefont {T.~W.}\ \bibnamefont
  {Baumgarte}}\ and\ \bibinfo {author} {\bibfnamefont {S.~L.}\ \bibnamefont
  {Shapiro}},\ }\href {\doibase 10.1103/PhysRevD.59.024007} {\bibfield
  {journal} {\bibinfo  {journal} {Phys.Rev.}\ }\textbf {\bibinfo {volume}
  {D59}},\ \bibinfo {pages} {024007} (\bibinfo {year} {1999})},\ \Eprint
  {http://arxiv.org/abs/gr-qc/9810065} {arXiv:gr-qc/9810065 [gr-qc]}
  \BibitemShut {NoStop}%
\bibitem [{\citenamefont {Etienne}\ \emph {et~al.}(2008)\citenamefont
  {Etienne}, \citenamefont {Faber}, \citenamefont {Liu}, \citenamefont
  {Shapiro}, \citenamefont {Taniguchi} \emph {et~al.}}]{Etienne:2007jg}%
  \BibitemOpen
  \bibfield  {author} {\bibinfo {author} {\bibfnamefont {Z.~B.}\ \bibnamefont
  {Etienne}}, \bibinfo {author} {\bibfnamefont {J.~A.}\ \bibnamefont {Faber}},
  \bibinfo {author} {\bibfnamefont {Y.~T.}\ \bibnamefont {Liu}}, \bibinfo
  {author} {\bibfnamefont {S.~L.}\ \bibnamefont {Shapiro}}, \bibinfo {author}
  {\bibfnamefont {K.}~\bibnamefont {Taniguchi}},  \emph {et~al.},\ }\href
  {\doibase 10.1103/PhysRevD.77.084002} {\bibfield  {journal} {\bibinfo
  {journal} {Phys.Rev.}\ }\textbf {\bibinfo {volume} {D77}},\ \bibinfo {pages}
  {084002} (\bibinfo {year} {2008})},\ \Eprint
  {http://arxiv.org/abs/astro-ph/0712.2460} {arXiv:astro-ph/0712.2460
  [astro-ph]} \BibitemShut {NoStop}%
\bibitem [{\citenamefont {Hinder}\ \emph {et~al.}(2014)\citenamefont {Hinder},
  \citenamefont {Buonanno}, \citenamefont {Boyle}, \citenamefont {Etienne},
  \citenamefont {Healy} \emph {et~al.}}]{Hinder:2013oqa}%
  \BibitemOpen
  \bibfield  {author} {\bibinfo {author} {\bibfnamefont {I.}~\bibnamefont
  {Hinder}}, \bibinfo {author} {\bibfnamefont {A.}~\bibnamefont {Buonanno}},
  \bibinfo {author} {\bibfnamefont {M.}~\bibnamefont {Boyle}}, \bibinfo
  {author} {\bibfnamefont {Z.~B.}\ \bibnamefont {Etienne}}, \bibinfo {author}
  {\bibfnamefont {J.}~\bibnamefont {Healy}},  \emph {et~al.},\ }\href {\doibase
  10.1088/0264-9381/31/2/025012} {\bibfield  {journal} {\bibinfo  {journal}
  {Class.Quant.Grav.}\ }\textbf {\bibinfo {volume} {31}},\ \bibinfo {pages}
  {025012} (\bibinfo {year} {2014})},\ \Eprint
  {http://arxiv.org/abs/astro-ph/1307.5307} {arXiv:astro-ph/1307.5307 [gr-qc]}
  \BibitemShut {NoStop}%
\bibitem [{\citenamefont {Lousto}\ and\ \citenamefont
  {Zlochower}(2011)}]{Lousto:2010ut}%
  \BibitemOpen
  \bibfield  {author} {\bibinfo {author} {\bibfnamefont {C.~O.}\ \bibnamefont
  {Lousto}}\ and\ \bibinfo {author} {\bibfnamefont {Y.}~\bibnamefont
  {Zlochower}},\ }\href {\doibase 10.1103/PhysRevLett.106.041101} {\bibfield
  {journal} {\bibinfo  {journal} {Phys.Rev.Lett.}\ }\textbf {\bibinfo {volume}
  {106}},\ \bibinfo {pages} {041101} (\bibinfo {year} {2011})},\ \Eprint
  {http://arxiv.org/abs/gr-qc/1009.0292} {arXiv:gr-qc/1009.0292 [gr-qc]}
  \BibitemShut {NoStop}%
\bibitem [{\citenamefont {M{\"u}ller}\ \emph {et~al.}(2010)\citenamefont
  {M{\"u}ller}, \citenamefont {Grigsby},\ and\ \citenamefont
  {Br{\"u}gmann}}]{Muller:2010zze}%
  \BibitemOpen
  \bibfield  {author} {\bibinfo {author} {\bibfnamefont {D.}~\bibnamefont
  {M{\"u}ller}}, \bibinfo {author} {\bibfnamefont {J.}~\bibnamefont {Grigsby}},
  \ and\ \bibinfo {author} {\bibfnamefont {B.}~\bibnamefont {Br{\"u}gmann}},\
  }\href {\doibase 10.1103/PhysRevD.82.064004} {\bibfield  {journal} {\bibinfo
  {journal} {Phys.Rev.}\ }\textbf {\bibinfo {volume} {D82}},\ \bibinfo {pages}
  {064004} (\bibinfo {year} {2010})},\ \Eprint
  {http://arxiv.org/abs/gr-qc/1003.4681} {arXiv:gr-qc/1003.4681 [gr-qc]}
  \BibitemShut {NoStop}%
\bibitem [{\citenamefont {{Shapiro}}\ and\ \citenamefont
  {{Teukolsky}}(1983)}]{shapiro_book_83}%
  \BibitemOpen
  \bibfield  {author} {\bibinfo {author} {\bibfnamefont {S.~L.}\ \bibnamefont
  {{Shapiro}}}\ and\ \bibinfo {author} {\bibfnamefont {S.~A.}\ \bibnamefont
  {{Teukolsky}}},\ }\href@noop {} {\emph {\bibinfo {title} {{Black holes, white
  dwarfs, and neutron stars: The physics of compact objects}}}}\ (\bibinfo
  {publisher} {Wiley, New York},\ \bibinfo {year} {1983})\BibitemShut {NoStop}%
\bibitem [{\citenamefont {{Peters}}(1964)}]{Peters:1964}%
  \BibitemOpen
  \bibfield  {author} {\bibinfo {author} {\bibfnamefont {P.~C.}\ \bibnamefont
  {{Peters}}},\ }\href {\doibase 10.1103/PhysRev.136.B1224} {\bibfield
  {journal} {\bibinfo  {journal} {Physical Review}\ }\textbf {\bibinfo {volume}
  {136}},\ \bibinfo {pages} {1224} (\bibinfo {year} {1964})}\BibitemShut
  {NoStop}%
\bibitem [{\citenamefont {Farris}\ \emph {et~al.}(2013)\citenamefont {Farris},
  \citenamefont {Duffell}, \citenamefont {MacFadyen},\ and\ \citenamefont
  {Haiman}}]{Farris:2013uqa}%
  \BibitemOpen
  \bibfield  {author} {\bibinfo {author} {\bibfnamefont {B.~D.}\ \bibnamefont
  {Farris}}, \bibinfo {author} {\bibfnamefont {P.}~\bibnamefont {Duffell}},
  \bibinfo {author} {\bibfnamefont {A.~I.}\ \bibnamefont {MacFadyen}}, \ and\
  \bibinfo {author} {\bibfnamefont {Z.}~\bibnamefont {Haiman}},\ }\href@noop {}
  {\  (\bibinfo {year} {2013})},\ \Eprint
  {http://arxiv.org/abs/astro-ph/1310.0492} {arXiv:astro-ph/1310.0492
  [astro-ph.HE]} \BibitemShut {NoStop}%
\bibitem [{\citenamefont {Yan}\ \emph {et~al.}(2014)\citenamefont {Yan},
  \citenamefont {Lu}, \citenamefont {Yu}, \citenamefont {Mao},\ and\
  \citenamefont {Wambsganss}}]{Yan:2014jva}%
  \BibitemOpen
  \bibfield  {author} {\bibinfo {author} {\bibfnamefont {C.-S.}\ \bibnamefont
  {Yan}}, \bibinfo {author} {\bibfnamefont {Y.}~\bibnamefont {Lu}}, \bibinfo
  {author} {\bibfnamefont {Q.}~\bibnamefont {Yu}}, \bibinfo {author}
  {\bibfnamefont {S.}~\bibnamefont {Mao}}, \ and\ \bibinfo {author}
  {\bibfnamefont {J.}~\bibnamefont {Wambsganss}},\ }\href {\doibase
  10.1088/0004-637X/784/2/100} {\bibfield  {journal} {\bibinfo  {journal}
  {Astrophys.J.}\ }\textbf {\bibinfo {volume} {784}},\ \bibinfo {pages} {100}
  (\bibinfo {year} {2014})},\ \Eprint {http://arxiv.org/abs/astro-ph/1402.2504}
  {arXiv:astro-ph/1402.2504 [astro-ph.CO]} \BibitemShut {NoStop}%
\bibitem [{\citenamefont {Farris}\ \emph
  {et~al.}(2014{\natexlab{b}})\citenamefont {Farris}, \citenamefont {Duffell},
  \citenamefont {MacFadyen},\ and\ \citenamefont {Haiman}}]{Farris:2014iga}%
  \BibitemOpen
  \bibfield  {author} {\bibinfo {author} {\bibfnamefont {B.~D.}\ \bibnamefont
  {Farris}}, \bibinfo {author} {\bibfnamefont {P.}~\bibnamefont {Duffell}},
  \bibinfo {author} {\bibfnamefont {A.~I.}\ \bibnamefont {MacFadyen}}, \ and\
  \bibinfo {author} {\bibfnamefont {Z.}~\bibnamefont {Haiman}},\ }\href@noop {}
  {\  (\bibinfo {year} {2014}{\natexlab{b}})},\ \Eprint
  {http://arxiv.org/abs/astro-ph/1406.0007} {arXiv:astro-ph/1406.0007
  [astro-ph.HE]} \BibitemShut {NoStop}%
\bibitem [{\citenamefont {McKinney}\ and\ \citenamefont
  {Gammie}(2004)}]{McKinney:2004ka}%
  \BibitemOpen
  \bibfield  {author} {\bibinfo {author} {\bibfnamefont {J.~C.}\ \bibnamefont
  {McKinney}}\ and\ \bibinfo {author} {\bibfnamefont {C.~F.}\ \bibnamefont
  {Gammie}},\ }\href {\doibase 10.1086/422244} {\bibfield  {journal} {\bibinfo
  {journal} {Astrophys.J.}\ }\textbf {\bibinfo {volume} {611}},\ \bibinfo
  {pages} {977} (\bibinfo {year} {2004})},\ \Eprint
  {http://arxiv.org/abs/astro-ph/0404512} {arXiv:astro-ph/0404512 [astro-ph]}
  \BibitemShut {NoStop}%
\bibitem [{\citenamefont {{Vlahakis}}\ and\ \citenamefont
  {{K{\"o}nigl}}(2003)}]{B2_over_2RHO_yields_target_Lorentz_factor}%
  \BibitemOpen
  \bibfield  {author} {\bibinfo {author} {\bibfnamefont {N.}~\bibnamefont
  {{Vlahakis}}}\ and\ \bibinfo {author} {\bibfnamefont {A.}~\bibnamefont
  {{K{\"o}nigl}}},\ }\href {\doibase 10.1086/378226} {\bibfield  {journal}
  {\bibinfo  {journal} {\apj}\ }\textbf {\bibinfo {volume} {596}},\ \bibinfo
  {pages} {1080} (\bibinfo {year} {2003})}\BibitemShut {NoStop}%
\bibitem [{\citenamefont {{Blandford}}\ and\ \citenamefont
  {{Payne}}(1982)}]{1982MNRAS.199..883B}%
  \BibitemOpen
  \bibfield  {author} {\bibinfo {author} {\bibfnamefont {R.~D.}\ \bibnamefont
  {{Blandford}}}\ and\ \bibinfo {author} {\bibfnamefont {D.~G.}\ \bibnamefont
  {{Payne}}},\ }\href@noop {} {\bibfield  {journal} {\bibinfo  {journal}
  {\mnras}\ }\textbf {\bibinfo {volume} {199}},\ \bibinfo {pages} {883}
  (\bibinfo {year} {1982})}\BibitemShut {NoStop}%
\bibitem [{\citenamefont {{Blandford}}\ and\ \citenamefont
  {{Znajek}}(1977)}]{Blandford1977}%
  \BibitemOpen
  \bibfield  {author} {\bibinfo {author} {\bibfnamefont {R.~D.}\ \bibnamefont
  {{Blandford}}}\ and\ \bibinfo {author} {\bibfnamefont {R.~L.}\ \bibnamefont
  {{Znajek}}},\ }\href@noop {} {\bibfield  {journal} {\bibinfo  {journal}
  {\mnras}\ }\textbf {\bibinfo {volume} {179}},\ \bibinfo {pages} {433}
  (\bibinfo {year} {1977})}\BibitemShut {NoStop}%
\bibitem [{\citenamefont {Milosavljevic}\ and\ \citenamefont
  {Phinney}(2005)}]{Milosavljevic:2004cg}%
  \BibitemOpen
  \bibfield  {author} {\bibinfo {author} {\bibfnamefont {M.}~\bibnamefont
  {Milosavljevic}}\ and\ \bibinfo {author} {\bibfnamefont {E.}~\bibnamefont
  {Phinney}},\ }\href {\doibase 10.1086/429618} {\bibfield  {journal} {\bibinfo
   {journal} {Astrophys.J.}\ }\textbf {\bibinfo {volume} {622}},\ \bibinfo
  {pages} {L93} (\bibinfo {year} {2005})},\ \Eprint
  {http://arxiv.org/abs/astro-ph/0410343} {arXiv:astro-ph/0410343 [astro-ph]}
  \BibitemShut {NoStop}%
\bibitem [{\citenamefont {Baker}\ \emph {et~al.}(2006)\citenamefont {Baker},
  \citenamefont {Centrella}, \citenamefont {Choi}, \citenamefont {Koppitz},
  \citenamefont {van Meter} \emph {et~al.}}]{Baker:2006vn}%
  \BibitemOpen
  \bibfield  {author} {\bibinfo {author} {\bibfnamefont {J.~G.}\ \bibnamefont
  {Baker}}, \bibinfo {author} {\bibfnamefont {J.}~\bibnamefont {Centrella}},
  \bibinfo {author} {\bibfnamefont {D.-I.}\ \bibnamefont {Choi}}, \bibinfo
  {author} {\bibfnamefont {M.}~\bibnamefont {Koppitz}}, \bibinfo {author}
  {\bibfnamefont {J.~R.}\ \bibnamefont {van Meter}},  \emph {et~al.},\ }\href
  {\doibase 10.1086/510448} {\bibfield  {journal} {\bibinfo  {journal}
  {Astrophys.J.}\ }\textbf {\bibinfo {volume} {653}},\ \bibinfo {pages} {L93}
  (\bibinfo {year} {2006})},\ \Eprint {http://arxiv.org/abs/astro-ph/0603204}
  {arXiv:astro-ph/0603204 [astro-ph]} \BibitemShut {NoStop}%
\bibitem [{\citenamefont {Megevand}\ \emph
  {et~al.}(2009{\natexlab{b}})\citenamefont {Megevand}, \citenamefont
  {Anderson}, \citenamefont {Frank}, \citenamefont {Hirschmann}, \citenamefont
  {Lehner}, \citenamefont {Liebling}, \citenamefont {Motl},\ and\ \citenamefont
  {Neilsen}}]{Megevand:2009}%
  \BibitemOpen
  \bibfield  {author} {\bibinfo {author} {\bibfnamefont {M.}~\bibnamefont
  {Megevand}}, \bibinfo {author} {\bibfnamefont {M.}~\bibnamefont {Anderson}},
  \bibinfo {author} {\bibfnamefont {J.}~\bibnamefont {Frank}}, \bibinfo
  {author} {\bibfnamefont {E.~W.}\ \bibnamefont {Hirschmann}}, \bibinfo
  {author} {\bibfnamefont {L.}~\bibnamefont {Lehner}}, \bibinfo {author}
  {\bibfnamefont {S.~L.}\ \bibnamefont {Liebling}}, \bibinfo {author}
  {\bibfnamefont {P.~M.}\ \bibnamefont {Motl}}, \ and\ \bibinfo {author}
  {\bibfnamefont {D.}~\bibnamefont {Neilsen}},\ }\href {\doibase
  10.1103/PhysRevD.80.024012} {\bibfield  {journal} {\bibinfo  {journal} {Phys.
  Rev. D}\ }\textbf {\bibinfo {volume} {80}},\ \bibinfo {pages} {024012}
  (\bibinfo {year} {2009}{\natexlab{b}})}\BibitemShut {NoStop}%
\bibitem [{\citenamefont {Gonzalez}\ \emph {et~al.}(2007)\citenamefont
  {Gonzalez}, \citenamefont {Sperhake}, \citenamefont {Bruegmann},
  \citenamefont {Hannam},\ and\ \citenamefont {Husa}}]{Gonzalez:2006md}%
  \BibitemOpen
  \bibfield  {author} {\bibinfo {author} {\bibfnamefont {J.~A.}\ \bibnamefont
  {Gonzalez}}, \bibinfo {author} {\bibfnamefont {U.}~\bibnamefont {Sperhake}},
  \bibinfo {author} {\bibfnamefont {B.}~\bibnamefont {Bruegmann}}, \bibinfo
  {author} {\bibfnamefont {M.}~\bibnamefont {Hannam}}, \ and\ \bibinfo {author}
  {\bibfnamefont {S.}~\bibnamefont {Husa}},\ }\href {\doibase
  10.1103/PhysRevLett.98.091101} {\bibfield  {journal} {\bibinfo  {journal}
  {Phys.Rev.Lett.}\ }\textbf {\bibinfo {volume} {98}},\ \bibinfo {pages}
  {091101} (\bibinfo {year} {2007})},\ \Eprint
  {http://arxiv.org/abs/gr-qc/0610154} {arXiv:gr-qc/0610154 [gr-qc]}
  \BibitemShut {NoStop}%
\bibitem [{\citenamefont {Bogdanovic}\ \emph {et~al.}(2007)\citenamefont
  {Bogdanovic}, \citenamefont {Reynolds},\ and\ \citenamefont
  {Miller}}]{Bogdanovic:2007hp}%
  \BibitemOpen
  \bibfield  {author} {\bibinfo {author} {\bibfnamefont {T.}~\bibnamefont
  {Bogdanovic}}, \bibinfo {author} {\bibfnamefont {C.~S.}\ \bibnamefont
  {Reynolds}}, \ and\ \bibinfo {author} {\bibfnamefont {M.~C.}\ \bibnamefont
  {Miller}},\ }\href {\doibase 10.1086/518769} {\bibfield  {journal} {\bibinfo
  {journal} {Astrophys.J.Lett.}\ }\textbf {\bibinfo {volume} {661}},\ \bibinfo
  {pages} {L147} (\bibinfo {year} {2007})},\ \Eprint
  {http://arxiv.org/abs/astro-ph/0703054} {arXiv:astro-ph/0703054 [astro-ph]}
  \BibitemShut {NoStop}%
\bibitem [{\citenamefont {Dotti}\ \emph {et~al.}(2010)\citenamefont {Dotti},
  \citenamefont {Volonteri}, \citenamefont {Perego}, \citenamefont {Colpi},
  \citenamefont {Ruszkowski},\ and\ \citenamefont {Haardt}}]{Dotti11022010}%
  \BibitemOpen
  \bibfield  {author} {\bibinfo {author} {\bibfnamefont {M.}~\bibnamefont
  {Dotti}}, \bibinfo {author} {\bibfnamefont {M.}~\bibnamefont {Volonteri}},
  \bibinfo {author} {\bibfnamefont {A.}~\bibnamefont {Perego}}, \bibinfo
  {author} {\bibfnamefont {M.}~\bibnamefont {Colpi}}, \bibinfo {author}
  {\bibfnamefont {M.}~\bibnamefont {Ruszkowski}}, \ and\ \bibinfo {author}
  {\bibfnamefont {F.}~\bibnamefont {Haardt}},\ }\href {\doibase
  10.1111/j.1365-2966.2009.15922.x} {\bibfield  {journal} {\bibinfo  {journal}
  {Monthly Notices of the Royal Astronomical Society}\ }\textbf {\bibinfo
  {volume} {402}},\ \bibinfo {pages} {682} (\bibinfo {year} {2010})},\ \Eprint
  {http://arxiv.org/abs/http://mnras.oxfordjournals.org/content/402/1/682.full.pdf+html}
  {http://mnras.oxfordjournals.org/content/402/1/682.full.pdf+html}
  \BibitemShut {NoStop}%
\bibitem [{\citenamefont {Ponce}\ \emph
  {et~al.}(2012{\natexlab{b}})\citenamefont {Ponce}, \citenamefont {Faber},\
  and\ \citenamefont {Lombardi}}]{Ponce-2012}%
  \BibitemOpen
  \bibfield  {author} {\bibinfo {author} {\bibfnamefont {M.}~\bibnamefont
  {Ponce}}, \bibinfo {author} {\bibfnamefont {J.~A.}\ \bibnamefont {Faber}}, \
  and\ \bibinfo {author} {\bibfnamefont {J.~C.}\ \bibnamefont {Lombardi}},\
  }\href {http://stacks.iop.org/0004-637X/745/i=1/a=71} {\bibfield  {journal}
  {\bibinfo  {journal} {The Astrophysical Journal}\ }\textbf {\bibinfo {volume}
  {745}},\ \bibinfo {pages} {71} (\bibinfo {year}
  {2012}{\natexlab{b}})}\BibitemShut {NoStop}%
\bibitem [{\citenamefont {{Rybicki}}\ and\ \citenamefont
  {{Lightman}}(1986)}]{rybicki86}%
  \BibitemOpen
  \bibfield  {author} {\bibinfo {author} {\bibfnamefont {G.~B.}\ \bibnamefont
  {{Rybicki}}}\ and\ \bibinfo {author} {\bibnamefont {{Lightman}}},\
  }\href@noop {} {\emph {\bibinfo {title} {{Radiative Processes in
  Astrophysics}}}}\ (\bibinfo  {publisher} {Wiley-VCH},\ \bibinfo {year}
  {1986})\BibitemShut {NoStop}%
\bibitem [{\citenamefont {{Kaplan}}\ \emph {et~al.}(2011)\citenamefont
  {{Kaplan}}, \citenamefont {{O'Shaughnessy}}, \citenamefont {{Sesana}},\ and\
  \citenamefont {{Volonteri}}}]{Kaplan:2011mz}%
  \BibitemOpen
  \bibfield  {author} {\bibinfo {author} {\bibfnamefont {D.~L.}\ \bibnamefont
  {{Kaplan}}}, \bibinfo {author} {\bibfnamefont {R.}~\bibnamefont
  {{O'Shaughnessy}}}, \bibinfo {author} {\bibfnamefont {A.}~\bibnamefont
  {{Sesana}}}, \ and\ \bibinfo {author} {\bibfnamefont {M.}~\bibnamefont
  {{Volonteri}}},\ }\href {\doibase 10.1088/2041-8205/734/2/L37} {\bibfield
  {journal} {\bibinfo  {journal} {\apjl}\ }\textbf {\bibinfo {volume} {734}},\
  \bibinfo {eid} {L37} (\bibinfo {year} {2011})},\ \Eprint
  {http://arxiv.org/abs/arXiv:1105.3653} {arXiv:arXiv:1105.3653 [astro-ph.HE]}
  \BibitemShut {NoStop}%
\bibitem [{\citenamefont {O'Shaughnessy}\ \emph {et~al.}(2011)\citenamefont
  {O'Shaughnessy}, \citenamefont {Kaplan}, \citenamefont {Sesana},\ and\
  \citenamefont {Kamble}}]{O'Shaughnessy:2011zz}%
  \BibitemOpen
  \bibfield  {author} {\bibinfo {author} {\bibfnamefont {R.}~\bibnamefont
  {O'Shaughnessy}}, \bibinfo {author} {\bibfnamefont {D.}~\bibnamefont
  {Kaplan}}, \bibinfo {author} {\bibfnamefont {A.}~\bibnamefont {Sesana}}, \
  and\ \bibinfo {author} {\bibfnamefont {A.}~\bibnamefont {Kamble}},\ }\href
  {\doibase 10.1088/0004-637X/743/2/136} {\bibfield  {journal} {\bibinfo
  {journal} {Astrophys.J.}\ }\textbf {\bibinfo {volume} {743}},\ \bibinfo
  {pages} {136} (\bibinfo {year} {2011})},\ \Eprint
  {http://arxiv.org/abs/arXiv:1109.1050} {arXiv:arXiv:1109.1050 [astro-ph.CO]}
  \BibitemShut {NoStop}%
\bibitem [{\citenamefont {{Sillanpaa}}\ \emph {et~al.}(1988)\citenamefont
  {{Sillanpaa}}, \citenamefont {{Haarala}}, \citenamefont {{Valtonen}},
  \citenamefont {{Sundelius}},\ and\ \citenamefont {{Byrd}}}]{Sillanpaa88}%
  \BibitemOpen
  \bibfield  {author} {\bibinfo {author} {\bibfnamefont {A.}~\bibnamefont
  {{Sillanpaa}}}, \bibinfo {author} {\bibfnamefont {S.}~\bibnamefont
  {{Haarala}}}, \bibinfo {author} {\bibfnamefont {M.~J.}\ \bibnamefont
  {{Valtonen}}}, \bibinfo {author} {\bibfnamefont {B.}~\bibnamefont
  {{Sundelius}}}, \ and\ \bibinfo {author} {\bibfnamefont {G.~G.}\ \bibnamefont
  {{Byrd}}},\ }\href {\doibase 10.1086/166033} {\bibfield  {journal} {\bibinfo
  {journal} {\apj}\ }\textbf {\bibinfo {volume} {325}},\ \bibinfo {pages} {628}
  (\bibinfo {year} {1988})}\BibitemShut {NoStop}%
\bibitem [{\citenamefont {Valtonen}\ and\ \citenamefont
  {Ciprini}(2011)}]{Valtonen:2011ny}%
  \BibitemOpen
  \bibfield  {author} {\bibinfo {author} {\bibfnamefont {M.}~\bibnamefont
  {Valtonen}}\ and\ \bibinfo {author} {\bibfnamefont {S.}~\bibnamefont
  {Ciprini}},\ }\href@noop {} {\  (\bibinfo {year} {2011})},\ \Eprint
  {http://arxiv.org/abs/astro-ph/1112.1162} {arXiv:astro-ph/1112.1162
  [astro-ph.HE]} \BibitemShut {NoStop}%
\bibitem [{\citenamefont {{Decarli}}\ \emph {et~al.}(2013)\citenamefont
  {{Decarli}}, \citenamefont {{Dotti}}, \citenamefont {{Fumagalli}},
  \citenamefont {{Tsalmantza}}, \citenamefont {{Montuori}}, \citenamefont
  {{Lusso}}, \citenamefont {{Hogg}},\ and\ \citenamefont
  {{Prochaska}}}]{Decarli:2013xwa}%
  \BibitemOpen
  \bibfield  {author} {\bibinfo {author} {\bibfnamefont {R.}~\bibnamefont
  {{Decarli}}}, \bibinfo {author} {\bibfnamefont {M.}~\bibnamefont {{Dotti}}},
  \bibinfo {author} {\bibfnamefont {M.}~\bibnamefont {{Fumagalli}}}, \bibinfo
  {author} {\bibfnamefont {P.}~\bibnamefont {{Tsalmantza}}}, \bibinfo {author}
  {\bibfnamefont {C.}~\bibnamefont {{Montuori}}}, \bibinfo {author}
  {\bibfnamefont {E.}~\bibnamefont {{Lusso}}}, \bibinfo {author} {\bibfnamefont
  {D.~W.}\ \bibnamefont {{Hogg}}}, \ and\ \bibinfo {author} {\bibfnamefont
  {J.~X.}\ \bibnamefont {{Prochaska}}},\ }\href {\doibase 10.1093/mnras/stt831}
  {\bibfield  {journal} {\bibinfo  {journal} {\mnras}\ }\textbf {\bibinfo
  {volume} {433}},\ \bibinfo {pages} {1492} (\bibinfo {year} {2013})},\ \Eprint
  {http://arxiv.org/abs/arXiv:1305.4941} {arXiv:arXiv:1305.4941 [astro-ph.CO]}
  \BibitemShut {NoStop}%
\bibitem [{\citenamefont {{Tanaka}}(2013)}]{Tanaka:2013cva}%
  \BibitemOpen
  \bibfield  {author} {\bibinfo {author} {\bibfnamefont {T.~L.}\ \bibnamefont
  {{Tanaka}}},\ }\href {\doibase 10.1093/mnras/stt1164} {\bibfield  {journal}
  {\bibinfo  {journal} {\mnras}\ }\textbf {\bibinfo {volume} {434}},\ \bibinfo
  {pages} {2275} (\bibinfo {year} {2013})},\ \Eprint
  {http://arxiv.org/abs/1303.6279} {arXiv:1303.6279 [astro-ph.CO]} \BibitemShut
  {NoStop}%
\bibitem [{\citenamefont {Moesta}\ \emph {et~al.}(2012)\citenamefont {Moesta},
  \citenamefont {Alic}, \citenamefont {Rezzolla}, \citenamefont {Zanotti},\
  and\ \citenamefont {Palenzuela}}]{Moesta:2011bn}%
  \BibitemOpen
  \bibfield  {author} {\bibinfo {author} {\bibfnamefont {P.}~\bibnamefont
  {Moesta}}, \bibinfo {author} {\bibfnamefont {D.}~\bibnamefont {Alic}},
  \bibinfo {author} {\bibfnamefont {L.}~\bibnamefont {Rezzolla}}, \bibinfo
  {author} {\bibfnamefont {O.}~\bibnamefont {Zanotti}}, \ and\ \bibinfo
  {author} {\bibfnamefont {C.}~\bibnamefont {Palenzuela}},\ }\href@noop {}
  {\bibfield  {journal} {\bibinfo  {journal} {Astrophys.J.}\ }\textbf {\bibinfo
  {volume} {749}},\ \bibinfo {pages} {L32} (\bibinfo {year} {2012})},\ \Eprint
  {http://arxiv.org/abs/astro-ph/1109.1177} {arXiv:astro-ph/1109.1177 [gr-qc]}
  \BibitemShut {NoStop}%
\bibitem [{\citenamefont {Loken}\ \emph {et~al.}(2010)\citenamefont {Loken},
  \citenamefont {Gruner}, \citenamefont {Groer}, \citenamefont {Peltier},
  \citenamefont {Bunn}, \citenamefont {Craig}, \citenamefont {Henriques},
  \citenamefont {Dempsey}, \citenamefont {Yu}, \citenamefont {Chen},
  \citenamefont {Dursi}, \citenamefont {Chong}, \citenamefont {Northrup},
  \citenamefont {Pinto}, \citenamefont {Knecht},\ and\ \citenamefont
  {Zon}}]{Scinet}%
  \BibitemOpen
  \bibfield  {author} {\bibinfo {author} {\bibfnamefont {C.}~\bibnamefont
  {Loken}}, \bibinfo {author} {\bibfnamefont {D.}~\bibnamefont {Gruner}},
  \bibinfo {author} {\bibfnamefont {L.}~\bibnamefont {Groer}}, \bibinfo
  {author} {\bibfnamefont {R.}~\bibnamefont {Peltier}}, \bibinfo {author}
  {\bibfnamefont {N.}~\bibnamefont {Bunn}}, \bibinfo {author} {\bibfnamefont
  {M.}~\bibnamefont {Craig}}, \bibinfo {author} {\bibfnamefont
  {T.}~\bibnamefont {Henriques}}, \bibinfo {author} {\bibfnamefont
  {J.}~\bibnamefont {Dempsey}}, \bibinfo {author} {\bibfnamefont {C.-H.}\
  \bibnamefont {Yu}}, \bibinfo {author} {\bibfnamefont {J.}~\bibnamefont
  {Chen}}, \bibinfo {author} {\bibfnamefont {L.~J.}\ \bibnamefont {Dursi}},
  \bibinfo {author} {\bibfnamefont {J.}~\bibnamefont {Chong}}, \bibinfo
  {author} {\bibfnamefont {S.}~\bibnamefont {Northrup}}, \bibinfo {author}
  {\bibfnamefont {J.}~\bibnamefont {Pinto}}, \bibinfo {author} {\bibfnamefont
  {N.}~\bibnamefont {Knecht}}, \ and\ \bibinfo {author} {\bibfnamefont {R.~V.}\
  \bibnamefont {Zon}},\ }\href {\doibase 10.1088/1742-6596/256/1/012026}
  {\bibfield  {journal} {\bibinfo  {journal} {J. Phys.: Conf. Ser.}\ }\textbf
  {\bibinfo {volume} {256}},\ \bibinfo {pages} {012026} (\bibinfo {year}
  {2010})}\BibitemShut {NoStop}%
\end{thebibliography}%


\end{document}